\documentclass[10pt,aps,showpacs,nofootinbib,prd,aps,epsf,floats,
               amsmath,amssymb,amsfonts,axodraw,twocolumn,floatfix]{revtex4}
\usepackage{amsmath, amssymb}
\newcommand{\mathsym}[1]{{}}

\usepackage{graphicx}
\usepackage{amsmath}
\usepackage{amssymb}
\usepackage{bm}
\newcommand{\be}{\begin{equation}}
\newcommand{\ee}{\end{equation}}
\newcommand{\bea}{\begin{eqnarray}}
\newcommand{\eea}{\end{eqnarray}}

\newcommand{\rem}[1]{}
\newsavebox{\PSLASH}
 \sbox{\PSLASH}{$p$\hspace{-1.8mm}/}
 
\renewcommand{\theequation}{\thesection.\arabic{equation}}
\newcounter{saveeqn}
\newcommand{\add}{\addtocounter{equation}{1}}
\newcommand{\alpheqn}{\setcounter{saveeqn}{\value{equation}}%
\setcounter{equation}{0}%
\renewcommand{\theequation}{\mbox{\thesection.\arabic{saveeqn}{\alph{equation}}}}}
\newcommand{\reseteqn}{\setcounter{equation}{\value{saveeqn}}%
\renewcommand{\theequation}{\thesection.\arabic{equation}}}

 \newsavebox{\notrightarrow}
 \sbox{\notrightarrow}{$\to$\hspace{-4mm}/}
 
 \newsavebox{\PARTIALSLASH}
 \sbox{\PARTIALSLASH}{$\partial$\hspace{-1.6mm}/}
 
 \newsavebox{\ASLASH}
 \sbox{\ASLASH}{$A$\hspace{-2.1mm}/}
 
 \newsavebox{\KSLASH}
 \sbox{\KSLASH}{$k$\hspace{-1.8mm}/}
 
 \newsavebox{\LSLASH}
 \sbox{\LSLASH}{$\ell$\hspace{-1.8mm}/}
 
 \newsavebox{\QSLASH}
 \sbox{\QSLASH}{$q$\hspace{-1.8mm}/}
 
 \newsavebox{\DSLASH}
 \sbox{\DSLASH}{$D$\hspace{-2.2mm}/}
 
 \newsavebox{\DbfSLASH}
 \sbox{\DbfSLASH}{${\mathbf D}$\hspace{-2.8mm}/}
 
 \newsavebox{\DELVECRIGHT}
 \sbox{\DELVECRIGHT}{$\stackrel{\rightarrow}{\partial}$}
 
 \newcommand{\blue}{\IfColor{\textCadetBlue}{}}
\newcommand{\black}{\IfColor{\textBlack}{}}
\newcommand{\red}{\IfColor{\textRed}{}}
\newcommand{\green}{\IfColor{\textOliveGreen}{}}
\newcommand{\lila}{\IfColor{\textRedViolet}{}}

\newcommand{\nc}{\newcommand}
\nc{\Gz}{\fullg}
\nc{\Gvc}{\boldsymbol{G}^c}
\nc{\Gam}{\boldsymbol{\Gamma}}
\nc{\Sig}{\boldsymbol{\Sigma}}
\nc{\TS}{\tilde{\Sig}}
\nc{\TG}{\tilde{\mbox{\boldmath $G$}}}
\nc{\gam}{\boldsymbol{\gamma}}
\nc{\alp}{\boldsymbol{\alpha}}
\nc{\hs}{\hspace*{1mm}}
\nc{\ti}{\tilde}
\nc{\p}{\mathcal{P}}
\nc{\si}{\Sigma}
\nc{\beq}{\begin{eqnarray}}
\nc{\eeq}{\end{eqnarray}}
\nc{\ba}{\begin{array}}
\nc{\ea}{\end{array}}
\nc{\la}{\label}
\nc{\no}{\nonumber}
\nc{\btab}{\begin{tabular}}
\nc{\etab}{\end{tabular}}
\nc{\ci}{\cite}
\nc{\ps}{\slash\hspace{-0.20cm}p}
\nc{\ells}{\slash\hspace{-0.23cm}\ell}
\nc{\ks}{\slash\hspace{-0.23cm}k}
\nc{\Ss}{\slash\hspace{-0.23cm}\Sigma}
\nc{\ms}{\slash\hspace{-0.25cm}\mathfrak{P}}
\nc{\trace}{{\rm Tr\,}}
\nc{\se}{\section}
\nc{\ptildes}{\slash\hspace{-0.20cm}\widetilde{p}}
\nc{\ktildes}{\slash\hspace{-0.18cm}\widetilde{k}}
\begin{document}
\title{
Magnetized plasminos in cold and hot QED plasmas}
\author{N. Sadooghi}\email{sadooghi@physics.sharif.ir}
\author{F. Taghinavaz}\email{taghinavaz@physics.sharif.ir}
\affiliation{Department of Physics, Sharif University of Technology,
P.O. Box 11155-9161, Tehran-Iran}
\begin{abstract}
The complete quasi-particle spectrum of a magnetized electromagnetic plasma is systematically explored at zero and nonzero temperatures. To this purpose, the general structure of the one-loop corrected propagator of magnetized fermions is determined, and the dispersion relations arising from the pole of this propagator are numerically solved. It turns out that in the lowest Landau level, where only one spin direction is allowed, the spectrum consists of one positively (negatively) charged fermionic mode with positive (negative) spin. In contrast, in higher Landau levels, as an indirect consequence of the double spin degeneracy of fermions, the spectrum consists of two massless collective modes with left- and right-chiralities. The mechanism through which these new collective excitations are created in a uniform magnetic field is similar to the production mechanism of dynamical holes (plasminos) at finite temperature and zero magnetic fields. Whereas cold magnetized plasminos appear for moderate magnetic fields and for all positive momenta of propagating fermions, hot magnetized plasminos appear only in the limit of weak magnetic fields and soft momenta.
\end{abstract}
  \maketitle
\section{Introduction}\label{introduction}\label{sec1}
Research on matter under extreme conditions has provided new insight in the physics of heavy-ion collisions (HIC) and the astrophysics of compact stars. Extreme conditions consist of high temperature, large density and/or the presence of intense magnetic fields. The latter affects, in particular, the phase diagram of Quantum Chromodynamics (QCD), and plays a significant role in the dynamics of relativistic fermions at zero and nonzero temperatures (for recent reviews, see \cite{kharzeev2013,andersen2014}). The phenomena driven by external magnetic fields have also various applications in quasi-relativistic condensed matter systems, such as graphene and Dirac semi-metals (see \cite{shovkovy2015} and the literature therein).
\par
Most theoretical studies deal with the idealized limit of constant and homogeneous magnetic fields. Standard field theoretical methods thus lead to the exact solution of the relativistic Dirac equation in a uniform magnetic field. One of the main consequences of the presence of constant magnetic fields is a certain dimensional reduction of the dynamics of propagating fermions in the lowest Landau level (LLL). The latter leads to a dynamical mass generation, and enhances the production of chiral condensates. This phenomenon, which is known as magnetic catalysis \cite{klimenko1992, miransky1995}, modifies, in particular, the phase diagram of QCD in the chiral and color superconductivity phase \cite{fayazbakhsh2010}. Another effect is the appearance of certain anisotropies in the dynamics of magnetized fermions in the longitudinal and transverse directions with respect to the direction of the external magnetic field. These anisotropies include those in the neutrino emission from magnetars \cite{duncan1992}, or anisotropies arising in the group velocities, refraction indices and decay constants of mesons in hot and magnetized quark matter \cite{fayazbakhsh2012}. Recently, the anisotropy appearing in the equation of states of magnetized quark matter is studied in \cite{menezes2015}.
\par
A similar privileged reference frame is also defined by a heat bath, and affects, in particular, the quasi-particle spectrum of electromagnetic and quark-gluon plasmas at finite temperature. Non-trivial bosonic and fermionic collective excitations, such as plasmons and plasminos are shown to be dynamically generated in hot QED and QCD plasmas, in addition to the normal bosonic and fermionic modes \cite{klimov1982, weldon1982}. In particular, plasminos are known to be collective excitations that arise as one of the poles of the one-loop corrected fermion propagator at finite temperature \cite{weldon1989}.  In the chiral limit, plasminos are characterized by their negative helicity to chirality ratio, in opposite to that of normal modes. They are intensively studied in the context of Yukawa theory, QED and QCD \cite{plasmino-rest, plasmino-transport, plasmino-damping, plasmino-production}. In \cite{plasmino-transport}, for instance, it is shown that the contributions from plasminos modify the transport properties of relativistic plasmas in- and out-of equilibrium. They also lead to the appearance of sharp structures (singularities and gaps) in the decay \cite{plasmino-damping} and production rates \cite{plasmino-production} of particles produced in relativistic and ultra-relativistic collisions. These structures provide unique signatures for the presence of deconfined collective quarks in the plasma of quarks and gluons.
\par
In principle, nontrivial collective modes can also be created in magnetized plasmas through the same mechanism as the one leading to the appearance of collective modes at finite temperature and zero magnetic fields. It is the purpose of the present paper to look for possible dynamical generation of fermionic excitations (plasminos) in electromagnetic plasmas in the presence of constant magnetic fields, and study their properties at zero and nonzero temperatures. Apart from various other applications, plasminos may play an important role in the physics of HICs. Very strong magnetic fields, which, according to recent experimental results, are believed to be created in early stage of non-central HICs \cite{mclerran2007} may affect, among others, the energy dispersion of deconfined quarks. The latter are believed to be produced in the quark-gluon plasma in the same stage as the magnetic fields. It is therefore important to explore the quasi-particle spectrum of Dirac equation in the presence of external magnetic fields at finite temperature, and study the properties of the potentially created collective modes under these conditions.
\par
In this paper, we will particularly focus on the mechanism of the production of plasminos in the presence of external magnetic fields. We will, in particular, determine the general structure of the one-loop corrected propagator of magnetized fermions, and, following the historical path that has led to plasminos at finite temperature \cite{weldon1989}, solve the dispersion relations arising from the pole of this propagator. To this purpose, we will first determine the general structure of the tree-level fermion propagator in the presence of a constant magnetic field by making use of the Ritus eigenfunction method \cite{ritus1972}. The free fermion propagator will then be combined with the one-loop self-energy of magnetized fermions. This will result in the desired one-loop corrected propagator of magnetized fermions. We will show that, in contrast to LLL, the dressed fermion propagator in higher Landau levels (HLL) can be decomposed into two parts, each of them leading to a separate energy dispersion relation for magnetized fermions. This fact, which eventually leads to the appearance of \textit{magnetized plasminos}, is an indirect consequence of the double spin degeneracy in HLL, in contrast to LLL, which is occupied with only one positive or negative fermion with positive or negative spin. We will show that at finite temperature and in the limit of weak magnetic fields, where HLLs have also to be taken into account, the spectrum consists of two massless collective modes with left- and right-chiralities. Moreover, it can be shown that whereas \textit{cold magnetized plasminos} appear for all positive momenta of propagating fermions and moderate magnetic fields, \textit{hot magnetized plasminos} appear only in the limit of soft momenta and weak magnetic fields.
\par
The organization of this paper is a follows: In Sect. \ref{sec2}, we will review two independent topics related to the main subject of the paper: In Sec. \ref{sec2a}, we will first show how thermal plasminos arise from the pole of the one-loop corrected fermion propagator in the Hard-Thermal Loop (HTL) approximation \cite{pisarski1990}. In Sec. \ref{sec2b}, we will then review the Ritus eigenfunction method \cite{ritus1972}, and present the general structure of the free propagator of magnetized fermions in the momentum space.
The general structure of one-loop fermion self-energy at zero and nonzero temperatures will be derived in Secs. \ref{sec3a} and \ref{sec3b}, respectively. In Sec. \ref{sec3b}, in particular, a certain HTL approximation in a constant magnetic field will be introduced, and the one-loop fermion self-energy will be presented in terms of a number of coefficients up to some integrations and a summation over Landau levels. In Sec. \ref{sec4}, the general structure of the one-loop corrected propagator of magnetized fermions will be determined by combining the free fermion propagator from Sec. \ref{sec2b} and the one-loop fermion self-energy from Sec. \ref{sec3}. Here, two different cases of massive fermions in LLL and massless fermions in HLL will be considered, and the properties of the fermionic excitations in these two cases will be systematically studied. In Sec. \ref{sec5}, after numerically determining the aforementioned coefficients, appearing in the one-loop fermion self-energy, we will study the spectrum of the fermionic excitations at finite [Sec. \ref{sec5a}] and zero [Sec. \ref{sec5b}] temperatures. Section \ref{sec6} is devoted to a brief summary of our results and a number of concluding remarks.
\section{Review material}\label{sec2}
The main goal of the present paper is to determine the spectrum of Dirac particles in a constant magnetic field at zero and nonzero temperatures. To this purpose, and in order to fix our notations, we will briefly review, in this section, two independent topics related to the main subject of this paper. In Sec. \ref{sec2a}, we will first repeat the computation presented in \cite{weldon1989}, and introduce the plasmino excitations in a hot QED plasma. To do this, we will compute the one-loop correction to the fermion self-energy at finite temperature in a HTL approximation \cite{lebellac}, and eventually determine the Dirac spectrum at finite temperature by solving the corresponding energy dispersion relation. In Sec. \ref{sec2b}, we will then briefly review the Ritus eigenfunction method \cite{ritus1972}, and after presenting the free propagator of magnetized fermions in the coordinate space, will determine its general structure in the momentum space. Our review will mainly base on the Ritus eigenfunction method, presented in \cite{fukushima2009}, and generalized in \cite{fayazbakhsh2012} for a multi-flavor system of charged fermions.
\subsection{Plasminos in a hot QED plasma}\label{sec2a}
In \cite{weldon1989}, the Dirac spectrum is determined for a hot QCD plasma. In this paper, however, we will focus, for simplicity, on the $U(1)$ subgroup of the $SU(N)$ gauge group. The difference between our $U(1)$ and the $SU(N)$ case, discussed in \cite{weldon1982, weldon1989}, is a certain prefactor $C_{F}$, the quadratic Casimir constant of the fermion representation, in the corresponding expression to the one-loop fermion self-energy. For the Abelian $U(1)$ gauge group, $C_{F}=1$ and for the fundamental representation of $SU(N)$ gauge group, $C_{F}=\frac{N^{2}-1}{2N}$.
\par
\begin{figure}[hbt]
\includegraphics[width=5cm,height=2cm]{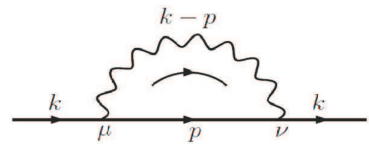}
\caption{One-loop fermion self-energy}\label{fig1}
\end{figure}
\par
\begin{figure}[h]
\includegraphics[width=8cm, height=5.5cm]{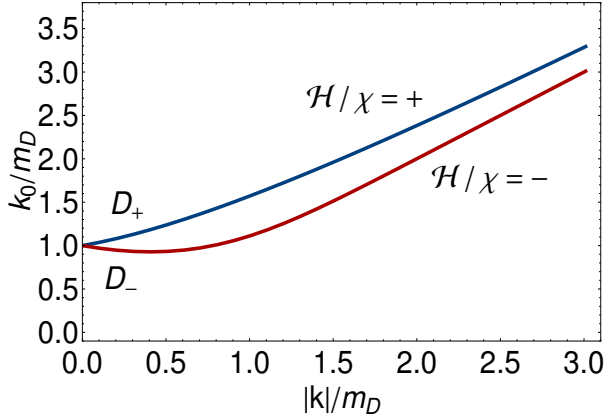}
\caption{(color online). Two different energy branches, corresponding to the energy dispersion relations ${D}_{+}=0$ of particles (blue curve) and ${D}_{-}=0$ of holes (red curve). The denominators ${D}_{\pm}$ are defined in (\ref{S13}). The particles/holes, whose energy dispersion relations are demonstrated by these two branches, have positive/negative helicity (${\cal{H}}$) to chirality ($\chi$) ratio.}\label{fig2}
\end{figure}

Let us start by determining the one-loop fermion self-energy of massless fermions at finite temperature.\footnote{At high enough temperature, the mass of fermions can be neglected. } Having in mind that the interaction term of fermions and photons in QED is given by ${\cal{L}}_{int}=-e\bar{\psi}\gamma_{\mu}A_{\mu}\psi$, the fermion self-energy $\Sigma(k)$ at finite temperature is given by [see Fig. \ref{fig1}]
\begin{eqnarray}\label{S1}
\Sigma(k)&=&e^{2}T\sum\limits_{n=-\infty}^{+\infty}\int \frac{d^{3}p}{(2\pi)^{3}}D_{\mu\nu}(k-p)\gamma^{\mu}S(p)\gamma^{\nu}\nonumber\\
&=&-2e^{2}T\sum\limits_{n=-\infty}^{+\infty}\int \frac{d^{3}p}{(2\pi)^{3}}\frac{\ps}{(k-p)^{2}p^{2}}.
\end{eqnarray}
Here, $D_{\mu\nu}(k)\equiv\frac{g^{\mu\nu}}{k^{2}}$ and $S(p)\equiv\frac{1}{\ps}$ are free photon and fermion propagators, respectively.\footnote{In what follows, we will denote the free and dressed fermion propagators with $S$ and ${\cal{S}}$, respectively.}  In the imaginary time formalism, $p_{0}$ is to be replaced by $i\omega_{n}$, where $\omega_{n}\equiv (2n+1)\pi T$ is the fermionic Matsubara frequency. Since in the presence of a hot medium, the relativistic Lorentz invariance is broken by introducing the reference frame corresponding to the heat bath, it is necessary to perform the following separation between the temporal and spacial components of $\Sigma(k)$
\begin{eqnarray}\label{S2}
\Sigma(k)=\gamma_{0}\Sigma_{t}(k)-\boldsymbol{\gamma}\cdot\boldsymbol{\Sigma}_{s}(k).
\end{eqnarray}
Introducing $E_{1}\equiv |\mathbf{p}|$, $E_{2}\equiv |\mathbf{p}-\mathbf{k}|$, and
using
\begin{eqnarray}\label{S3}
\Delta_{f}(p_{0}, E_{1})&\equiv&\frac{1}{p_{0}^{2}+E_{1}^{2}},\nonumber\\
\Delta_{b}(k_{0}-p_{0},E_{2})&\equiv&\frac{1}{(k_{0}-p_{0})^{2}+E_{2}^{2}},
\end{eqnarray}
the temporal and spacial components of $\Sigma(k)$ are given by
\begin{eqnarray}\label{S4}
\hspace{-0.4cm}\Sigma_{t}(k)&=&-2e^{2}\int\frac{d^{3}p}{(2\pi)^{3}}~{\cal{I}}_{t}(k,\mathbf{p}),\nonumber\\
\hspace{-0.4cm}\boldsymbol{\Sigma}_{s}(k)&=&-2e^{2}\int\frac{d^{3}p}{(2\pi)^{3}}~\mathbf{p}\ {\cal{I}}_{s}(k,\mathbf{p}),
\end{eqnarray}
with
\begin{eqnarray}\label{S5}
\hspace{-0.5cm}{\cal{I}}_{t}&\equiv& T\sum\limits_{n=-\infty}^{\infty}i\omega_{n}\Delta_{f}(i\omega_{n},E_{1})\Delta_{b}(k_{0}-i\omega_{n},E_{2}),\nonumber\\
\hspace{-0.5cm}{\cal{I}}_{s}&\equiv&T\sum\limits_{n=-\infty}^{\infty}\Delta_{f}(i\omega_{n},E_{1})\Delta_{b}(k_{0}-i\omega_{n},E_{2}).
\end{eqnarray}
Following the standard method introduced in \cite{lebellac}, the summation over Matsubara frequencies can be performed by making use of
\begin{eqnarray}\label{S6}
{\cal{I}}_{t}(k,\mathbf{p})&=&-\sum_{s_{1},s_{2}=\pm}\frac{s_{2}}{4E_{2}}\frac{1+f_{b}(s_{2}E_{2})-f_{f}(s_{1}E_{1})}{k_{0}-s_{1}E_{1}-s_{2}E_{2}},\nonumber\\
{\cal{I}}_{s}(k,\mathbf{p})&=&-\sum_{s_{1},s_{2}=\pm}\frac{s_{1}s_{2}}{4E_{1}E_{2}}\frac{1+f_{b}(s_{2}E_{2})-f_{f}(s_{1}E_{1})}{k_{0}-s_{1}E_{1}-s_{2}E_{2}}.\nonumber\hspace{-0.3cm}\\
\end{eqnarray}
Here, the fermionic and bosonic distribution functions are defined by
\begin{eqnarray}\label{S7}
f_{f}(\ell_{0})&\equiv&\frac{1}{e^{\beta \ell_{0}}+1},\nonumber\\
f_{b}(\ell_{0})&\equiv&\frac{1}{e^{\beta \ell_{0}}-1},\
\end{eqnarray}
with $\beta\equiv T^{-1}$, and $\ell_{0}=\pm|\boldsymbol{\ell}|$ for massless fermions. Replacing the above expressions in (\ref{S4}), performing an appropriate expansion in ${|\mathbf{k}|}\ll {|\mathbf{p}|}$ with $|\mathbf{p}|\sim T$,\footnote{In this approximation $E_{2}$ is, in particular, replaced by $E_{2}\approx |\mathbf{p}|-|\mathbf{k}|\cos\theta$, with $\theta$ the angle between $\mathbf{k}$ and $\mathbf{p}$.} and keeping only the leading $T^{2}$ contributions from the resulting expressions, $\Sigma_{t}$ and $\boldsymbol{\Sigma}_{s}$ in this HTL approximation are given by
\begin{eqnarray}\label{S8}
\hspace{-0.4cm}\Sigma_{t}(k)&=&\frac{m_{D}^{2}}{2|\mathbf{k}|}\ln\left(\frac{k_{0}+|\mathbf{k}|}{k_{0}-|\mathbf{k}|}\right),\nonumber\\
\hspace{-0.4cm}\boldsymbol{\Sigma}_{s}(k)&=&\frac{m_{D}^{2}}{|\mathbf{k}|}\left(1-\frac{k_{0}}{2|\mathbf{k}|}\ln\left(\frac{k_{0}+|\mathbf{k}|}{k_{0}-|\mathbf{k}|}\right)\right)\widehat{\mathbf{k}}.
\end{eqnarray}
Here, $m_{D}^{2}\equiv \frac{e^{2}T^{2}}{8}$ is the Debye mass. To perform the integration over $\mathbf{p}$ (here denoted by $p$), the following integrals are used
\begin{eqnarray}\label{S9}
\int_{0}^{\infty}dp~ p f_{f}(p)&=&\frac{\pi^{2}T^{2}}{6},\nonumber\\
\int_{0}^{\infty}dp~ p f_{b}(p)&=&\frac{\pi^{2}T^{2}}{12}.
\end{eqnarray}
Plugging at this stage the results from (\ref{S8}) in (\ref{S2}), and combining the resulting expression for $\Sigma(k)$ with the inverse of the free fermion propagator, $S^{-1}(k)=\gamma_{0}k_{0}-\boldsymbol{\gamma}\cdot\widehat{\mathbf{k}}$,  the inverse of the one-loop corrected fermion propagator, ${\cal{S}}^{-1}(k)=S^{-1}(k)-\Sigma(k)$ reads
\begin{eqnarray}\label{S10}
{\cal{S}}^{-1}(k)=\gamma_{0}A_{0}-\boldsymbol{\gamma}\cdot\widehat{\mathbf{k}}~ A_{s},
\end{eqnarray}
with
\begin{eqnarray}\label{S11}
\hspace{-0.6cm}A_{0}&=&k_{0}-\frac{m_{D}^{2}}{2|\mathbf{k}|}\ln\left(\frac{k_{0}+|\mathbf{k}|}{k_{0}-|\mathbf{k}|}\right),\nonumber\\
\hspace{-0.6cm}A_{s}&=&|\mathbf{k}|+\frac{m_{D}^{2}}{|\mathbf{k}|}\left(1-\frac{k_{0}}{2|\mathbf{k}|}\ln\left(\frac{k_{0}+|\mathbf{k}|}{k_{0}-|\mathbf{k}|}\right)\right).
\end{eqnarray}
The dressed fermion propagator up to one-loop order is therefore given by
\begin{eqnarray}\label{S12}
{\cal{S}}(k)=\frac{\gamma_{0}-
\boldsymbol{\gamma}\cdot\widehat{\mathbf{k}}}{2{D}_{+}}+\frac{\gamma_{0}+\boldsymbol{\gamma}\cdot\widehat{\mathbf{k}}}{2{D}_{-}},
\end{eqnarray}
with
\begin{eqnarray}\label{S13}
{D}_{\pm}\equiv A_{0}\mp A_{s},
\end{eqnarray}
and $A_{0/s}$ from (\ref{S11}). To determine the spectrum of Dirac particles, we set either ${D}_{\pm}=0$ or $\mbox{det}[{\cal{S}}^{-1}(k)]=0$, and arrive at two solutions $A_{0}=\pm A_{s}$, or more explicit\-ly at two different energy branches for particles ($p$) and dynamical holes ($h$) \cite{weldon1989},
\begin{eqnarray}\label{S14}
E_{p}&=&|\mathbf{k}|+\frac{m_{D}^{2}}{|\mathbf{k}|}\bigg[1+\frac{1}{2}\left(1-z_{p}\right)\ln\left(\frac{z_{p}+1}{z_{p}-1}\right)\bigg], \nonumber\\
E_{h}
&=&-\bigg\{|\mathbf{k}|+\frac{m_{D}^{2}}{|\mathbf{k}|}\bigg[1-\frac{1}{2}\left(1+z_{h}\right)\ln\left(\frac{z_{h}+1}{z_{h}-1}\right)\bigg]\bigg\}, \nonumber\\
\end{eqnarray}
where $z_{p/h}\equiv \frac{E_{p/h}}{|\mathbf{k}|}$.\footnote{Here, $k_{0}$ is replaced by $E_{p}$ [first expression in (\ref{S14})] and $E_{h}$ [second expression in (\ref{S14})].} Replacing the expression on the right hand side (r.h.s.) of $E_{h}$ with $$E_{h}=|\mathbf{k}|\coth\left(\mathbf{k}^{2}+\frac{|\mathbf{k}|}{|\mathbf{k}|+E_{h}}\right),$$ the dependence of dimensionless quantities $k_{0}/m_{D}$ for $k_{0}=E_{p}$ and $k_{0}=E_{h}$ on $|\mathbf{k}|/m_{D}$ can be determined numerically. Two different energy branches corresponding to ${D}_{+}$ and ${D}_{-}$ arise. They are demonstrated in Fig. \ref{fig2}.
\par
In what follows, we will argue that the energy branch arising from ${D}_{+}=0$ (${D}_{-}=0$) [blue (red) curve in Fig. \ref{fig2}] corresponds to a particle (hole) with positive (negative) helicity (${\cal{H}}$) to chirality ($\chi$) ratio. To do this, let us determine the eigenvectors of the numerators in (\ref{S12}),
\begin{eqnarray}\label{S15}
\hspace{-0.2cm}{\cal{N}}_{\pm}\equiv\gamma_{0}\mp\boldsymbol{\gamma}\cdot\widehat{\mathbf{k}}=\left(
\begin{array}{cc}
0&1\mp\boldsymbol{\sigma}\cdot\widehat{\mathbf{k}}\\
1\pm\boldsymbol{\sigma}\cdot\widehat{\mathbf{k}}&0
\end{array}
\right).
\end{eqnarray}
Here, $\boldsymbol{\sigma}=(\sigma_{1},\sigma_{2},\sigma_{3})$ are the three Pauli matrices.\footnote{To derive (\ref{S15}), we have chosen the following chiral representation of Dirac $\gamma$-matrices
\begin{eqnarray*}
\gamma^{\mu}=\left(
\begin{array}{cc}
0&\sigma^{\mu}\\
\bar{\sigma}^{\mu}&0
\end{array}
\right),\qquad \gamma^{5}=\left(
\begin{array}{cc}
-1&0\\
0&1
\end{array}
\right),
\end{eqnarray*}
with $\sigma^{\mu}=(1,\boldsymbol{\sigma})$ and $\bar{\sigma}^{\mu}=(1,-\boldsymbol{\sigma})$.
} For ${D}_{+}$ branch, ${\cal{N}}_{+}$ has the following two nontrivial eigenvectors
\begin{eqnarray}\label{S16}
W_{+}^{(1)}&=&\left(0,0,\frac{k_{3}+|\mathbf{k}|}{k_{1}+ik_{2}},1\right),\nonumber\\
W_{+}^{(2)}&=&\left(\frac{k_{3}-|\mathbf{k}|}{k_{1}+ik_{2}},1,0,0\right).
\end{eqnarray}
Defining the helicity
\begin{eqnarray}\label{S17}
{\cal{H}}=\boldsymbol{\Sigma}\cdot\widehat{\mathbf{k}},
\end{eqnarray}
with $\boldsymbol{\Sigma}\equiv \mbox{diag}\left(\boldsymbol{\sigma},\boldsymbol{\sigma}\right)$, and right- ($R$) as well as left- ($L$) chirality operators
\begin{eqnarray}\label{S18}
{\cal{P}}_{R}\equiv\frac{1+\gamma_{5}}{2},\qquad
{\cal{P}}_{L}\equiv\frac{1-\gamma_{5}}{2},
\end{eqnarray}
it turns out that $W_{+}^{(i)}, i=1,2$ satisfy
\begin{eqnarray}\label{S19}
{\cal{P}}_{R}W_{+}^{(1)}=W_{+}^{(1)},\qquad {\cal{H}}W_{+}^{(1)}=+W_{+}^{(1)},\nonumber\\
{\cal{P}}_{L}W_{+}^{(2)}=W_{+}^{(2)},\qquad {\cal{H}}W_{+}^{(2)}=-W_{+}^{(2)}.
\end{eqnarray}
Having in mind that right (left)-handed particles have positive (negative) chirality, $\chi=+1$ ($\chi=-1$), the relations in (\ref{S19}) show that fermions, whose energy dispersion relation is given by ${D}_{+}$ branch in Fig. \ref{fig2}, have a positive helicity to chirality ratio, $\frac{{\cal{H}}}{\chi}=+1$.  They are therefore particles. Similarly, it can be shown that for ${D}_{-}$, the eigenvectors of ${\cal{N}}_{-}$ are given by
\begin{eqnarray}\label{S20}
W_{-}^{(1)}&=&\left(0,0,\frac{k_{3}-|\mathbf{k}|}{k_{1}+ik_{2}},1\right),\nonumber\\
W_{-}^{(2)}&=&\left(\frac{k_{3}+|\mathbf{k}|}{k_{1}+ik_{2}},1,0,0\right),
\end{eqnarray}
which satisfy
\begin{eqnarray}\label{S21}
{\cal{P}}_{R}W_{-}^{(1)}=W_{-}^{(1)},\qquad {\cal{H}}W_{-}^{(1)}=-W_{-}^{(1)},\nonumber\\
{\cal{P}}_{L}W_{-}^{(2)}=W_{-}^{(2)},\qquad {\cal{H}}W_{-}^{(2)}=+W_{+}^{(2)}.
\end{eqnarray}
These relations indicate that fermionic mode, whose energy dispersion relation is given by ${D}_{-}$ branch in Fig. \ref{fig2}, have a negative helicity to chirality ratio, $\frac{{\cal{H}}}{\chi}=-1$.  According to the standard terminology of relativistic quantum mechanics, they are therefore holes. In conclusion, the spectrum of Dirac particles at high enough temperature, consists of two different soft excitations with positive and negative helicity to chirality ratio. In Sec. \ref{sec3}, the above HTL approximation will be used to determine the general structure of one-loop self-energy of  fermions in the limit of soft momenta and weak magnetic fields. This will eventually lead to hot magnetized plasminos in these limits [see Sec. \ref{sec5}].
\subsection{Magnetized fermions in a cold QED plasma; The Ritus eigenfunction method}\label{sec2b}
In this section, we will solve the Dirac equation of positively and negatively charged fermions in the presence of an external magnetic field
\begin{eqnarray}\label{S22}
(\gamma\cdot \Pi^{(q)}-m_{q})\psi=0,
\end{eqnarray}
with $\Pi_{\mu}^{(q)}\equiv i\partial_{\mu}+eq A_{\mu}^{\mbox{\tiny{ext}}}$, and $m_{q}$ the fermionic mass. Here, $e>0$ is the unit electric charge, and $q=\pm 1$ indicate the positive and negative charges of Dirac fermions.  The gauge field  $A_{\mu}^{\mbox{\tiny{ext}}}=(0,0,Bx_{1},0)$ is chosen so that it yields a magnetic field $\mathbf{B}$, aligned in the positive third direction, $\mathbf{B}=B\mathbf{e}_{3}$ with $B>0$.
In the method originally introduced by V. I. Ritus in \cite{ritus1972}, and generalized recently in \cite{fukushima2009, fayazbakhsh2012} for multi-flavor systems, (\ref{S22}) can be solved by making use of the Ansatz $\psi^{(q)}=\mathbb{E}_{p}^{(q)}u_{\tilde{p}}$ for a Dirac fermion with charge $q$. The Ritus eigenfunction $\mathbb{E}_{p}^{(q)}$ then satisfies
\begin{eqnarray}\label{S23}
(\gamma\cdot \Pi)\mathbb{E}_{p}^{(q)}=\mathbb{E}_{p}^{(q)}(\gamma\cdot \widetilde{p}_{p}),
\end{eqnarray}
and the free Dirac spinor $u_{\tilde{p}}$ satisfies $(\ptildes_{p}-m )u_{\tilde{p}}=0$. As it turns out, the Ritus momentum $\tilde{p}_{p}$ is given by
\begin{eqnarray}\label{S24}
\widetilde{p}_{p}\equiv (p_{0},0,-s_{q}\sqrt{2p|q eB|},p_{3}),
\end{eqnarray}
where $p$ labels the Landau levels in the external magnetic field $B$, and $s_{q}\equiv \mbox{sgn}(qeB)$. The Ritus eigenfunction for Dirac fermions with charge $q$ can be derived from (\ref{S23}) and (\ref{S24}), and reads
\begin{eqnarray}\label{S25}
\mathbb{E}_{p}^{(q)}(x,p)=e^{-i\bar{p}\cdot \bar{x}}P_{p}^{(q)}(x_{1},p_{2}),
\end{eqnarray}
where $\bar{p}\equiv (p_{0},0,p_{2},p_{3})$, $\bar{x}\equiv (x_{0},0,x_{2},x_{3})$, and
\begin{eqnarray}\label{S26}
P_{p}^{(q)}\equiv {\cal{P}}_{+}^{(q)}f_{p}^{+s_{q}}+\Pi_{p}{\cal{P}}_{-}^{(q)}f_{p}^{-s_{q}},
\end{eqnarray}
with ${\cal{P}}_{+}^{(\pm)}={\cal{P}}_{\pm}$ and ${\cal{P}}_{-}^{(\pm)}={\cal{P}}_{\mp}$, as well as
\begin{eqnarray}\label{S27}
{\cal{P}}_{\pm}\equiv\frac{1\pm i\gamma_{1}\gamma_{2}}{2}.
\end{eqnarray}
In (\ref{S26}), $\Pi_{p}\equiv 1-\delta_{p0}$ considers the spin degeneracy in the Landau level $p$ (see below). For $eB>0$, $P_{p}^{(\pm)}$ for positively ($q=+1$) and negatively ($q=-1$) charged particles  read
\begin{eqnarray}\label{S28}
P_{p}^{(\pm)}&=&{\cal{P}}_{\pm}f_{p}^{\pm}+\Pi_{p}{\cal{P}}_{\mp}f_{p}^{\mp},
\end{eqnarray}
respectively. Here, $f_{p}^{\pm}(x_{1},p_{2})$ are given by
\begin{eqnarray}\label{S29}
\begin{array}{rclcccrcl}
f_{p}^{+}&=&\Phi_{p}(x_{1}-\ell_{q}^{2}p_{2}),&~~&\mbox{for}&~~&p&=&0,1,2,\cdots,\\
f_{p}^{-}&=&\Phi_{p-1}(x_{1}-\ell_{q}^{2}p_{2}),&~~&\mbox{for}&~~&p&=&1,2,\cdots,\\
\end{array}\hspace{-0.3cm}\nonumber\\
\end{eqnarray}
with
\begin{eqnarray}\label{S30}
\Phi_{p}(x)\equiv a_{p}\exp\left(-\frac{x^{2}}{2\ell_{q}^{2}}\right)H_{p}\left(\frac{x}{\ell_{q}}\right),
\end{eqnarray}
$\ell_{q}\equiv (|qeB|)^{-1/2}$ and the normalization factor $a_{p}\equiv \left(2^{p}p!\sqrt{\pi}\ell_{q}\right)^{-1/2}$.
In (\ref{S30}), $H_{p}(x)$ is the Hermite polynomial of order $p$. Using the above results, it is easy to determine the free propagator of Dirac fermions in the coordinate space at zero temperature \cite{fukushima2009, fayazbakhsh2012},
\begin{eqnarray}\label{S31}
\lefteqn{\hspace{-0.7cm}S^{(q)}(x,y)=\sum\limits_{p=0}^{\infty}\int\frac{d^{3}\bar{p}}{(2\pi)^{3}}e^{i\bar{p}\cdot(\bar{x}-\bar{y})}
}\nonumber\\
&&\hspace{-0.7cm}\times P_{p}^{(q)}(x_{1},p_{2})~\frac{1}{\ptildes_{p}-m_{q}}~P_{p}^{(q)}(y_{1},p_{2}).
\end{eqnarray}
Here, $\widetilde{p}_{p}$ and $P_{p}^{(q)}$ for Dirac fermions with charge $q$ are given in (\ref{S24}) and (\ref{S26}), respectively.
\par
In what follows, we will derive the general structure of free fermion propagator in the presence of a constant magnetic field at zero temperature in the momentum space. To do this, let us first consider the free fermion propagator (\ref{S31}) in the coordinate space. The free fermion propagator in the momentum space is given by performing an appropriate inverse Fourier transformation, where, instead of the standard plane wave basis, the Ritus basis (\ref{S25}) is used
\begin{eqnarray}\label{S32}
S_{n}^{(q)}(k,k')=\int d^{4}x d^{4}y~\mathbb{E}_{n}^{(q)}(x,k)S^{(q)}(x,y)\mathbb{E}_{n}^{(q)\dagger}(y,k').\hspace{-0.4cm}\nonumber\\
\end{eqnarray}
Plugging $\mathbb{E}_{n}^{(q)}(x,k)$ from (\ref{S26}) on the r.h.s. of (\ref{S32}), and performing the integrations over $\bar{x}=(x_{0},0,x_{2},x_{3})$ as well as $\bar{y}=(y_{0},0,y_{2},y_{3})$, we arrive first at
\begin{eqnarray}\label{S33}
S_{n}^{(q)}(k,k')=(2\pi)^{3}\delta^{3}(\bar{k}-\bar{k'})S_{n}^{(q)}(\tilde{k}),
\end{eqnarray}
with
\begin{eqnarray}\label{S34}
S_{n}^{(q)}(\tilde{k})=\sum_{p=0}^{\infty}W_{pn}^{(q)}~\frac{1}{\ktildes_{p}-m_{q}}~W_{pn}^{(q)}.
\end{eqnarray}
Here, $W_{pn}^{(q)}$ is defined by
\begin{eqnarray}\label{S35}
W_{pn}^{(q)}\equiv \int_{-\infty}^{+\infty} dz_{1}P_{p}^{(q)}(z_{1},k_{2})P_{n}^{(q)}(z_{1},k_{2}).
\end{eqnarray}
Using at this stage the definition of $P_{p}^{(q)}$ from (\ref{S26}), the properties of the projectors ${\cal{P}}_{\pm}^{(q)}{\cal{P}}_{\pm}^{(q)}={\cal{P}}_{\pm}^{(q)}$ and ${\cal{P}}_{\pm}^{(q)}{\cal{P}}_{\mp}^{(q)}=0$, and the integration
\begin{eqnarray}\label{S36}
\int_{-\infty}^{+\infty}  dx_{1}f_{p}^{\pm s_{q}}(x_{1},k_{2})f_{n}^{\pm s_{q}}(x_{1},k_{2})=\delta_{pn},
\end{eqnarray}
which arises from the standard integration over Hermite polynomials
\begin{eqnarray}\label{S37}
\int_{-\infty}^{+\infty} dx~e^{-x^{2}}H_{p}(x)H_{n}(x)=2^{p}p!\sqrt{\pi}\delta_{pn},
\end{eqnarray}
we arrive, after some calculation, at
\begin{eqnarray}\label{S38}
W_{pn}^{(q)}=[{\cal{P}}_{+}^{(q)}+\Pi_{p}\Pi_{n}{\cal{P}}_{-}^{(q)}]\delta_{pn}.
\end{eqnarray}
Plugging then (\ref{S38}) in (\ref{S3}), splitting $\ktildes$ into longitudinal and transverse parts, $\ktildes=\mathbf{{\ktildes}}_{\|}+\mathbf{{\ktildes}}_{\perp}$
with~$\mathbf{{\ktildes}}_{\|}\equiv\boldsymbol{\gamma}^{\|}\cdot\mathbf{k}_{\|}$ and~$\mathbf{{\ktildes}}_{\perp}\equiv\boldsymbol{\gamma}^{\perp}
\cdot\widetilde{\mathbf{k}}_{\perp}$,
and $\mathbf{k}_{\|}\equiv (k_{0},0,0,k_{3})$ and $\widetilde{\mathbf{k}}_{\perp}=(0,0,-s_{q}\sqrt{2p|qeB|},0)$ as well as  $\boldsymbol{\gamma}_{\|}\equiv (\gamma_{0},0,0,\gamma_{3})$ and $\boldsymbol{\gamma}_{\perp}\equiv (0,\gamma_{1},\gamma_{2},0)$, and using eventually\footnote{In (\ref{S39}), $\mu_{\|}$ and $\mu_{\perp}$ denote $\mu=0,3$  and  $\mu=1,2$, respectively. }
\begin{eqnarray}\label{S39}
{\cal{P}}_{\pm}^{(q)}\gamma^{\mu_\|}=\gamma^{\mu_\|}{\cal{P}}_{\pm}^{(q)},\qquad
{\cal{P}}_{\pm}^{(q)}\gamma^{\mu_\perp}=\gamma^{\mu_\perp}{\cal{P}}_{\mp}^{(q)},
\end{eqnarray}
the free fermion propagator (\ref{S31}) in the momentum space reads
\begin{eqnarray}\label{S40}
\hspace{-0.3cm}S_{n}^{(q)}(\tilde{k})=\frac{({\cal{P}}_{+}^{(q)}+\Pi_{n}{\cal{P}}_{-}^{(q)})(\mathbf{\ktildes}_{\|}+m_{q})+\Pi_{n}\mathbf{\ktildes}_{\perp}}{\widetilde{k}_{n}^{2}-m_{q}^{2}}.
\end{eqnarray}
Here, $\Pi_{n}^{2}=\Pi_{n}$ and ${\cal{P}}_{+}^{(q)}+{\cal{P}}_{-}^{(q)}=1$ are also used. According to (\ref{S40}), the free fermion propagator for positively and negatively charged particles in the LLL and HLL at zero temperature reads
\begin{eqnarray}\label{S41}
\left\{
\begin{array}{lcrcl}
\mbox{For}~n=0&\qquad&S_{0}^{(\pm)}(\tilde{k})&=&\frac{{\cal{P}}_{\pm}}{\mathbf{\ktildes}_{\|}-m_{q}},\\
\mbox{For}~n\neq 0&\qquad&S_{n}^{(\pm)}(\tilde{k})&=&\frac{1}{\ktildes_{n}-m_{q}},\\
\end{array}
\right.
\end{eqnarray}
where the superscripts $\pm$ on $S_{n}^{(\pm)}$ stand for $q=\pm 1$. The appearance of the projectors ${\cal{P}}_{\pm}$ in $S_{0}^{\pm}$ reflects the fact that in the LLL, the charged fermions have either a positive (positively charged fermions) or a negative spin (negatively charged fermions), while in HLL fermions a certain double spin degeneracy occurs. In other words, in HLL positively (negatively) charged particles have both positive and negative spin orientations. Let us also notice that, the appearance of~ $\mathbf{\ktildes}_{\|}$ in the denominator of $S_{0}^{(\pm)}$ is related with the expected dimensional reduction from $D=4$ to $D=2$ in the LLL \cite{miransky1995}. In the next section, we will derive the general structure of the one-loop perturbative correction to the fermion self-energy of positively and negatively charged particles for $B\neq 0$ and at $T\neq 0$. The results will then be combined with (\ref{S41}), and lead to the general structure of the one-loop corrected propagator of magnetized fermions at zero and nonzero temperatures.
\section{One-loop self-energy of magnetized fermions at zero and nonzero temperatures}\label{sec3}
\setcounter{equation}{0}
In this section, we will determine the one-loop self-energy contribution of charged fermions in the presence of a constant magnetic field at zero and nonzero temperatures. In Sec. \ref{sec3a}, we will first consider the zero temperature case, and will show that the one-loop fermion self-energy, $\Sigma_{n}^{(q)}(\widetilde{k})$, has the following general structure
\begin{eqnarray}\label{A1}
\Sigma_{n}^{(q)}&=&\mathbf{\ktildes}_{\|}({\cal{P}}_{+}A_{+}^{(q)}+{\cal{P}}_{-}A_{-}^{(q)})+\widetilde{\mathbf{\ks}}_{\perp}D^{(q)}
\nonumber\\
&&+m_{q}({\cal{P}}_{+}C_{+}^{(q)}+{\cal{P}}_{-}C_{-}^{(q)}),
\end{eqnarray}
where the superscripts $(q)$ stand for positively ($q=+1$) and negatively ($q=-1$)
charged fermions. According to the notations introduced in the previous section,  ${\mathbf{k}}_{\|}=(k_{0},0,0,k_{3})$ and $\widetilde{\mathbf{k}}_{\perp}=(0,0,-s_{q}\sqrt{2n|qeB|},0)$.
\par
In Sec. \ref{sec3b}, we will then show that at finite temperature an additional splitting between the zero and the third components of $\mathbf{k}_{\|}$ occurs. The  general structure of $\Sigma_{n}^{(q)}(\widetilde{k})$ in a hot and magnetized QED plasma therefore reads
\begin{eqnarray}\label{A2}
\Sigma_{n}^{(q)}&=&\ks_{0}({\cal{P}}_{+}{A}_{+}^{(q)}+{\cal{P}}_{-}{A}_{-}^{(q)})
+\ks_{3}({\cal{P}}_{+}{B}_{+}^{(q)}+{\cal{P}}_{-}{B}_{-}^{(q)})\nonumber\\
&&
+\widetilde{\mathbf{\ks}}_{\perp}{D}^{(q)}+m_{q}({\cal{P}}_{+}{C}_{+}^{(q)}+{\cal{P}}_{-}{C}_{-}^{(q)}).
\end{eqnarray}
Let us notice that the additional splitting between the zero and third components occurs because of broken Lorentz invariance induced by the heat bath.
\par
In this section, the coefficients $A_{\pm}^{(q)}, D^{(q)}$ and $C_{\pm}^{(q)}$ from (\ref{A1}) and ${A}_{\pm}^{(q)}, {B}_{\pm}^{(q)}, {D}^{(q)}$ and ${C}_{\pm}^{(q)}$ from (\ref{A2}) will be analytically determined up to a  summation over Landau levels and a number of integrations. The latter will be then numerically performed in Secs. \ref{sec5a} and \ref{sec5b}, where the spectrum of Dirac fermions in a magnetized QED plasma will be determined at zero and nonzero temperatures. The results will eventually be compared with the results presented in Sec. \ref{sec2a} for hot QED plasma and for vanishing magnetic fields.
\subsection{One-loop self-energy of magnetized fermions at zero temperature}\label{sec3a}
The one-loop self-energy of magnetized fermions at zero temperature is given by combining the free fermion propagator (\ref{S31}), the free photon propagator $D^{\mu\nu}(p')=\frac{g^{\mu\nu}}{{p'}^{2}}$ in the Feynman gauge and the corresponding vertex $-eq\gamma^{\mu}$ of a photon and a magnetized fermion pair,
\begin{eqnarray}\label{A3}
\lefteqn{\hspace{-0.5cm}i\Sigma_{n}^{(q)}(x,y)=e^{2}q^{2}\sum\limits_{\ell=0}^{\infty}\int \frac{d^{3}\bar{p}}{(2\pi)^{3}}\frac{d^{4}p'}{(2\pi)^{4}}e^{i\bar{p}\cdot(\bar{x}-\bar{y})}e^{ip'\cdot(x-y)}
}\nonumber\\
&&\hspace{-0.cm}\times \gamma^{\mu}P_{\ell}^{(q)}(x_{1},q_{2})\frac{(\ptildes_{\ell}+m_{q})}{(\widetilde{p}_{\ell}^{2}-m_{q}^{2})}P_{\ell}^{(q)}(y_{1},p_{2})\gamma^{\nu}\frac{g_{\mu\nu}}{{p'}^{2}}.\nonumber\\
\end{eqnarray}
To derive the fermion self-energy in the momentum space, we use, as in the previous section, an appropriate inverse Fourier transformation, using the Ritus eigenfunctions $\mathbb{E}_{n}$ from (\ref{S25}),
\begin{eqnarray}\label{A4}
\lefteqn{\hspace{-0.8cm}\Sigma_{n}^{(q)}(k,k')}\nonumber\\
&&\hspace{-1cm}=\int d^{4}x d^{4}y~\mathbb{E}_{n}^{(q)}(x,k)\Sigma_{n}^{(q)}(x,y)\mathbb{E}^{(q)\dagger}(y,k').
\end{eqnarray}
After performing the integration over $\bar{x}$ and $\bar{y}$, we arrive at
\begin{eqnarray}\label{A5}
\Sigma_{n}^{(q)}(k,k')=(2\pi)^{3}\delta^{3}(\bar{k}-\bar{k'})\Sigma_{n}^{(q)}(\widetilde{k}),
\end{eqnarray}
with
\begin{eqnarray}\label{A6}
\lefteqn{i\Sigma_{n}^{(q)}(\widetilde{k})=e^{2}q^{2}\sum\limits_{\ell=0}^{\infty}\int \frac{d^{3}\bar{p}}{(2\pi)^{3}}\frac{dp'_{1}}{2\pi}~dx_{1}dy_{1}e^{-ip'_{1}(x_{1}-y_{1})}
}\nonumber\\
&&\hspace{-0.3cm}\times\frac{{\cal{N}}_{n\ell}^{(q)}(x_{1},y_{1};k_{2},p_{2})}{[(\mathbf{k}_{\|}-\mathbf{p}_{\|})^{2}-{p'}_{1}^{2}-(k_{2}-p_{2})^{2}][\mathbf{p}_{\|}^{2}-2\ell|qeB|-m_{q}^{2}]},\nonumber\\
\end{eqnarray}
and the numerator ${\cal{N}}_{n\ell}^{(q)}(x_{1},y_{1};k_{2},p_{2})$ defined by
\begin{eqnarray}\label{A7}
\lefteqn{
{\cal{N}}_{n\ell}^{(q)}(x_{1},y_{1};k_{2},p_{2})}\nonumber\\
&&\equiv W_{n\ell}^{\mu(q)}(x_{1};k_{2},p_{2})(\ptildes_{\ell}+m_{q})
W_{\ell n}^{(q)\nu}(y_{1};p_{2},k_{2})g_{\mu\nu},\nonumber\\
\end{eqnarray}
and
\begin{eqnarray}\label{A8}
\hspace{-0.3cm}W_{mn}^{(q)\mu}(z_{1};\ell_{1},\ell_{2})\equiv P_{m}^{(q)}(z_{1},\ell_{1})\gamma^{\mu}P_{n}^{(q)}(z_{1},\ell_{2}).
\end{eqnarray}
To evaluate the numerator ${\cal{N}}_{n\ell}^{(q)}(x_{1},y_{1};k_{2},p_{2})$, we use the definition of $P_{n}^{(q)}$ from (\ref{S26}), the properties of ${\cal{P}}_{\pm}^{(q)}$ projectors, ${\cal{P}}_{\pm}^{(q)}{\cal{P}}_{\pm}^{(q)}={\cal{P}}_{\pm}^{(q)}$ and ${\cal{P}}_{\pm}^{(q)}{\cal{P}}_{\mp}^{(q)}=0$, the relations from (\ref{S39})
and the following Dirac algebra
\begin{eqnarray}\label{A9}
\boldsymbol{\gamma}^{\|}\cdot\boldsymbol{\gamma}_{\perp}&=&0,\nonumber\\
\hspace{-1cm}\boldsymbol{\gamma}^{\|}\cdot\boldsymbol{\gamma}_{\|}&=&
\boldsymbol{\gamma}_{\perp}\cdot\boldsymbol{\gamma}^{\perp}=2,\nonumber\\
\hspace{-2cm}\gamma^{\mu_{\|}}\gamma^{\alpha}\gamma_{\mu_{\|}}=
-2\gamma^{\alpha_{\perp}},&~~~&
\gamma^{\mu_{\perp}}\gamma^{\alpha}\gamma_{\mu_\perp}=
-2\gamma^{\alpha_{\|}},
\end{eqnarray}
and arrive after some computation at
\begin{widetext}
\begin{eqnarray}\label{A10}
\lefteqn{\hspace{-0.5cm}{\cal{N}}_{n\ell}^{(q)}(x_{1},y_{1};k_{2},p_{2})}\nonumber\\
&&\hspace{-0.5cm}=-2\bigg\{
\mathbf{\ptildes}_{\|}\bigg[{\cal{P}}_{+}^{(q)}
\Pi_{\ell}f_{n}^{+s_{q}}(x_{1},k_{2})f_{\ell}^{-s_{q}}(x_{1},p_{2})f_{\ell}^{-s_{q}}(y_{1},p_{2})f_{n}^{+s_{q}}(y_{1},k_{2})\nonumber\\
&&\hspace{1.2cm}+{\cal{P}}_{-}^{(q)}\Pi_{n}f_{n}^{-s_{q}}(x_{1},k_{2})f_{\ell}^{+s_{q}}(x_{1},p_{2})f_{\ell}^{+s_{q}}(y_{1},p_{2})f_{n}^{-s_{q}}(y_{1},k_{2})\bigg]\nonumber\\
&&\hspace{0.4cm}+\mathbf{\ptildes}_{\perp}\bigg[{\cal{P}}_{+}^{(q)}
\Pi_{n}\Pi_{\ell}f_{n}^{-s_{q}}(x_{1},k_{2})f_{\ell}^{-s_{q}}(x_{1},p_{2})f_{\ell}^{+s_{q}}(y_{1},p_{2})f_{n}^{+s_{q}}(y_{1},k_{2})\nonumber\\
&&\hspace{1.cm}+{\cal{P}}_{-}^{(q)}
\Pi_{n}\Pi_{\ell}f_{n}^{+s_{q}}(x_{1},k_{2})f_{\ell}^{+s_{q}}(x_{1},p_{2})
f_{\ell}^{-s_{q}}(y_{1},p_{2})f_{n}^{-s_{q}}(y_{1},k_{2})\bigg]\nonumber\\
&&\hspace{0.4cm}-m_{q}\bigg[
{\cal{P}}_{+}^{(q)}f_{n}^{+s_{q}}(x_{1},k_{2})\left(f_{\ell}^{+s_{q}}(x_{1},p_{2})f_{\ell}^{+s_{q}}(y_{1},p_{2})+\Pi_{\ell}f_{\ell}^{-s_{q}}(x_{1},p_{2})f_{\ell}^{-s_{q}}(y_{1},p_{2})\right)f_{n}^{+s_{q}}(y_{1},k_{2})\nonumber\\
&&\hspace{1.cm}+{\cal{P}}_{-}^{(q)}
\Pi_{n}f_{n}^{-s_{q}}(x_{1},k_{2})\left(\Pi_{\ell}f_{\ell}^{-s_{q}}(x_{1},p_{2})f_{\ell}^{-s_{q}}(y_{1},p_{2})+
f_{\ell}^{+s_{q}}(x_{1},p_{2})f_{\ell}^{+s_{q}}(y_{1},p_{2})\right)f_{n}^{-s_{q}}(y_{1},k_{2})
\bigg]\bigg\}.\nonumber\\
\end{eqnarray}
\end{widetext}
Plugging at this stage ${\cal{N}}_{n\ell}^{(q)}(x_{1},y_{1};k_{2},p_{2})$ from (\ref{A10}) in (\ref{A6}), using ${\cal{P}}_{+}^{(\pm)}={\cal{P}}_{\pm}$ and ${\cal{P}}_{-}^{(\pm)}={\cal{P}}_{\mp}$, performing the integration over $x_{1}$ and $y_{1}$ by making use of the method introduced in the Appendix, and evaluating the integrals
\begin{eqnarray}\label{A11}
\hspace{-0.8cm}I_{\ell}&\equiv&\int \frac{d^{2}p_{\|}}{(2\pi)^{3}}\frac{1}{(\mathbf{p}_{\|}^{2}-M_{\ell}^{2})((\mathbf{k}_{\|}-\mathbf{p}_{\|})^{2}-\mathbf{p'}_{\perp}^{2})},\nonumber\\
\hspace{-0.8cm}J_{\ell}&\equiv&\int \frac{d^{2}p_{\|}}{(2\pi)^{3}}\frac{\mathbf{\ptildes}_{\|}}{(\mathbf{p}_{\|}^{2}-M_{\ell}^{2})((\mathbf{k}_{\|}-\mathbf{p}_{\|})^{2}-\mathbf{p'}_{\perp}^{2})},
\end{eqnarray}
using the standard Feynman integration method, we arrive after a lengthy but straightforward computation at [see also (\ref{A1})]
\begin{widetext}
\begin{eqnarray}\label{A12}
\Sigma_{n}^{(q)}(\widetilde{k})=\mathbf{\ktildes}_{\|}({\cal{P}}_{+}A_{+}^{(q)}+{\cal{P}}_{-}A_{-}^{(q)})+\widetilde{\mathbf{\ks}}_{\perp}D^{(q)}
+m_{q}({\cal{P}}_{+}C_{+}^{(q)}+{\cal{P}}_{-}C_{-}^{(q)}),
\end{eqnarray}
with the coefficients given by
\begin{eqnarray}\label{A13}
\hspace{-0.3cm}A_{+}^{(+)}&=&-2e^{2}q^{2}\sum\limits_{\ell=0}^{\infty}\int\frac{d^{2}p'_{\perp}}{(2\pi)^{2}}\frac{1}{n!\ell !}e^{-\kappa}\ell\Pi_{\ell} G_{1}(\kappa;M,m){\cal{J}}_{\ell}(\mathbf{p}'_{\perp}, \mathbf{k}_{\|}),\nonumber\\
\hspace{-0.3cm}A_{-}^{(+)}&=&-2e^{2}q^{2}\sum\limits_{\ell=0}^{\infty}\int\frac{d^{2}p'_{\perp}}{(2\pi)^{2}}\frac{1}{n!\ell !}e^{-\kappa}n \Pi_{n} G_{1}(\kappa;M',m'){\cal{J}}_{\ell}(\mathbf{p}'_{\perp}, \mathbf{k}_{\|}),\nonumber\\
\hspace{-0.3cm}D^{(+)}&=&+2e^{2}q^{2}\sum\limits_{\ell=0}^{\infty}\int\frac{d^{2}p'_{\perp}}{(2\pi)^{2}}\frac{1}{n!\ell !}e^{-\kappa}\ell\Pi_{n}\Pi_{\ell} G_{2}(\kappa;M'',m'',M''-1,m''-1){\cal{I}}_{\ell}(\mathbf{p}'_{\perp}, \mathbf{k}_{\|}),\nonumber\\
\hspace{-0.3cm}C_{+}^{(+)}&=&+2e^{2}q^{2}\sum\limits_{\ell=0}^{\infty}\int\frac{d^{2}p'_{\perp}}{(2\pi)^{2}}\frac{1}{n!\ell !}e^{-\kappa}
[\ell\Pi_{\ell}G_{1}(\kappa;M,m)+G_{1}(\kappa;M'',m'')] {\cal{I}}_{\ell}(\mathbf{p}'_{\perp}, \mathbf{k}_{\|}),\nonumber\\
\hspace{-0.3cm}C_{-}^{(+)}&=&+2e^{2}q^{2}\sum\limits_{\ell=0}^{\infty}\int\frac{d^{2}p'_{\perp}}{(2\pi)^{2}}\frac{1}{n!\ell !}e^{-\kappa}
\Pi_{n}[n\ell\Pi_{\ell}G_{1}(\kappa;M''-1,m''-1)+n G_{1}(\kappa;M',m')] {\cal{I}}_{\ell}(\mathbf{p}'_{\perp}, \mathbf{k}_{\|}),
\end{eqnarray}
for positively charged particles, and
\begin{eqnarray}\label{A14}
\hspace{-0.3cm}A_{+}^{(-)}&=&-2e^{2}q^{2}\sum\limits_{\ell=0}^{\infty}\int\frac{d^{2}p'_{\perp}}{(2\pi)^{2}}\frac{1}{n!\ell !}e^{-\kappa}\ell\Pi_{n}G_{1}(\kappa;M,m) {\cal{J}}_{\ell}(\mathbf{p}'_{\perp}, \mathbf{k}_{\|}),\nonumber\\
\hspace{-0.3cm}A_{-}^{(-)}&=&-2e^{2}q^{2}\sum\limits_{\ell=0}^{\infty}\int\frac{d^{2}p'_{\perp}}{(2\pi)^{2}}\frac{1}{n!\ell !}e^{-\kappa}n\Pi_{\ell}G_{1}(\kappa;M',m') {\cal{J}}_{\ell}(\mathbf{p}'_{\perp}, \mathbf{k}_{\|}),\nonumber\\
\hspace{-0.3cm}D^{(-)}&=&+2e^{2}q^{2}\sum\limits_{\ell=0}^{\infty}\int\frac{d^{2}p'_{\perp}}{(2\pi)^{2}}\frac{1}{n!\ell !}e^{-\kappa}\ell\Pi_{n}\Pi_{\ell} G_{2}(\kappa;M'',m'',M''-1,m''-1){\cal{I}}_{\ell}(\mathbf{p}'_{\perp}, \mathbf{k}_{\|}),\nonumber\\
\hspace{-0.3cm}C_{+}^{(-)}&=&+2e^{2}q^{2}\sum\limits_{\ell=0}^{\infty}\int\frac{d^{2}p'_{\perp}}{(2\pi)^{2}}\frac{1}{n!\ell !}e^{-\kappa}
\Pi_{n}[\ell G_{1}(\kappa;M,m)+\Pi_{\ell}G_{1}(\kappa;M'',m'')] {\cal{I}}_{\ell}(\mathbf{p}'_{\perp}, \mathbf{k}_{\|}),\nonumber\\
\hspace{-0.3cm}C_{-}^{(-)}&=&+2e^{2}q^{2}\sum\limits_{\ell=0}^{\infty}\int\frac{d^{2}p'_{\perp}}{(2\pi)^{2}}\frac{1}{n!\ell !}e^{-\kappa}
[n\ell G_{1}(\kappa;M''-1,m''-1)+n\Pi_\ell G_{1}(\kappa;M',m')] {\cal{I}}_{\ell}(\mathbf{p}'_{\perp}, \mathbf{k}_{\|}),
\end{eqnarray}
for negatively charged particles.
\end{widetext}
In the above relations, $\mathbf{k}_{\|}=(k_{0},0,0,k_{3})$, $\mathbf{p}'_{\perp}\equiv \mathbf{k}_{\perp}-\mathbf{p}_{\perp}$ with $\mathbf{k}_{\perp}=(0,k_{1},k_{2},0)$, $M_{\ell}^{2}\equiv 2\ell|qeB|+m_{q}^{2}$, $\kappa\equiv \frac{\ell_{q}^{2}\mathbf{p'}_{\perp}^{2}}{2}$ with $\ell_{q}=|qeB|^{-1/2}$, and $G_{i}(\kappa;\alpha,\beta), i=1,2$ are defined by
\begin{eqnarray}\label{A15}
G_{1}(\kappa;\alpha,\beta)&\equiv&\kappa^{\alpha-\beta}[{\cal{U}}_{\alpha-\beta+1}^{-\beta}(\kappa)]^{2},\nonumber\\
G_{2}(\kappa;\alpha,\beta,\gamma,\delta)&=&\kappa^{\alpha-\beta}[{\cal{U}}_{\alpha-\beta+1}^{-\beta}(\kappa)][{\cal{U}}_{\gamma-\delta+1}^{-\delta}(\kappa)],\nonumber\\
\end{eqnarray}
where ${\cal{U}}_{a}^{b}(\kappa)$ is the confluent hypergeometric function of the second kind \cite{mathworld}. Moreover, we have
\begin{eqnarray}\label{A16}
\begin{array}{lclclcl}
m&=&\mbox{min}(n,\ell-1), &\qquad&M&=&\mbox{max}(n,\ell-1),\\
m'&=&\mbox{min}(n-1,\ell), &\qquad&M'&=&\mbox{max}(n-1,\ell),\\
m''&=&\mbox{min}(n,\ell), &\qquad&M''&=&\mbox{max}(n,\ell).\nonumber\\
\end{array}\hspace{-0.3cm}\\
\end{eqnarray}
and
\begin{widetext}
\begin{eqnarray}\label{A17}
{\cal{I}}_{\ell}(\mathbf{p}'_{\perp},\mathbf{k}_{\|})&\equiv&\frac{1}{2\pi}\frac{1}{
\sqrt{4\mathbf{k}_{\|}^{2}M_{\ell}^{2}-(\mathbf{p'}^{2}_{\perp}-
\mathbf{k}_{\|}^{2}
-M_{\ell}^{2})^{2}}}\mbox{arctan}\left(\frac{\sqrt{4\mathbf{k}_{\|}^{2}
M_{\ell}^{2}
-(\mathbf{p'}^{2}_{\perp}-\mathbf{k}_{\|}^{2}
-M_{\ell}^{2})^{2}}}{\mathbf{p'}^{2}_{\perp}-\mathbf{k}_{\|}^{2}
+M_{\ell}^{2}}\right),\nonumber\\
{\cal{J}}_{\ell}(\mathbf{p}'_{\perp},\mathbf{k}_{\|})&\equiv&\frac{1}{4\pi\mathbf{k}_{\perp}^{2}}\bigg\{
-\frac{(\mathbf{p'}^{2}_{\perp}-\mathbf{k}_{\|}^{2}
-M_{\ell}^{2})}{\sqrt{4\mathbf{k}_{\|}^{2}M_{\ell}^{2}-(\mathbf{p'}^{2}_{\perp}-
\mathbf{k}_{\|}^{2}
-M_{\ell}^{2})^{2}}}\mbox{arctan}\left(\frac{\sqrt{4\mathbf{k}_{\|}^{2}M_{\ell}^{2}-(\mathbf{p'}^{2}_{\perp}-
\mathbf{k}_{\|}^{2}
-M_{\ell}^{2})^{2}}}{\mathbf{p'}^{2}_{\perp}-\mathbf{k}_{\|}^{2}
+M_{\ell}^{2}}\right)
-\frac{1}{2}\ln\frac{M_{\ell}^{2}}{\mathbf{p'}_{\perp}^{2}}
\bigg\}.\nonumber\\
\end{eqnarray}
\end{widetext}
These integrals arise from $I_{\ell}$ and $J_{\ell}$ in (\ref{A11}). As aforementioned, the summation over Landau levels $\ell$ and the two-dimensional integration over $\mathbf{p}'_{\perp}$ appearing in (\ref{A13}) and (\ref{A14}) will be numerically performed in Sec. \ref{sec5a}. At this stage, let us only notice that
in the LLL ($n=0$) $A_{-}^{(+)}, D^{(+)}$ and $C_{-}^{(+)}$ from (\ref{A13}) as well as $A_{\pm}^{(-)}, D^{(-)}$ and $C_{\pm}^{(-)}$ from (\ref{A14}) vanish.  The general structure of $\Sigma_{n}^{(\pm)}$ for positively and negatively charged fermions in the lowest and higher Landau levels is therefore given by
\begin{widetext}
\begin{eqnarray}\label{A18}
\Sigma_{n}^{(q)}=\left\{
\begin{array}{lcll}
{\cal{P}}_{+}\left(\mathbf{\ktildes}_{\|}A_{+}^{(+)}+m_{q}
C_{+}^{(+)}\right),&\qquad&\mbox{for}&n=0,~ q=+1,\\
0,&\qquad&\mbox{for}&n=0,~ q=-1,\\
\mathbf{\ktildes}_{\|}({\cal{P}}_{+}A_{+}^{(q)}+{\cal{P}}_{-}A_{-}^{(q)})+\widetilde{\mathbf{\ks}}_{\perp}D^{(q)}
+m_{q}({\cal{P}}_{+}C_{+}^{(q)}+{\cal{P}}_{-}C_{-}^{(q)}),&\qquad&\mbox{for}&n\neq 0,~ q=\pm 1.\\
\end{array}
\right.
\end{eqnarray}
\end{widetext}
In Sec. \ref{sec4}, we will combine the above result from (\ref{A18}) with the general structure of the free fermion propagator from (\ref{S41}), and will derive the general structure of the one-loop corrected fermion propagator for nonvanishing magnetic fields at zero temperature.
\subsection{One-loop self-energy of magnetized fermions at finite temperature}\label{sec3b}
In this section, we will show that the general structure of $\Sigma_{n}^{(q)}$ for $B\neq 0$ and at $T\neq 0$ is given by (\ref{A2}). We will, in particular, determine the coefficients ${A}_{\pm}^{(q)}, {B}_{\pm}^{(q)},{D}^{(q)}$ and ${C}_{\pm}^{(q)}$ appearing in (\ref{A2}) for $B\neq 0$ and at finite temperature $T$ and chemical potential $\mu$.
\par
To start, let us consider $i\Sigma_{n}^{(q)}$ from (\ref{A6}) with ${\cal{N}}_{n\ell}^{(q)}(x_{1},y_{1};k_{2},p_{2})$ presented in (\ref{A10}). To determine $\Sigma_{n}^{(q)}$ at finite temperature and chemical potential, all we have to do is to evaluate the integrals appearing in (\ref{A11}) at finite $T$ and $\mu$.
We thus focus only on the integrals
\begin{eqnarray}\label{A19}
I_{\ell}^{T}&=&iT\sum\limits_{n=-\infty}^{+\infty}\int \frac{dp_{3}}{2\pi}\frac{1}{(\mathbf{p}_{\|}^{2}-M_{\ell}^{2})((\mathbf{k}_{\|}-\mathbf{p}_{\|})^{2}-\mathbf{p'}_{\perp}^{2})},\nonumber\\
J_{\ell}^{T}&=&iT\sum\limits_{n=-\infty}^{+\infty}\int \frac{dp_{3}}{2\pi}\frac{\mathbf{\ptildes}_{\|}}{(\mathbf{p}_{\|}^{2}-M_{\ell}^{2})((\mathbf{k}_{\|}-\mathbf{p}_{\|})^{2}-\mathbf{p'}_{\perp}^{2})},\nonumber\\
\end{eqnarray}
that arise from (\ref{A11}) with the standard replacement
\begin{eqnarray}\label{A20}
\int \frac{d^{2}p_{\|}}{(2\pi)^{2}}\to iT\sum\limits_{n=-\infty}^{+\infty}\int \frac{dp_{3}}{2\pi}.
\end{eqnarray}
Here, $p_{0}=i\omega_{n}-\mu$ with the fermionic Matsubara frequencies $\omega_{n}=(2n+1)\pi T$. To compute these integrals, we will use two different methods. First,  using the method presented in \cite{fayazbakhsh2012}, we will evaluate them without any approximation  [see Sec. \ref{sec3b1}]. We will then perform the HTL approximation in a weak magnetic field, characterized by $k_{3}\ll T$ as well as by
\begin{eqnarray}\label{A21}
\sqrt{eB}\ll |\mathbf{p}'_{\perp}|\ll p_{3}\sim T,
\end{eqnarray}
and determine $I_{\ell}^{T}$ and $J_{\ell}^{T}$ from (\ref{A19}) within this approximation for $\mu=0$ [see Sec. \ref{sec3b2}].
\subsubsection{First method: Exact results}\label{sec3b1}
Using the standard Feynman parametrization, the integrals $I_{\ell}^{T}$ and $J_{\ell}^{T}$ from (\ref{A19}) are given by
\begin{eqnarray}\label{A22}
I_{\ell}^{T}&=&i\int_{0}^{1}dx\int\frac{dp_{3}}{2\pi}{\cal{S}}_{2}^{(0)}(\omega_{x}),\nonumber\\
J_{\ell}^{T}&=&i\int_{0}^{1}dx\int\frac{dp_{3}}{2\pi}[\gamma_{0}\bar{\cal{S}}_{2}^{(0)}(\omega_{x})+x~\mathbf{\ktildes}_{\|}{\cal{S}}_{2}^{(0)}(\omega_{x})]. \nonumber\\
\end{eqnarray}
where ${\cal{S}}_{\ell}^{(m)}(\omega)$ as well as $\bar{\cal{S}}_{\ell}^{(m)}(\omega)$ are defined by \cite{fayazbakhsh2012}
\begin{eqnarray}\label{A23}
{\cal{S}}_{\ell}^{(m)}(\omega)&\equiv&T\sum\limits_{n=-\infty}^{+\infty}\frac{(\ell_{0}^{2})^{m}}{(\ell_{0}^{2}-\omega^{2})^{\ell}},\nonumber\\
\bar{\cal{S}}_{\ell}^{(m)}(\omega)&\equiv&T\sum\limits_{n=-\infty}^{+\infty}\frac{(\ell_{0}^{2})^{m+\frac{1}{2}}}{(\ell_{0}^{2}-\omega^{2})^{\ell}},
\end{eqnarray}
with $\ell\geq 1$ and $m\geq 0$. Moreover, $\omega_{x}^{2}\equiv p_{3}^{2}+x\mathbf{p'}_{\perp}^{2}-x(1-x)\mathbf{k}_{\|}^{2}+(1-x)M_{\ell}^{2}$ with $M_{\ell}^{2}=2\ell|qeB|+m_{q}^{2}$. To proceed, we use
\begin{eqnarray}\label{A24}
{\cal{S}}_{1}^{(0)}(\omega)&=& \frac{1}{2\omega}(1-N_{f}(\omega)),\nonumber\\
\bar{\cal{S}}_{1}^{(0)}(\omega)&=&-\frac{1}{2}\bar{N}_{f}(\omega),
\end{eqnarray}
with $N_{f}(\omega)=n_{f}^{+}(\omega)+n_{f}^{-}(\omega)$ and $\bar{N}_{f}(\omega)=n_{f}^{+}(\omega)-n_{f}^{-}(\omega)$, with the fermion distribution function
\begin{eqnarray}\label{A25}
n_{f}^{\pm}(\omega)=\frac{1}{e^{\beta(\omega\mp\mu)}+1},
\end{eqnarray}
and the following relations for $\ell\geq 2$ \cite{fayazbakhsh2012}
\begin{eqnarray}\label{A26}
{\cal{S}}_{\ell}^{(0)}(\omega)&=&\frac{1}{2(\ell-1)\omega}\frac{d{\cal{S}}_{\ell-1}^{(0)}(\omega)}{d\omega},\nonumber\\
\bar{\cal{S}}_{\ell}^{(0)}(\omega)&=&\frac{1}{2(\ell-1)\omega}\frac{d\bar{\cal{S}}_{\ell-1}^{(0)}(\omega)}{d\omega}.
\end{eqnarray}
We arrive therefore at
\begin{eqnarray}\label{A27}
{\cal{S}}_{2}^{(0)}(\omega)&=& -\frac{1}{4\omega^{3}}+\frac{N_{f}(\omega)}{4\omega^{3}}-\frac{N'_{f}(\omega)}{4\omega^{2}},\nonumber\\
\bar{\cal{S}}_{2}^{(0)}(\omega)&=&-\frac{\bar{N}'_{f}(\omega)}{4\omega}.
\end{eqnarray}
The general structure of $\Sigma_{n}^{(q)}(\tilde{k})$ at finite temperature and chemical potential is thus given by [see also (\ref{A2})]
\begin{widetext}
\begin{eqnarray}\label{A28}
\Sigma_{n}^{(q)}(\tilde{k})=\ks_{0}({\cal{P}}_{+}{A}_{+}^{(q)}+{\cal{P}}_{-}{A}_{-}^{(q)})
+\ks_{3}({\cal{P}}_{+}{B}_{+}^{(q)}+{\cal{P}}_{-}{B}_{-}^{(q)})
+\widetilde{\mathbf{\ks}}_{\perp}{D}^{(q)}+m_{q}({\cal{P}}_{+}{C}_{+}^{(q)}+{\cal{P}}_{-}{C}_{-}^{(q)}),
\end{eqnarray}
with the coefficients given by
\begin{eqnarray}\label{A29}
\hspace{-0.5cm}{A}_{+}^{(+)}&=&-2e^{2}q^{2}\sum\limits_{\ell=0}^{\infty}\int\frac{d^{2}p'_{\perp}}{(2\pi)^{2}}\frac{1}{n!\ell !}e^{-\kappa}\ell\Pi_{\ell}G_{1}(\kappa;M,m) {\cal{J}}_{\ell}^{(0)T}(\mathbf{p}'_{\perp},\mathbf{k}_{\|}),\nonumber\\
\hspace{-0.5cm}{A}_{-}^{(+)}&=&-2e^{2}q^{2}\sum\limits_{\ell=0}^{\infty}\int\frac{d^{2}p'_{\perp}}{(2\pi)^{2}}\frac{1}{n!\ell !}e^{-\kappa}n\Pi_{n}G_{1}(\kappa;M',m') {\cal{J}}_{\ell}^{(0)T}(\mathbf{p}'_{\perp},\mathbf{k}_{\|}),\nonumber\\
\hspace{-0.5cm}{B}_{+}^{(+)}&=&-2e^{2}q^{2}\sum\limits_{\ell=0}^{\infty}\int\frac{d^{2}p'_{\perp}}{(2\pi)^{2}}\frac{1}{n!\ell !}e^{-\kappa}\ell\Pi_{\ell}G_{1}(\kappa;M,m) {\cal{J}}_{\ell}^{(3)T}(\mathbf{p}'_{\perp},\mathbf{k}_{\|}),\nonumber\\
\hspace{-0.5cm}{B}_{-}^{(+)}&=&-2e^{2}q^{2}\sum\limits_{\ell=0}^{\infty}\int\frac{d^{2}p'_{\perp}}{(2\pi)^{2}}\frac{1}{n!\ell !}e^{-\kappa}n\Pi_{n}G_{1}(\kappa;M',m') {\cal{J}}_{\ell}^{(3)T}(\mathbf{p}'_{\perp},\mathbf{k}_{\|}),\nonumber\\
\hspace{-0.5cm}{D}^{(+)}&=&+2e^{2}q^{2}\sum\limits_{\ell=0}^{\infty}\int\frac{d^{2}p'_{\perp}}{(2\pi)^{2}}\frac{1}{n!\ell !}e^{-\kappa}\ell\Pi_{n}\Pi_{\ell}G_{2}(\kappa;M'',m'',M''-1,m''-1) {\cal{I}}_{\ell}^{T}(\mathbf{p}'_{\perp},\mathbf{k}_{\|}),\nonumber\\
\hspace{-0.5cm}{C}_{+}^{(+)}&=&+2e^{2}q^{2}\sum\limits_{\ell=0}^{\infty}\int\frac{d^{2}p'_{\perp}}{(2\pi)^{2}}\frac{1}{n!\ell !}e^{-\kappa}
[\ell\Pi_{\ell}G_{1}(\kappa;M,m)+G_{1}(\kappa;M'',m'')] {\cal{I}}_{\ell}^{T}(\mathbf{p}'_{\perp},\mathbf{k}_{\|}),\nonumber\\
\hspace{-0.5cm}{C}_{-}^{(+)}&=&+2e^{2}q^{2}\sum\limits_{\ell=0}^{\infty}\int\frac{d^{2}p'_{\perp}}{(2\pi)^{2}}\frac{1}{n!\ell !}e^{-\kappa}
\Pi_{n}[n\ell\Pi_{\ell}G_{1}(\kappa;M''-1,m''-1)+n G_{1}(\kappa;M',m')] {\cal{I}}_{\ell}^{T}(\mathbf{p}'_{\perp},\mathbf{k}_{\|}),
\end{eqnarray}
for the positively charged particles, and
\begin{eqnarray}\label{A30}
\hspace{-0.5cm}{A}_{+}^{(-)}&=&-2e^{2}q^{2}\sum\limits_{\ell=0}^{\infty}\int\frac{d^{2}p'_{\perp}}{(2\pi)^{2}}\frac{1}{n!\ell !}e^{-\kappa}\ell\Pi_{n}G_{1}(\kappa;M,m) {\cal{J}}_{\ell}^{(0)T}(\mathbf{p}'_{\perp},\mathbf{k}_{\|}),\nonumber\\
\hspace{-0.5cm}{A}_{-}^{(-)}&=&-2e^{2}q^{2}\sum\limits_{\ell=0}^{\infty}\int\frac{d^{2}p'_{\perp}}{(2\pi)^{2}}\frac{1}{n!\ell !}e^{-\kappa}n\Pi_{\ell}G_{1}(\kappa;M',m') {\cal{J}}_{\ell}^{(0)T}(\mathbf{p}'_{\perp},\mathbf{k}_{\|}),\nonumber\\
\hspace{-0.5cm}{B}_{+}^{(-)}&=&-2e^{2}q^{2}\sum\limits_{\ell=0}^{\infty}\int\frac{d^{2}p'_{\perp}}{(2\pi)^{2}}\frac{1}{n!\ell !}e^{-\kappa}\ell\Pi_{n}G_{1}(\kappa;M,m) {\cal{J}}_{\ell}^{(3)T}(\mathbf{p}'_{\perp},\mathbf{k}_{\|}),\nonumber\\
\hspace{-0.5cm}{B}_{-}^{(-)}&=&-2e^{2}q^{2}\sum\limits_{\ell=0}^{\infty}\int\frac{d^{2}p'_{\perp}}{(2\pi)^{2}}\frac{1}{n!\ell !}e^{-\kappa}n\Pi_{\ell} G_{1}(\kappa;M',m'){\cal{J}}_{\ell}^{(3)T}(\mathbf{p}'_{\perp},\mathbf{k}_{\|}),\nonumber\\
\hspace{-0.5cm}{D}^{(-)}&=&+2e^{2}q^{2}\sum\limits_{\ell=0}^{\infty}\int\frac{d^{2}p'_{\perp}}{(2\pi)^{2}}\frac{1}{n!\ell !}e^{-\kappa}\ell\Pi_{n}\Pi_{\ell} G_{2}(\kappa;M'',m'',M''-1,m''-1){\cal{I}}_{\ell}^{T}(\mathbf{p}'_{\perp},\mathbf{k}_{\|}),\nonumber\\
\hspace{-0.5cm}{C}_{+}^{(-)}&=&+2e^{2}q^{2}\sum\limits_{\ell=0}^{\infty}\int\frac{d^{2}p'_{\perp}}{(2\pi)^{2}}\frac{1}{n!\ell !}e^{-\kappa}
\Pi_{n}[\ell G_{1}(\kappa;M,m)+\Pi_{\ell}G_{1}(\kappa;M'',m'')] {\cal{I}}_{\ell}^{T}(\mathbf{p}'_{\perp},\mathbf{k}_{\|}),\nonumber\\
\hspace{-0.5cm}{C}_{-}^{(-)}&=&+2e^{2}q^{2}\sum\limits_{\ell=0}^{\infty}\int\frac{d^{2}p'_{\perp}}{(2\pi)^{2}}\frac{1}{n!\ell !}e^{-\kappa}
[n\ell G_{1}(\kappa;M''-1,m''-1)+n\Pi_\ell G_{1}(\kappa;M',m')] {\cal{I}}_{\ell}^{T}(\mathbf{p}'_{\perp},\mathbf{k}_{\|}),
\end{eqnarray}
for negatively charged particles.
\end{widetext}
In the above relations, $\kappa$ is again given by  $\kappa=\frac{\ell_{q}^{2}\mathbf{p'}_{\perp}^{2}}{2}$, $G_{i}(\kappa;\alpha,\beta), i=1,2$ are defined in (\ref{A15}), and $m,m',m''$ as well as $M,M'$ and $M''$ are presented in (\ref{A16}). Moreover, ${\cal{I}}_{\ell}^{T}(\mathbf{p'}_{\perp},\mathbf{k}_{\|})$ and ${\cal{J}}_{\ell}^{(i)T}(\mathbf{p'}_{\perp},\mathbf{k}_{\|}), i=0,3$ are given by
\begin{eqnarray}\label{A31}
{\cal{I}}_{\ell}^{T}&\equiv&\int_{0}^{1}
dx\int\frac{dp_{3}}{2\pi}~{\cal{S}}_{2}^{(0)}(\omega_{x}),\nonumber\\
{\cal{J}}_{\ell}^{(0)T}&\equiv&\int_{0}^{1}dx\int\frac{dp_{3}}{2\pi}~\bigg[\frac{1}{k_{0}}\bar{\cal{S}}_{2}^{(0)}(\omega_{x})+x~{\cal{S}}_{2}^{(0)}(\omega_{x})\bigg],\nonumber\\
{\cal{J}}_{\ell}^{(3)T}&\equiv&\int_{0}^{1} dx\int\frac{dp_{3}}{2\pi}~x~{\cal{S}}_{2}^{(0)}(\omega_{x}),
\end{eqnarray}
with ${\cal{S}}_{2}^{(0)}(\omega)$ and $\bar{\cal{S}}_{2}^{(0)}(\omega)$ from (\ref{A27}). Let us notice that, for $\mu=0$, because of ${\cal{J}}_{\ell}^{(0)T}={\cal{J}}_{\ell}^{(3)T}$, we obtain $A_{\pm}^{(q)}=B_{\pm}^{(q)}$ for both $q=+ 1$ and $q=-1$. As in $T=0$ case, the summation over Landau levels and the two-dimensional over $\mathbf{p}'_{\perp}$, appearing in (\ref{A29}) and (\ref{A30}), as well as the integration over $p_{3}$ and $x$, appearing in (\ref{A31}), are to be numerically performed. Here, similar to the $T=0$ case, in the LLL ($n=0$), all the coefficients ${A}_{\pm}^{(-)},{B}_{\pm}^{(-)},{D}^{(-)}$ and ${C}_{\pm}^{(-)}$ for negatively charged particles vanish. Similarly, for positively charged particles, the coefficients ${A}_{-}^{(+)},{B}_{-}^{(+)},{D}^{(+)}$ and ${C}_{-}^{(+)}$ vanish for $n=0$. The general structure of $\Sigma_{n}^{(\pm)}$ in the lowest as well as higher Landau levels is therefore given by
\begin{widetext}
\begin{eqnarray}\label{A32}
\Sigma_{n}^{(q)}=\left\{
\begin{array}{lcll}
{\cal{P}}_{+}\left(\ks_{0}{A}_{+}^{(+)}
+\ks_{3}{B}_{+}^{(+)}
+m_{q}{C}_{+}^{(+)}\right),&&\mbox{for}&n=0,~ q=+1,\\
0,&&\mbox{for}&n=0,~ q=-1,\\
\ks_{0}({\cal{P}}_{+}{A}_{+}^{(q)}+{\cal{P}}_{-}{A}_{-}^{(q)})
+\ks_{3}({\cal{P}}_{+}{B}_{+}^{(q)}+{\cal{P}}_{-}{B}_{-}^{(q)})
+\widetilde{\mathbf{\ks}}_{\perp}{D}^{(q)}+m_{q}({\cal{P}}_{+}{C}_{+}^{(q)}+{\cal{P}}_{-}{C}_{-}^{(q)}),&&\mbox{for}&n\neq 0,~ q=\pm 1.\nonumber\\
\end{array}
\right.\hspace{-0.8cm}\\
\end{eqnarray}
\end{widetext}
In Sec. \ref{sec4}, we will combine these results with the general structure of the free propagator from (\ref{S41}), and will derive the general structure of the one-loop corrected fermion propagator for nonvanishing magnetic field at finite temperature.
\subsubsection{Second method: HTL approximation}\label{sec3b2}
In this section, we will evaluate the integrals (\ref{A19}), using the modified HTL approximation (\ref{A21}). For simplicity, we consider the $\mu=0$ case. Let us introduce, as in Sec. \ref{sec2a},
\begin{eqnarray}\label{A33}
E_{1}^{2}&\equiv& p_{3}^{2}+M_{\ell}^{2},\nonumber\\
E_{2}^{2}&\equiv& (p_{3}-k_{3})^{2}+\mathbf{p'}_{\perp}^{2},
\end{eqnarray}
and rewrite the integrals appearing in (\ref{A19}) as
\begin{eqnarray}\label{A34}
I_{\ell}^{T}(\mathbf{p'}_{\perp},\mathbf{k}_{\|})&=&i\int\frac{dp_{3}}{2\pi}~\Xi_{\ell}^{(3)}(p_{3};\mathbf{p'}_{\perp},\mathbf{k}_{\|}),\nonumber\\
J_{\ell}^{T}(\mathbf{p'}_{\perp},\mathbf{k}_{\|})&=&i\int\frac{dp_{3}}{2\pi}[\gamma_{0}\Xi_{\ell}^{(0)}(p_{3};\mathbf{p'}_{\perp},\mathbf{k}_{\|})\nonumber\\
&&-\gamma_{3}p_{3} \Xi_{\ell}^{(3)}(p_{3};\mathbf{p'}_{\perp},\mathbf{k}_{\|})],
\end{eqnarray}
with
\begin{eqnarray}\label{A35}
\hspace{-0.5cm}\Xi_{\ell}^{(0)}&\equiv&
T\sum\limits_{n=-\infty}^{+\infty}i\omega_{n}
\Delta_{f}(i\omega_{n},E_{1})\Delta_{b}(k_{0}-i\omega_{n},E_{2}),\nonumber\\
\hspace{-0.5cm}\Xi_{\ell}^{(3)}&\equiv &T\sum\limits_{n=-\infty}^{+\infty}\Delta_{f}(i\omega_{n},E_{1})\Delta_{b}(k_{0}-i\omega_{n},E_{2}),
\end{eqnarray}
similar to (\ref{S5}), with $\Delta_{f}(p_{0},E_{1})$ and $\Delta_{b}(k_{0}-p_{0},E_{2})$ defined in (\ref{S3}), and $E_{i}, i=1,2$ given in (\ref{A33}). The summation over Matsubara frequencies can be performed using the same method as described in Sec. \ref{sec2a}. Similar to (\ref{S6}), we therefore have
\begin{eqnarray}\label{A36}
\Xi_{\ell}^{(0)}&=&-\sum_{s_{1},s_{2}=\pm}\frac{s_{2}}{4E_{2}}\frac{1+f_{b}(s_{2}E_{2})-f_{f}(s_{1}E_{1})}{k_{0}-s_{1}E_{1}-s_{2}E_{2}},\nonumber\\
\Xi_{\ell}^{(3)}&=&-\sum_{s_{1},s_{2}=\pm}\frac{s_{1}s_{2}}{4E_{1}E_{2}}\frac{1+f_{b}(s_{2}E_{2})-f_{f}(s_{1}E_{1})}{k_{0}-s_{1}E_{1}-s_{2}E_{2}},\nonumber\hspace{-0.3cm}\\
\end{eqnarray}
with the distribution functions $f_{f/b}(E)$ defined in (\ref{S7}), and $E_{i}, i=1,2$ given in (\ref{A33}). Under the modified HTL approximation introduced in (\ref{A21}), we set
\begin{eqnarray}\label{A37}
E_{1}&\approx& p_{3}+\frac{M_{\ell}^{2}}{2p_{3}},\nonumber\\
E_{2}&\approx&p_{3}-k_{3}+\frac{\mathbf{p'}_{\perp}^{2}}{2p_{3}},
\end{eqnarray}
and have consequently
\begin{eqnarray}\label{A38}
\hspace{-0.5cm}f_{f}(E_{1})&\approx&f_{f}(p_{3})+\frac{M_{\ell}^{2}}{2p_{3}}\frac{df_{f}(p_{3})}{dp_{3}},\nonumber\\
\hspace{-0.5cm}f_{b}(E_{2})&\approx&f_{b}(p_{3})+\left(\frac{\mathbf{p'}_{\perp}^{2}}{2p_{3}}-k_{3}\right)\frac{df_{b}(p_{3})}{dp_{3}}.
\end{eqnarray}
Using furthermore the relation $f_{f}(-\omega)=1-f_{f}(\omega)$ and $f_{b}(-\omega)=-1-f_{b}(\omega)$, we arrive after some computation at
\begin{eqnarray}\label{A39}
I_{\ell}^{T}&=&i{\cal{I}}_{\ell}^{T},\nonumber\\
J_{\ell}^{T}&=&i\ \ks_{0}{\cal{J}}_{\ell}^{(0)T}+i\ \ks_{3}{\cal{J}}_{\ell}^{(3)T},
\end{eqnarray}
with
\begin{widetext}
\begin{eqnarray}\label{A40}
{\cal{I}}_{\ell}^{T}(\mathbf{p'}_{\perp},\mathbf{k}_{\|})&=&\frac{1}{8\pi T^{2}}\bigg\{\mathbf{I}_{0}^{(3)}
-\frac{2k_{3}T}{\mathbf{k}_{\|}^{2}}[\mathbf{I}_{b}^{(2)}+\mathbf{I}_{f}^{(2)}]
-\frac{M_{\ell}^{2}k_{3}}{T\mathbf{k}_{\|}^{2}}\bar{\mathbf{I}}_{f}^{(3)}
+\frac{2k_{3}^{2}}{\mathbf{k}_{\|}^{2}}\bar{\mathbf{I}}_{b}^{(2)}+\frac{M_{\ell}^{2}}{2T^{2}}\bigg[\frac{(k_{0}^{2}+k_{3}^{2})(\mathbf{p'}_{\perp}^{2}-M_{\ell}^{2})}{(\mathbf{k}_{\|}^{2})^{2}}-1\bigg]\bar{\mathbf{I}}_{f}^{(4)}
\nonumber\\
&&+
\bigg[\frac{(k_{0}^{2}+k_{3}^{2})(\mathbf{p'}_{\perp}^{2}-M_{\ell}^{2})}
{(\mathbf{k}_{\|}^{2})^{2}}-1\bigg]\mathbf{I}_{f}^{(3)}+
\bigg[\frac{(k_{0}^{2}+k_{3}^{2})(\mathbf{p'}_{\perp}^{2}-M_{\ell}^{2})}
{(\mathbf{k}_{\|}^{2})^{2}}+1\bigg]\mathbf{I}_{b}^{(3)}\nonumber\\
&&+
\frac{\mathbf{p'}_{\perp}^{2}}{2T^{2}}\bigg[\frac{(k_{0}^{2}+k_{3}^{2})(\mathbf{p'}_{\perp}^{2}-M_{\ell}^{2})}
{(\mathbf{k}_{\|}^{2})^{2}}+1\bigg]\bar{\mathbf{I}}_{b}^{(4)}+\frac{k_{3}}{T}\bigg[\frac{-2k_{0}^{2}\mathbf{p'}_{\perp}^{2}+M_{\ell}^{2}(k_{0}^{2}+k_{3}^{2})}{(\mathbf{k}_{\|}^{2})^{2}}-1\bigg]\bar{\mathbf{I}}_{b}^{(3)}\bigg\},\nonumber\\
{\cal{J}}_{\ell}^{(0)T}(\mathbf{p'}_{\perp},\mathbf{k}_{\|})&=&\frac{1}{4\pi T}\bigg\{
-\frac{T}{\mathbf{k}_{\|}^{2}}[\mathbf{I}_{b}^{(1)}+\mathbf{I}_{f}^{(1)}]+\frac{k_{3}(\mathbf{p'}_{\perp}^{2}-M_{\ell}^{2})}{(\mathbf{k}_{\|}^{2})^{2}}[\mathbf{I}_{b}^{(2)}+\mathbf{I}_{f}^{(2)}]-\frac{M_{\ell}^{2}}{2T\mathbf{k}_{\|}^{2}}\bar{\mathbf{I}}_{f}^{(2)}
+\frac{M_{\ell}^{2}k_{3}(\mathbf{p'}_{\perp}^{2}-M_{\ell}^{2})}{2T^{2}(\mathbf{k}_{\|}^{2})^{2}}\bar{\mathbf{I}}_{f}^{(3)}\nonumber\\
&&
+\frac{k_{3}}{\mathbf{k}_{\|}^{2}}\bar{\mathbf{I}}_{b}^{(1)}
-\bigg[\frac{\mathbf{p'}_{\perp}^{2}(k_{0}^{2}+k_{3}^{2})-2k_{3}^{2}M_{\ell}^{2}}{2T(\mathbf{k}_{\|}^{2})^{2}}\bigg]\bar{\mathbf{I}}_{b}^{(2)}+\frac{k_{3}\mathbf{p}_{\perp}^{'2}(\mathbf{p'}_{\perp}^{2}-M_{\ell}^{2})}{2T^{2}(\mathbf{k}_{\|}^{2})^{2}}\bar{\mathbf{I}}_{b}^{(3)}
\bigg\},\nonumber\\
{\cal{J}}_{\ell}^{(3)T}(\mathbf{p'}_{\perp},\mathbf{k}_{\|})&=&\frac{1}{8\pi T}\bigg\{\frac{\mathbf{I}_{0}^{(2)}}{k_{3}}
-\frac{2T}{\mathbf{k}_{\|}^{2}}[\mathbf{I}_{b}^{(1)}+\mathbf{I}_{f}^{(1)}]
-\frac{M_{\ell}^{2}}{T\mathbf{k}_{\|}^{2}}\bar{\mathbf{I}}_{f}^{(2)}
+\frac{2k_{3}}{\mathbf{k}_{\|}^{2}}\bar{\mathbf{I}}_{b}^{(1)}+\frac{M_{\ell}^{2}}{2T^{2}k_{3}}\bigg[\frac{(k_{0}^{2}+k_{3}^{2})(\mathbf{p'}_{\perp}^{2}-M_{\ell}^{2})}{(\mathbf{k}_{\|}^{2})^{2}}-1\bigg]\bar{\mathbf{I}}_{f}^{(3)}
\nonumber\\
&&+
\frac{1}{k_{3}}\bigg[\frac{(k_{0}^{2}+k_{3}^{2})(\mathbf{p'}_{\perp}^{2}-M_{\ell}^{2})}
{(\mathbf{k}_{\|}^{2})^{2}}-1\bigg]\mathbf{I}_{f}^{(2)}+\frac{1}{k_{3}}\bigg[\frac{(k_{0}^{2}+k_{3}^{2})(\mathbf{p'}_{\perp}^{2}-M_{\ell}^{2})}
{(\mathbf{k}_{\|}^{2})^{2}}+1\bigg]\mathbf{I}_{b}^{(2)}
\nonumber\\
&&+
\frac{\mathbf{p'}_{\perp}^{2}}{2T^{2}k_{3}}\bigg[\frac{(k_{0}^{2}+k_{3}^{2})(\mathbf{p'}_{\perp}^{2}-M_{\ell}^{2})}
{(\mathbf{k}_{\|}^{2})^{2}}+1\bigg]\bar{\mathbf{I}}_{b}^{(3)}+\frac{1}{T}\bigg[\frac{-2k_{0}^{2}\mathbf{p'}_{\perp}^{2}+M_{\ell}^{2}(k_{0}^{2}+k_{3}^{2})}{(\mathbf{k}_{\|}^{2})^{2}}-1\bigg]\bar{\mathbf{I}}_{b}^{(2)}\bigg\}.
\end{eqnarray}
\end{widetext}
In the above relations
\begin{eqnarray}\label{A41}
\mathbf{I}_{0}^{(n)}(z)&\equiv&\int_{z}^{\infty}\frac{dy}{y^{n}},\nonumber\\
\mathbf{I}_{b/f}^{(n)}(z)&\equiv&\int_{z}^{\infty}\frac{dy}{y^{n}}f_{b/f}(y),\nonumber\\
\bar{\mathbf{I}}_{b/f}^{(n)}(z)&\equiv&\int_{z}^{\infty}\frac{dy}{y^{n}}\frac{df_{b/f}(y)}{dy},
\end{eqnarray}
with  $z\equiv\frac{\sqrt{eB}}{T}$, and
\begin{eqnarray}\label{A42}
f_{b}(y)=\frac{1}{e^{y}-1},\qquad f_{f}(y)=\frac{1}{e^{y}+1}.
\end{eqnarray}
The general structure of $\Sigma_{n}^{(q)}(\tilde{k})$ for hot and magnetized fermions in the above HTL approximation is therefore given by (\ref{A32}) with the coefficients ${A}_{\pm}^{(q)},{B}_{\pm}^{(q)},{D}^{(q)}$ and ${C}_{\pm}^{(q)}$ from (\ref{A29}) as well as (\ref{A30}), and the integrals ${\cal{I}}_{\ell}^{T}$ as well as ${\cal{J}}_{\ell}^{(i)T}, i=1,2$ given in  (\ref{A40}).
\section{General structure of the dressed propagator of hot and magnetized fermions}\label{sec4}
\setcounter{equation}{0}
In (\ref{S41}), the free propagator of magnetized fermions is presented in the momentum space. The general structure of the one-loop self-energy of these fermions at finite temperature is presented in (\ref{A32}). In this section, we will combine these two results, and determine the one-loop corrected propagator of hot and magnetized fermions up to one-loop level. The case $T=0$ will be then considered as a special case of $T\neq 0$ case, because by comparing $\Sigma_{n}^{(q)}$ from (\ref{A18}) for $T=0$ with (\ref{A32}) for $T\neq 0$, it turns out that $\Sigma_{n}^{(q)}$ for $T=0$ can be determined from (\ref{A32}) by setting ${B}_{\pm}^{(q)}={A}_{\pm}^{(q)}$, i.e. by removing the anisotropy in the components of $\mathbf{k}_{\|}=(k_{0},0,0,k_{3})$, appearing in (\ref{A32}) in comparison to (\ref{A18}) for $n=0$ and $n\neq 0$. In this section, we are not interested on the numerical values of the coefficients. Our main goal is to use the general structure of the dressed propagator of hot and magnetized fermions, and, performing an analysis similar to what is presented in Sec. (\ref{sec2a}), to determine the properties of possible excitations arising from the poles of this propagator.
\par
To start, let us therefore consider the series expansion of the full fermion propagator in the momentum space ${\cal{S}}_{n}^{(q)}(\tilde{k})$,
\begin{eqnarray}\label{D1}
{\cal{S}}_{n}^{(q)}(\tilde{k})=S_{n}^{(q)}(\tilde{k})+S_{n}^{(q)}(\tilde{k})\Sigma_{n}^{(q)}S_{n}^{(q)}+\cdots.
\end{eqnarray}
Truncating this series after the one-loop contribution in the second term on the r.h.s., and using (\ref{S41}) as well as (\ref{A32}), the general structure of ${\cal{S}}_{n}^{(q)}$ of magnetized fermions at $T\neq 0$ in the lowest ($n=0$) and higher ($n\neq 0$) Landau levels up to one-loop level reads
\begin{eqnarray}\label{D2}
\hspace{-0.3cm}
{\cal{S}}_{n}^{(q)}=
\left\{
\begin{array}{ccll}
\frac{{\cal{P}}_{+}}{\mathbf{\ktildes}_{\|}-m_{q}-\widetilde{\Sigma}_{0}^{(+)}},&~~&\mbox{for}&n=0,~q=+1,\\
\frac{{\cal{P}}_{-}}{\mathbf{\ktildes}_{\|}-m_{q}},&~~&\mbox{for}&n=0,~q=-1,\\
\frac{1}{{\ktildes}_{n}-m_{q}-{\Sigma}_{n}^{(q)}},&~~&\mbox{for}&n\neq 0,~q=\pm 1,\\
\end{array}
\right.
\end{eqnarray}
where
\begin{eqnarray}\label{D3}
\widetilde{\Sigma}_{0}^{(+)}&\equiv& \ks_{0}{A}_{+}^{(+)}
+\ks_{3}{B}_{+}^{(+)}+m_{q}{C}_{+}^{(+)},\nonumber\\
{\Sigma}_{n}^{(q)}&\equiv&\ks_{0}({\cal{P}}_{+}{A}_{+}^{(q)}+{\cal{P}}_{-}{A}_{-}^{(q)})
+\ks_{3}({\cal{P}}_{+}{B}_{+}^{(q)}+{\cal{P}}_{-}{B}_{-}^{(q)})\nonumber\\
&&
+\widetilde{\mathbf{\ks}}_{\perp}{D}^{(q)}+m_{q}({\cal{P}}_{+}{C}_{+}^{(q)}+{\cal{P}}_{-}{C}_{-}^{(q)}),
\end{eqnarray}
are obtained from (\ref{A32}). At $T=0$, ${\cal{S}}_{n}^{(q)}$ is given by (\ref{D2}) and (\ref{D3}) with $B_{\pm}^{(q)}=A_{\pm}^{(q)}$ for $q=\pm 1$. The poles of the fermion propagator (\ref{D2}) can be determined by computing
\begin{eqnarray}\label{D4}
0=
\left\{
\begin{array}{ccll}
\mbox{det}(\mathbf{\ktildes}_{\|}-m_{q}-\widetilde{\Sigma}_{0}^{(+)}),&~&\mbox{for}&n=0,~q=+1,\\
\mbox{det}(\mathbf{\ktildes}_{\|}-m_{q}),&~&\mbox{for}&n=0,~q=-1,\\
\mbox{det}({\ktildes}_{n}-m_{q}-{\Sigma}_{n}^{(q)}),&~&\mbox{for}&n\neq 0,~q=\pm 1.\nonumber\\
\end{array}
\right.\hspace{-0.5cm}\\
\end{eqnarray}
This will be done numerically in Sec. \ref{sec5a} for $T\neq 0$ [hot magnetized QED plasma], and in Sec. \ref{sec5b}  for $T=0$ [cold magnetized QED plasma].
To study the properties of possible excitations arising from the pole of the dressed fermion propagator, let us consider ${\cal{S}}_{n}^{(q)}$ from (\ref{D2}). Defining
\begin{eqnarray}\label{D5}
a_{\pm}^{(q)}&\equiv& k_{0}(1-A_{\pm}^{(q)}),\nonumber\\
b_{\pm}^{(q)}&\equiv& k_{3}(1-B_{\pm}^{(q)}),\nonumber\\
c_{\pm}^{(q)}&\equiv& m_{q}(1+C_{\pm}^{(q)}), \nonumber\\
d^{(q)}&\equiv&s_{q}\sqrt{2n|qeB|}(1-D^{(q)}),
\end{eqnarray}
${\cal{S}}_{n}^{(q)}$ for $n=0$ and $n\neq 0$ can be simplified as\footnote{The $\gamma$-matrices in the Dirac representation are given by
\begin{eqnarray*}
\gamma_{0}=\left(
\begin{array}{cc}
1&0\\
0&-1
\end{array}
\right),~~
\boldsymbol{\gamma}=\left(
\begin{array}{cc}
0&\boldsymbol{\sigma}\\
-\boldsymbol{\sigma}&0
\end{array}
\right),~~
\gamma_{5}=\left(
\begin{array}{cc}
0&1\\
1&0
\end{array}
\right),
\end{eqnarray*}
and $\boldsymbol{\Sigma}=\mbox{diag}\left(\boldsymbol{\sigma},\boldsymbol{\sigma}\right)$.
}
\begin{eqnarray}\label{D6}
\hspace{-0.5cm}{\cal{S}}_{0}^{(q)}=
\left\{
\begin{array}{ccll}
\frac{{\cal{P}}_{+}\left(\gamma_{0}a_{+}^{(+)}
-\gamma_{3}b_{+}^{(+)}+c_{+}^{(+)}\right)}{{\cal{D}}_{0}^{(+)}},&&\mbox{for}&q=+1,\\
\frac{{\cal{P}}_{-}\left(\gamma_{0}k_{0}
-\gamma_{3}k_{3}+m_{q}\right)}{{\cal{D}}_{0}^{(-)}},&&\mbox{for}
&q=-1,\\
\end{array}
\right.
\end{eqnarray}
with
\begin{eqnarray}\label{D7}
{\cal{D}}_{0}^{(+)}&\equiv&a_{+}^{(+)2}-b_{+}^{(+)2}+c_{+}^{(+)2},\nonumber\\
{\cal{D}}_{0}^{(-)}&\equiv&\mathbf{k}_{\|}^{2}-m_{q}^{2},
\end{eqnarray}
and
\begin{eqnarray}\label{D8}
{\cal{S}}_{n}^{(q)}(\tilde{k})=\frac{{\cal{N}}_{n}^{(q)}(\tilde{k})}{{\cal{D}}_{n}^{(q)}(\tilde{k})},
\end{eqnarray}
with the numerator\footnote{Here, the relations ${\cal{P}}_{+}+{\cal{P}}_{-}=1$ and ${\cal{P}}_{+}-{\cal{P}}_{-}=\Sigma_{3}$ are used. }
\begin{eqnarray}\label{D9}
{\cal{N}}_{n}^{(q)}
 &\equiv&\gamma_{0}\left({\cal{C}}^{(q)}_{1}\p_{+}+ {\cal{C}}^{(q)}_{2}\p_{-}\right)+\gamma_{2}{\cal{C}}^{(q)}_{3} \nonumber\\
&&+\gamma_{3}\left({\cal{C}}^{(q)}_{4}\p_{+}+{\cal{C}}^{(q)}_{5}\p_{-}\right)\nonumber\\
&&+\left({\cal{C}}^{(q)}_{6}\gamma_{0}\gamma_{2}+{\cal{C}}^{(q)}_{7} \gamma_{2}\gamma_{3}+{\cal{C}}^{(q)}_{8}\gamma_{0}\gamma_{2}
\gamma_{3}\right)\Sigma_{3}\nonumber\\
&&+{\cal{C}}^{(q)}_{9}\p_{+} +{\cal{C}}^{(q)}_{10}\p_{-},
\end{eqnarray}
and the denominator
\begin{eqnarray}\label{D10}
\lefteqn{
{\cal{D}}_{n}^{(q)}\equiv d^{(q)4} +2d^{(q)2} \left(c_{-}^{(q)}c_{+}^{(q)}+b_{-}^{(q)}b_{+}^{(q)}-a_{-}^{(q)}a_{+}^{(q)}\right)
}\nonumber\\
&&+\left(a_{-}^{(q)2} -b_{-}^{(q)2}-c_{-}^{(q)2}\right)
 \left(a_{+}^{(q)2} -b_{+}^{(q)2}-c_{+}^{(q)2}\right).\nonumber\\
\end{eqnarray}
In (\ref{D9}), the coefficients ${\cal{C}}_{i}^{(q)}, i=1,\cdots,10$ are given by
\begin{eqnarray}\label{D11}
\hspace{-0.4cm}
{\cal{C}}_{1}^{(q)}&\equiv&{a_{+}^{(q)}\left(a_{-}^{(q)2} -b_{-}^{(q)2}-c_{-}^{(q)2}\right)- a_{-}^{(q)}d^{(q)2}},\nonumber\\
\hspace{-0.4cm}
{\cal{C}}_{2}^{(q)}&\equiv&{a_{-}^{(q)}\left(a_{+}^{(q)2} -b_{+}^{(q)2}-c_{+}^{(q)2}\right)- a_{+}^{(q)}d^{(q)2}},\nonumber\\
\hspace{-0.4cm}
{\cal{C}}_{3}^{(q)}&\equiv&{\left(d^{(q)2} +c_{-}^{(q)}c_{+}^{(q)} +b_{-}^{(q)}b_{+}^{(q)} -a_{-}^{(q)}a_{+}^{(q)}\right)d^{(q)}},\nonumber\\
\hspace{-0.4cm}
{\cal{C}}_{4}^{(q)}&\equiv&{b_{+}^{(q)}\left(b_{-}^{(q)2} -a_{-}^{(q)2}-c_{-}^{(q)2}\right)+ b_{-}^{(q)}d^{(q)2}},\nonumber\\
\hspace{-0.4cm}
{\cal{C}}_{5}^{(q)}&\equiv&{b_{-}^{(q)}\left(b_{+}^{(q)2} -a_{+}^{(q)2}-c_{+}^{(q)2}\right)+ b_{+}^{(q)}d^{(q)2}},\nonumber\\
\hspace{-0.4cm}
{\cal{C}}_{6}^{(q)}&\equiv&{\left(c_{+}^{(q)}a_{-}^{(q)} -c_{-}^{(q)}a_{+}^{(q)}\right)d^{(q)}},\nonumber\\
\hspace{-0.4cm}
{\cal{C}}_{7}^{(q)}&\equiv&{\left(c_{+}^{(q)}b_{-}^{(q)} -c_{-}^{(q)}b_{+}^{(q)}\right)d^{(q)}},\nonumber\\
\hspace{-0.4cm}
{\cal{C}}_{8}^{(q)}&\equiv&{\left(a_{+}^{(q)} b_{-}^{(q)} -a_{-}^{(q)}b_{+}^{(q)}\right)d^{(q)}},\nonumber\\
\hspace{-0.4cm}
{\cal{C}}_{9}^{(q)}&\equiv&{c_{+}^{(q)}\left(a_{-}^{(q)2} -b_{-}^{(q)2}-c_{-}^{(q)2}\right)- c_{-}^{(q)}d^{(q)2}},\nonumber\\
\hspace{-0.4cm}
{\cal{C}}_{10}^{(q)}&\equiv&{c_{-}^{(q)}\left(a_{+}^{(q)2} -b_{+}^{(q)2}-c_{+}^{(q)2}\right)- c_{+}^{(q)}d^{(q)2}}.
\end{eqnarray}
In what follows, we will determine the eigenvalues and eigenvectors of the numerators of  ${\cal{S}}_{n}^{(q)}$ for two special cases:
\begin{eqnarray*}
\mbox{Case 1:}&\qquad&n=0,~q=\pm 1,~ m_{q}\neq 0,\nonumber\\
\mbox{Case 2:}&\qquad&n\neq 0,~q=\pm 1,~ m_{q}=0.
\end{eqnarray*}
\subsubsection*{Case 1: Properties of massive and positively charged fermionic excitations in LLL}
Let us consider the one-loop corrected propagator of positively charged massive fermions in the LLL from (\ref{D6})
\begin{eqnarray}\label{D12}
{\cal{S}}_{0}^{(+)}(\tilde{k})=\frac{{\cal{P}}_{+}\left(\gamma_{0}a_{+}^{(+)}
-\gamma_{3}b_{+}^{(+)}+c_{+}^{(+)}\right)}{{\cal{D}}_{0}^{(+)}},
\end{eqnarray}
with the coefficients given in (\ref{D5}) and the denominator ${\cal{D}}_{0}^{(+)}$ given in (\ref{D7}). The eigenvectors of the numerator of ${\cal{S}}_{0}^{(+)}$ are
\begin{eqnarray}\label{D13}
\psi_{1}^{(+)}&=&(0,0,0,1),\nonumber\\
\psi_{2}^{(+)}&=&(0,1,0,0),\nonumber\\
\psi_{3}^{(+)}&=&\frac{1}{b_{+}^{(+)}}\left(a_{+}^{(+)}-\sqrt{a_{+}^{(+)2}-b_{+}^{(+)2}},0,b_{+}^{(+)},0\right),\nonumber\\
\psi_{4}^{(+)}&=&\frac{1}{b_{+}^{(+)}}\left({a_{+}^{(+)}+\sqrt{a_{+}^{(+)2}-b_{+}^{(+)2}}},0,b_{+}^{(+)},0\right),\nonumber\\
\end{eqnarray}
with the eigenvalues
\begin{eqnarray}\label{D14}
\left\{0,0,c_{+}^{(+)}-\sqrt{a_{+}^{(+)2}-b_{+}^{(+)2}}, c_{+}^{(+)}+
\sqrt{a_{+}^{(+)2}-b_{+}^{(+)2}}\right\}.\hspace{-0.5cm}\nonumber\\
\end{eqnarray}
The eigenvectors from (\ref{D13}) satisfy
\begin{eqnarray}\label{D15}
\hspace{-0.3cm}\begin{array}{ccccc}
 {\cal{P}}_{-}\psi_{1}^{(+)}=\psi_{1}^{(+)},&
~~&
\mbox{or}&
~~&
\Sigma_{3}\psi_{1}^{(+)}=-\psi_{1}^{(+)},\\
 {\cal{P}}_{-}\psi_{2}^{(+)}=\psi_{2}^{(+)},&
~~&
\mbox{or}&
~~&
\Sigma_{3}\psi_{2}^{(+)}=-\psi_{2}^{(+)},\\
 {\cal{P}}_{+}\psi_{3}^{(+)}=\psi_{3}^{(+)},&
~~&
\mbox{or}&
~~&
\Sigma_{3}\psi_{3}^{(+)}=+\psi_{3}^{(+)},\\
 {\cal{P}}_{+}\psi_{4}^{(+)}=\psi_{4}^{(+)},&
~~&
\mbox{or}&
~~&
\Sigma_{3}\psi_{4}^{(+)}=+\psi_{4}^{(+)}.\\
\end{array}
\end{eqnarray}
From the above four eigenvectors only $\psi_{i}^{(+)}, i=3,4$ are acceptable. They correspond to positively charged fermions with positive spins (spin up) in the LLL.
\subsubsection*{Case 1: Properties of massive and negatively charged fermionic excitations in LLL}
According to (\ref{D6}), the one-loop corrected propagator of negatively charged and massive fermions does not receive any contribution from the one-loop fermion self-energy
\begin{eqnarray}\label{D16}
{\cal{S}}_{0}^{(-)}(\tilde{k})=\frac{{\cal{P}}_{-}\left(\gamma_{0}k_{0}
-\gamma_{3}k_{3}+m_{q}\right)}{{\cal{D}}_{0}^{(-)}}.
\end{eqnarray}
Here, ${\cal{D}}_{0}^{(-)}$ is given in (\ref{D7}). The eigenvectors of the numerator of ${\cal{S}}_{0}^{(-)}$ are
\begin{eqnarray}\label{D17}
\psi_{1}^{(-)}&=&(0,0,1,0),\nonumber\\
\psi_{2}^{(-)}&=&(1,0,0,0),\nonumber\\
\psi_{3}^{(-)}&=&\frac{1}{k_{3}}\left(0,|\mathbf{k}_{\|}|-k_{0},0,1\right),\nonumber\\
\psi_{4}^{(-)}&=&\frac{1}{k_{3}}\left(0,-|\mathbf{k}_{\|}|-k_{0},0,1\right),
\end{eqnarray}
with the eigenvalues
\begin{eqnarray}\label{D18}
\{0,0,m_{q}-|\mathbf{k}_{\|}|, m_{q}+|\mathbf{k}_{\|}|\}.
\end{eqnarray}
The eigenvectors from (\ref{D17}) satisfy
\begin{eqnarray}\label{D19}
\hspace{-0.3cm}\begin{array}{ccccc}
 {\cal{P}}_{+}\psi_{1}^{(-)}=\psi_{1}^{(-)},&
~~&
\mbox{or}&
~~&
\Sigma_{3}\psi_{1}^{(-)}=+\psi_{1}^{(-)},\\
 {\cal{P}}_{+}\psi_{2}^{(-)}=\psi_{2}^{(-)},&
~~&
\mbox{or}&
~~&
\Sigma_{3}\psi_{2}^{(-)}=+\psi_{2}^{(-)},\\
 {\cal{P}}_{-}\psi_{3}^{(-)}=\psi_{3}^{(-)},&
~~&
\mbox{or}&
~~&
\Sigma_{3}\psi_{3}^{(-)}=-\psi_{3}^{(-)},\\
 {\cal{P}}_{-}\psi_{4}^{(-)}=\psi_{4}^{(-)},&
~~&
\mbox{or}&
~~&
\Sigma_{3}\psi_{4}^{(-)}=-\psi_{4}^{(-)}.\\
\end{array}
\end{eqnarray}
Similar to the previous case, only $\psi_{i}^{(-)}, i=3,4$ corresponding to negatively charged fermions with negative spins (spin down) in the LLL, are acceptable.
In contrast to the case discussed in Sec. \ref{sec2a},  $\psi_{i}^{(+)}, i=3,4$ and $\psi_{i}^{(-)}, i=3,4$ are neither eigenvalues of the helicity nor those of the chirality operators. These operators are defined in (\ref{S17}) and (\ref{S18}), respectively.
\subsubsection*{Case 2: Properties of massless fermionic excitations in HLL}
The one-loop corrected fermion self-energy of massless fermions for $n\neq 0$ and $q=\pm 1$ is given by (\ref{D8})-(\ref{D11}) with $c_{\pm}^{(q)}=0$. To study the properties of possible fermionic excitations, let us simplify ${\cal{S}}_{n}^{(q)}$, in analogy to the results presented in Sec. \ref{sec2a}, as
\begin{eqnarray}\label{D20}
{\cal{S}}^{(q)}_{n}(\tilde{k})&=&\frac{{\cal{N}}_{L}^{(q)}(\tilde{k})}{2{\cal{D}}_{L}^{(q)}(\tilde{k})}+\frac{{\cal{N}}_{R}^{(q)}(\tilde{k})}{2{\cal{D}}_{R}^{(q)}(\tilde{k})},
\end{eqnarray}
with the numerators
\begin{eqnarray}\label{D21}
\lefteqn{
{\cal{N}}_{L}^{(q)}=\gamma_{0}[\p_{+}(a_{-}^{(q)}-b_{-}^{(q)})+\p_{-}(a_{+}^{(q)}+b_{+}^{(q)})]
}\nonumber\\
&&-2d^{(q)} \gamma_{2} \p_{R} +\gamma_{3}[\p_{+}(a_{-}^{(q)}-b_{-}^{(q)})-\p_{-}(a_{+}^{(q)}+b_{+}^{(q)})],\nonumber\\
\lefteqn{
{\cal{N}}_{R}^{(q)}=\gamma_{0}[\p_{+}(a_{-}^{(q)}+b_{-}^{(q)})+\p_{-}(a_{+}^{(q)}-b_{+}^{(q)})]
}\nonumber\\
&&- 2d^{(q)}\gamma_{2} \p_{L} -\gamma_{3}[\p_{+}(a_{-}^{(q)}+b_{-}^{(q)})-\p_{-}(a_{+}^{(q)}-b_{+}^{(q)})],\nonumber\\
\end{eqnarray}
and the denominators
\begin{eqnarray}\label{D22}
{\cal{D}}_{L/R}^{(q)}\equiv (a_{+}^{(q)}\pm b_{+}^{(q)})(a_{-}^{(q)}\mp b_{-}^{(q)})-d^{(q)2}.
\end{eqnarray}
Nontrivial eigenvectors of the numerators ${\cal{N}}_{L}^{(q)}$ and ${\cal{N}}_{R}^{(q)}$ are given by
\begin{eqnarray}\label{D23}
{\cal{V}}_{1}^{L}&=&(1,0,0,0),\nonumber\\
{\cal{V}}_{2}^{L}&=&(0,1,0,0),
\end{eqnarray}
and
\begin{eqnarray}\label{D24}
{\cal{V}}_{1}^{R}&=&(0,0,1,0),\nonumber\\
{\cal{V}}_{2}^{R}&=&(0,0,0,1),
\end{eqnarray}
respectively. They have trivial eigenvalues, and satisfy
\begin{eqnarray}\label{D25}
\begin{array}{rclcccrcl}
{\cal{P}}_{L}{\cal{V}}_{1}^{L}&=&{\cal{V}}_{1}^{L},&~&\mbox{and}&~&
\Sigma_{3}{\cal{V}}_{1}^{L}&=&+{\cal{V}}_{1}^{L},\\
{\cal{P}}_{L}{\cal{V}}_{2}^{L}&=&{\cal{V}}_{2}^{L},&~&\mbox{and}&~&
\Sigma_{3}{\cal{V}}_{2}^{L}&=&-{\cal{V}}_{2}^{L},\\
{\cal{P}}_{R}{\cal{V}}_{1}^{R}&=&{\cal{V}}_{1}^{R},&~&\mbox{and}&~&
\Sigma_{3}{\cal{V}}_{1}^{R}&=&+{\cal{V}}_{1}^{R},\\
{\cal{P}}_{R}{\cal{V}}_{2}^{R}&=&{\cal{V}}_{2}^{R},&~&\mbox{and}&~&
\Sigma_{3}{\cal{V}}_{2}^{R}&=&-{\cal{V}}_{2}^{R}.\\
\end{array}
\end{eqnarray}
As yet the above results (\ref{D23})-(\ref{D25}) are independent of the choice of electric charges $q$. They are valid for both positively and negatively charged particles, and can thus be summarized as
\begin{eqnarray}\label{D26}
{\cal{D}}_{L}^{(q)}:\left\{
\begin{array}{cccccc}
{\cal{V}}_{1}^{L}&~&\mbox{with}&q=+1,&s_{3}=\pm 1,&\chi=-1,\\
{\cal{V}}_{2}^{L}&~&\mbox{with}&q=-1,&s_{3}=\pm 1,&\chi=-1,\\
\end{array}
\right.\nonumber\\
{\cal{D}}_{R}^{(q)}:\left\{
\begin{array}{cccccc}
{\cal{V}}_{1}^{R}&~&\mbox{with}&q=+1,&s_{3}=\pm 1,&\chi=+1,\\
{\cal{V}}_{2}^{R}&~&\mbox{with}&q=-1,&s_{3}=\pm 1,&\chi=+1.\nonumber\\
\end{array}
\right.\\
\end{eqnarray}
Here, $s_{3}$ and $\chi$ denote the eigenvalues of the spin and chirality operators, $\Sigma_{3}$ and ${\cal{P}}_{L/R}$, respectively.  Let us notice that the appearance of two energy branches, arising from the poles of two denominators ${\cal{D}}_{L}^{(q)}$ and ${\cal{D}}_{R}^{(q)}$ in (\ref{D20}), can be regarded as an evidence of the appearance of additional fermionic excitations. In the next section, we will study the spectrum of Dirac particles at finite temperature and for nonvanishing magnetic fields. We will show, that in the limit of soft momenta and weak magnetic fields, new excitations appear, which will be referred to as hot and magnetized plasminos, in analogy to the excitations appearing at finite $T$ and vanishing $B$.
\section{Numerical results}\label{sec5}
\setcounter{equation}{0}
In this section, we will numerically solve the energy dispersion relations arising from the one-loop corrected fermion propagator, ${\cal{S}}_{n}^{(q)}$, for a number of special cases. In Sec. \ref{sec5a}, we will first consider ${\cal{S}}_{n}^{(q)}$ at nonzero $T$ and $B$ in the lowest ($n=0$) and higher ($n\neq 0$) Landau levels, separately. We will focus on both massive and massless case. In Sec. \ref{sec5b}, we will then determine the spectrum of massive fermions in a cold and magnetized electromagnetic plasma.
\subsection{Plasminos in a hot and magnetized QED plasma}\label{sec5a}
In the previous section, we have analytically determined the general structure of the one-loop corrected fermion propagator ${\cal{S}}_{n}^{(q)}$ in lowest and higher Landau levels. For $n=0$, ${\cal{S}}_{n}^{(q)}$ is given in (\ref{D6}), and for $n\neq 0$ in (\ref{D8})-(\ref{D11}) as well as in (\ref{D20})-(\ref{D22}) for the special case of massless fermions. Except for negatively charged fermions in the LLL, the one-loop corrected fermion propagator is, in particular, given in terms of nontrivial coefficients $a_{\pm}^{(q)}, b_{\pm}^{(q)}, c_{\pm}^{(q)}$ and $d^{(q)}$, which are defined in (\ref{D5}), and are given in terms of the coefficients $A_{\pm}^{(q)}, B_{\pm}^{(q)}, C_{\pm}^{(q)}$ and $D^{(q)}$, arising from the one-loop self-energy of charged fermions. In Sec. \ref{sec3b}, we have analytically determined $A_{\pm}^{(q)}, B_{\pm}^{(q)}, C_{\pm}^{(q)}$ and $D^{(q)}$ at finite $T$ and $B$, using two different methods: In Sec. \ref{sec3b1}, the exact expressions of these coefficients at finite $T$ and for nonvanishing $B$ are presented in (\ref{A29}) and (\ref{A30}), with ${\cal{I}}_{\ell}^{T}$ and ${\cal{J}}_{\ell}^{(i)T}, i=0,3$ from (\ref{A31}), and in Sec. \ref{sec3b2}, they are evaluated using a HTL approximation in a weak magnetic field, and are given by the same (\ref{A29}) and (\ref{A30}), with ${\cal{I}}_{\ell}^{T}$ and ${\cal{J}}_{\ell}^{(i)T}, i=0,3$ from (\ref{A40}).
\par
In order to explore the spectrum of fermionic excitations from the poles of ${\cal{S}}_{n}^{(q)}$ in the lowest and higher Landau levels, the coefficients $A_{\pm}^{(q)}, B_{\pm}^{(q)}, C_{\pm}^{(q)}$ and $D^{(q)}$ are first to be determined as functions of $k_{0}$ and $k_{3}$. To do this, we have numerically evaluated the integration over $\mathbf{p'}_{\perp}, x$ and $p_{3}$ appearing in (\ref{A29})-(\ref{A31}), as well as the integration over $y$ appearing in $\mathbf{I}_{0}, \mathbf{I}_{b/f}$ and $\bar{\mathbf{I}}_{b/f}$ from (\ref{A41}) for a large number of fixed $k_{0}$ and $k_{3}$. The summation over Landau levels appearing in (\ref{A29})-(\ref{A30}), has also been performed numerically. In this way, it was possible to find the best fits for $A_{\pm}^{(q)}, B_{\pm}^{(q)}, C_{\pm}^{(q)}$ and $D^{(q)}$ as functions of $k_{0}$ and $k_{3}$. We have then considered the fermion propagators in LLL ($n=0$) and HLL ($n=1$) separately,\footnote{For HLL, we considered only $n=1$ case. All $n\geq 1$ cases can be evaluated in a similar way.} and solved the energy dispersion relations, arising from the poles of the propagators. In what follows, we will only report the corresponding numerical results for a number of special cases.
\subsubsection{Special case: $T\neq 0$, $n=0$ for $m_{q}\neq 0$ and $m_{q}=0$}
\begin{figure}[hbt]
\includegraphics[width=8cm,height=6cm]{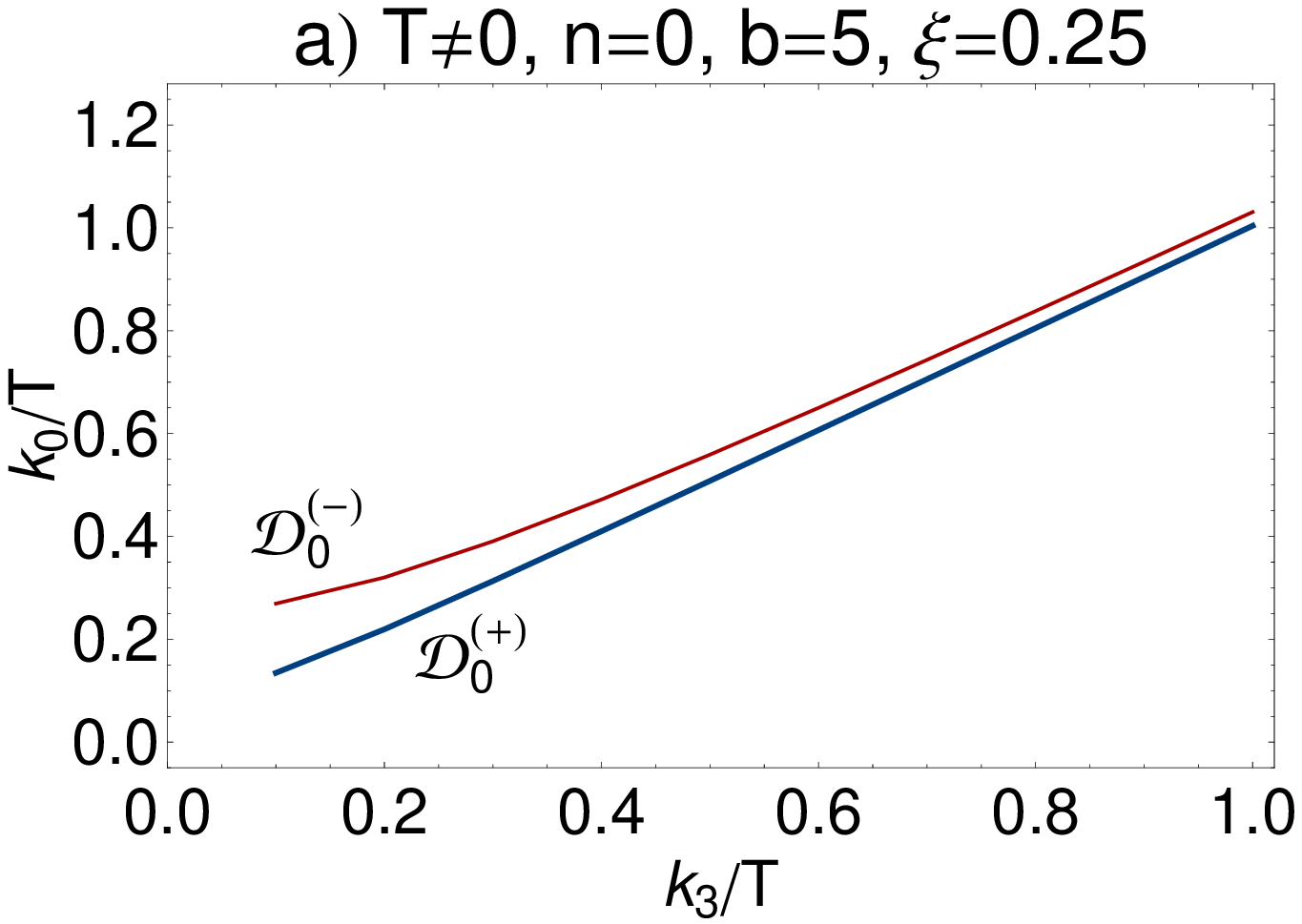}
\includegraphics[width=8cm,height=6cm]{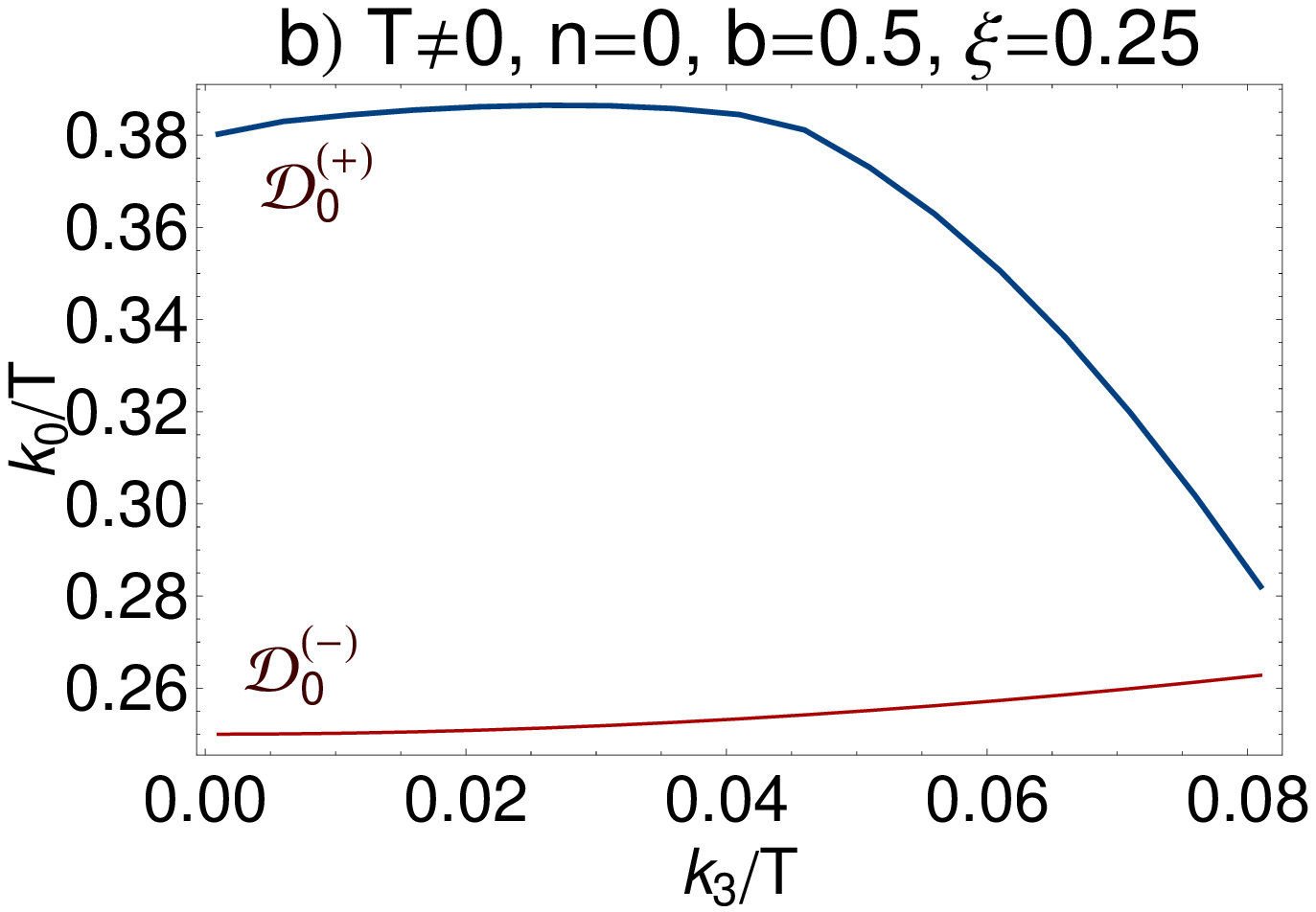}
\caption{(color online). a) The exact $k_{3}/T$  dependence of $k_{0}/T$ for positively (thick blue curve) and negatively (red thin curve) charged fermions in the LLL ($n=0$) for $b=eB/T^{2}=5$ and $\xi=m_{q}/T=0.25$, in the regime $k_{3}/T\geq 0.2$, which arises from (\ref{U1}) for $q=+1$ and (\ref{U2}) for $q=-1$. b) The HTL-approximated $k_{3}/T$ dependence of $k_{0}/T$ for positively (thick blue curve) and negatively (thin red curve) charged fermions in the LLL ($n=0$) for $b=\frac{eB}{T^{2}}=0.5$ and $\xi=\frac{m_{q}}{T}=0.25$, in the regime $k_{3}/T<0.2$.
 }\label{fig3}
\end{figure}
\begin{figure}[hbt]
\includegraphics[width=8cm,height=6cm]{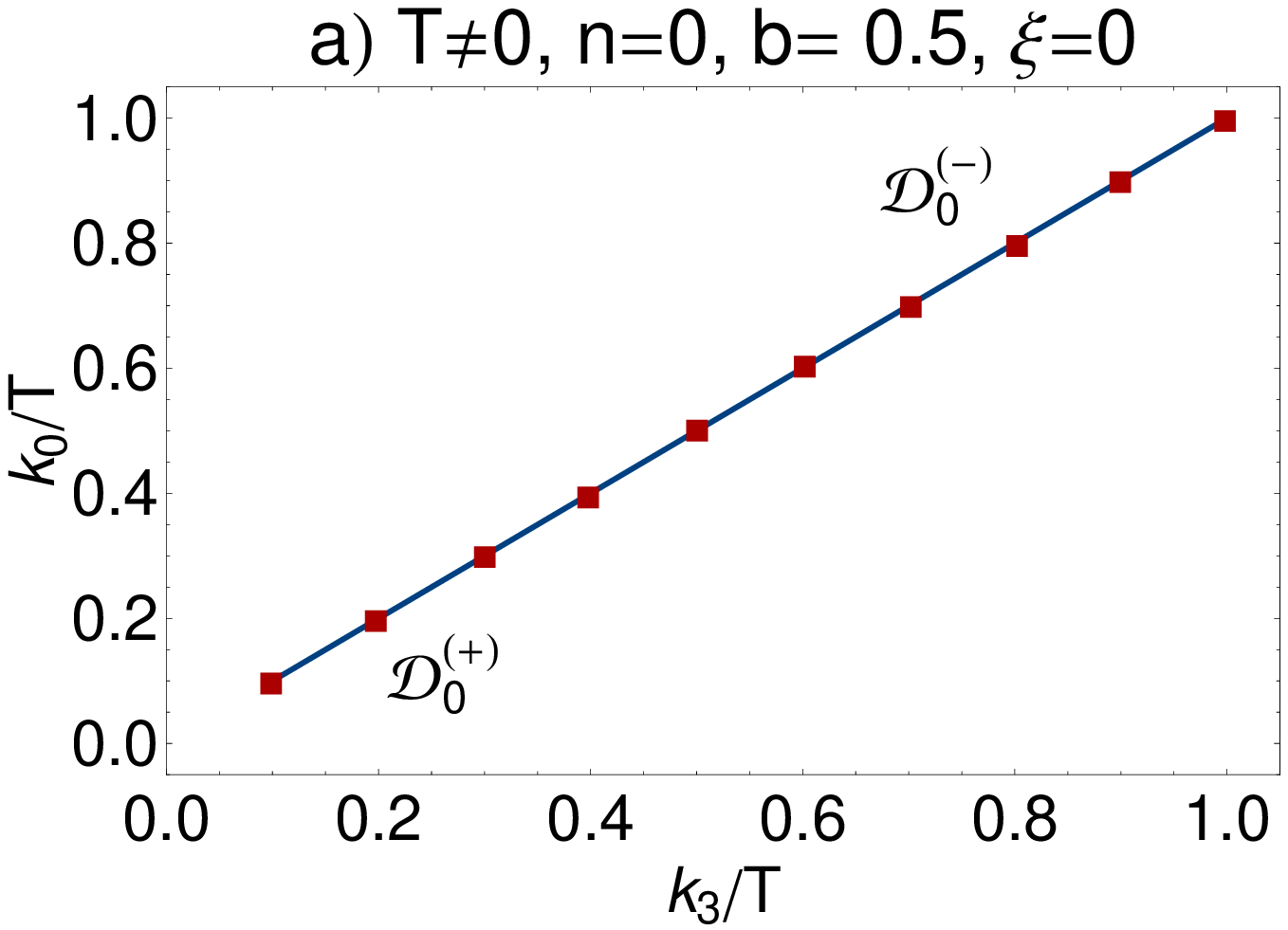}
\includegraphics[width=8cm,height=6cm]{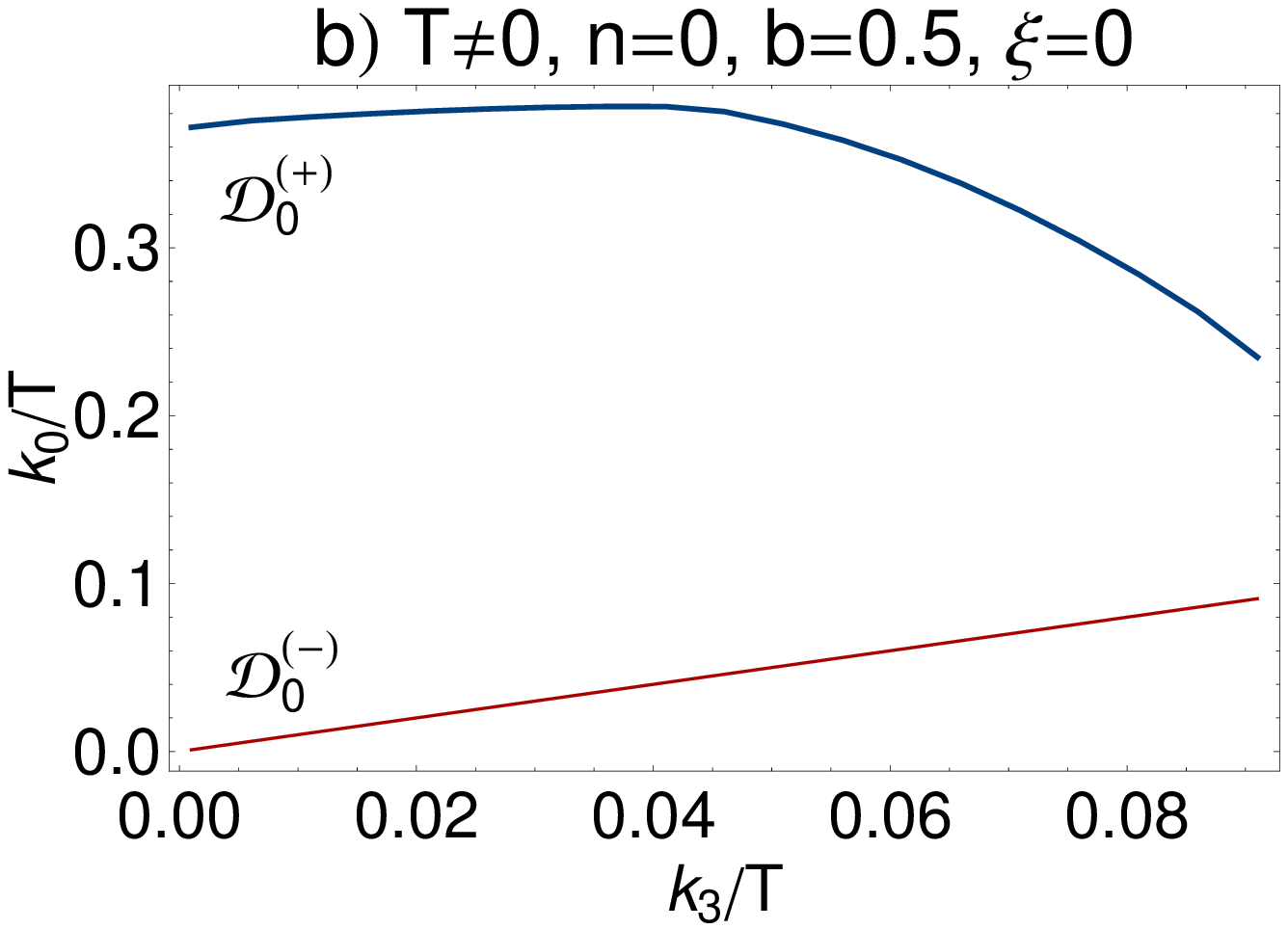}
\caption{
a) The exact $k_{3}/T$  dependence of $k_{0}/T$ for positively (blue curve) and negatively (red squares) charged fermions in the LLL ($n=0$) for $b=eB/T^{2}=0.5$ and $\xi=m_{q}/T=0$, in the regime $k_{3}/T\geq 0.2$, which arises from (\ref{U3}) for $q=+1$ and (\ref{U4}) for $q=-1$. b) The HTL-approximated $k_{3}/T$ dependence of $k_{0}/T$ for positively (thick blue curve) and negatively (thin red curve) charged fermions in the LLL ($n=0$) for $b=eB/T^{2}=0.5$ and $\xi=m_{q}/T=0$, in the regime $k_{3}/T<0.2$.
 }\label{fig4}
\end{figure}
\begin{figure}[hbt]
\includegraphics[width=8cm,height=6cm]{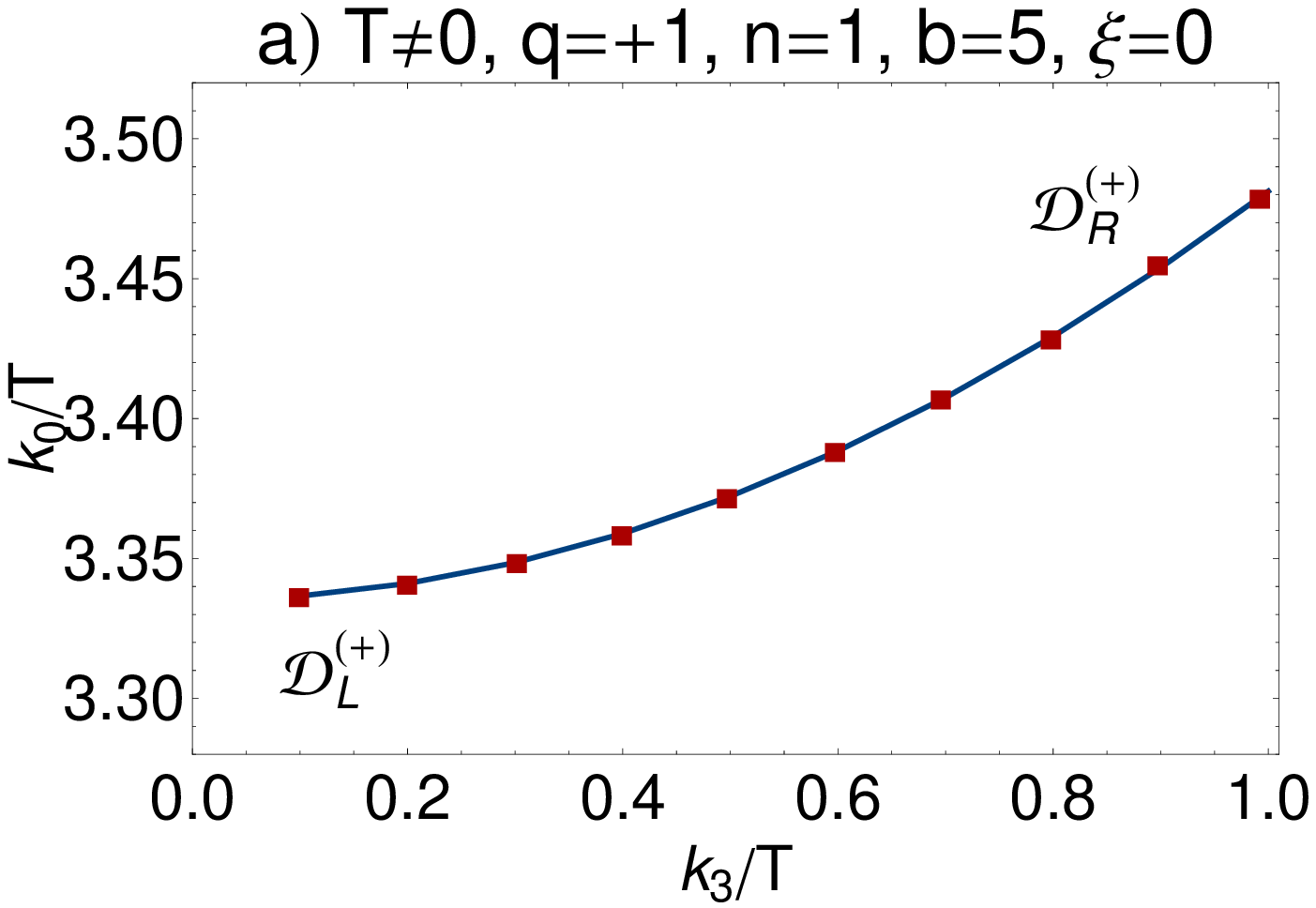}
\includegraphics[width=8cm,height=6cm]{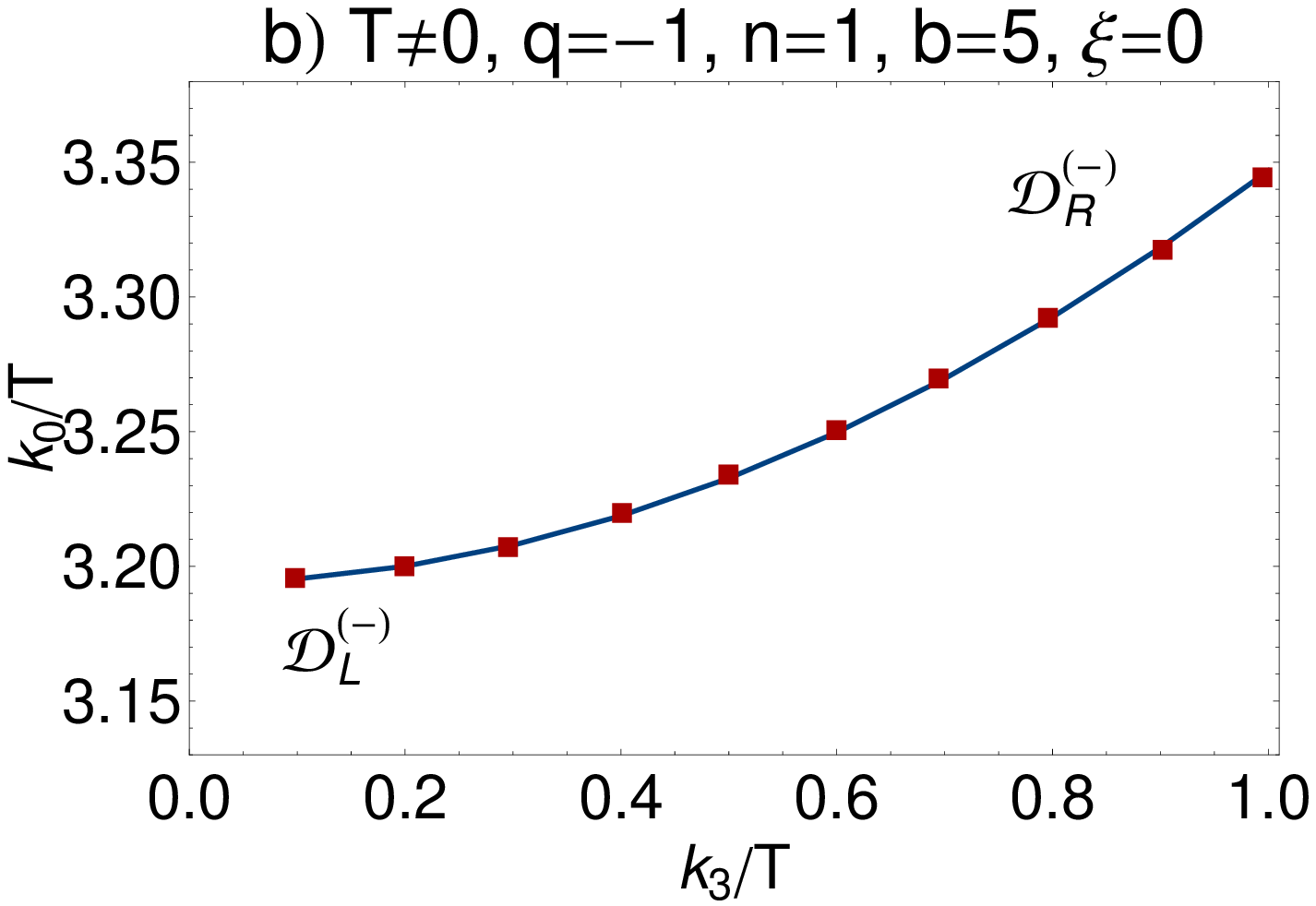}
\caption{(color online). The exact $k_{3}/T$ dependence of $k_{0}/T$ for positively (panel a) and negatively (panel b) charged fermions for $n=1, b=eB/T^{2}=5$ and $\xi=m_{q}/T=0$ in the regime $k_{3}/T\geq 0.2$. The results arise from the dispersion relations (\ref{U5}) and (\ref{U6}) for left- and right-handed massless fermions. Blue curves and red squares denote the solutions of ${\cal{D}}_{L}^{(\pm)}=0$ and ${\cal{D}}_{R}^{(\pm)}=0$, respectively.
 }\label{fig5}
\end{figure}
\begin{figure}[hbt]
\includegraphics[width=8cm,height=6cm]{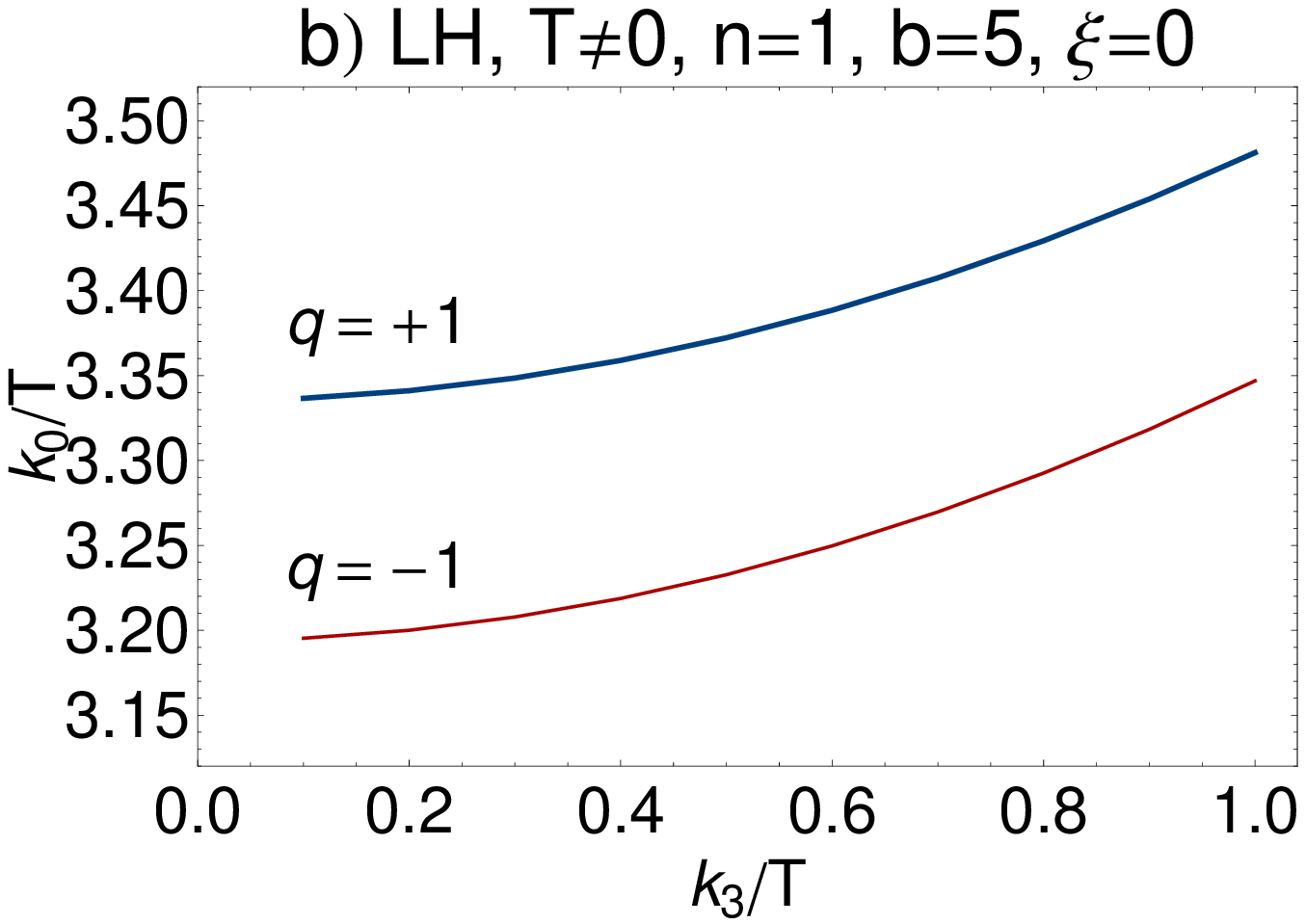}
\includegraphics[width=8cm,height=6cm]{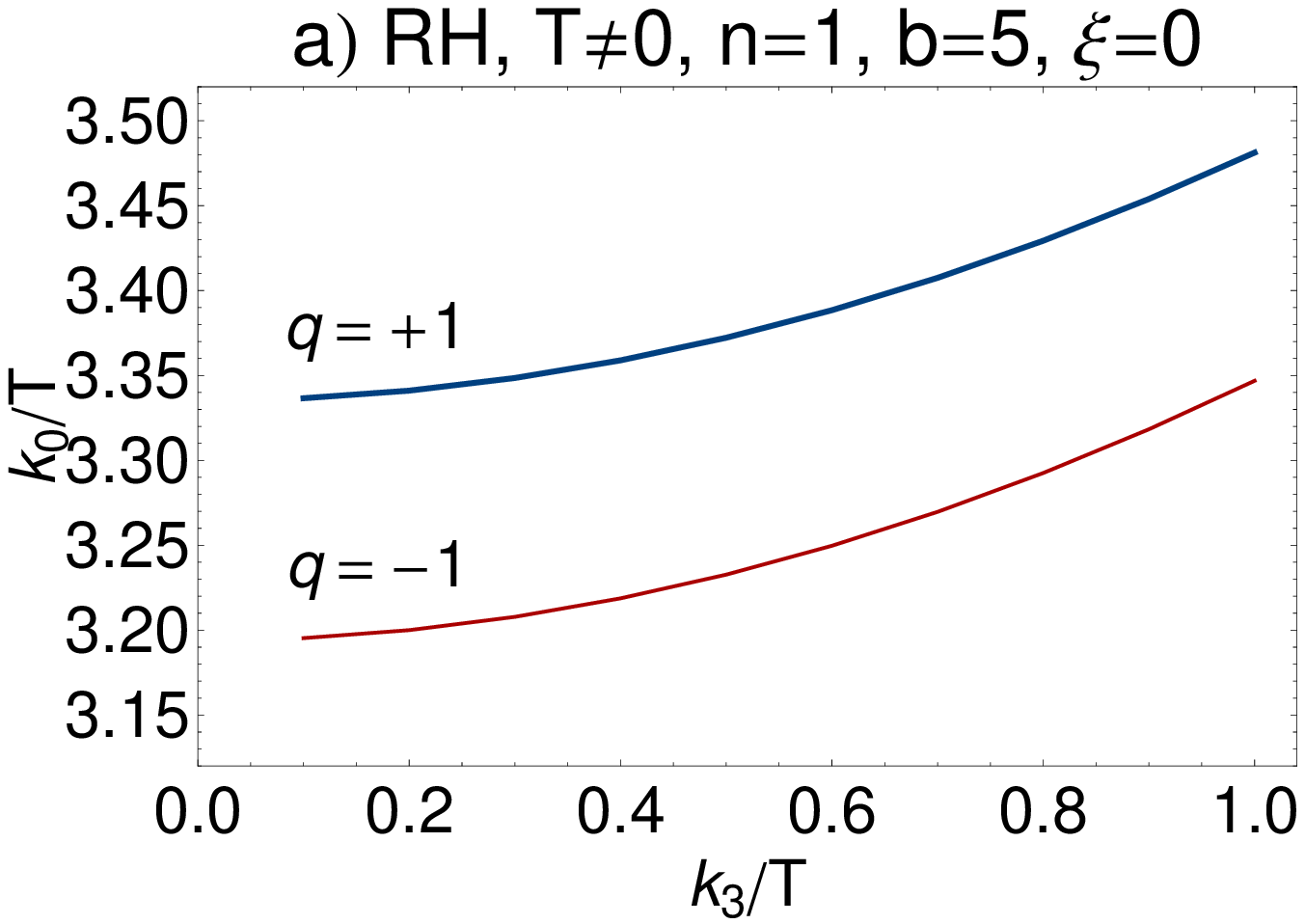}
\caption{(color online). The exact $k_{3}/T$ dependence of $k_{0}/T$ for left-handed (panel a) and right-handed (panel b) fermions. Thick blue and thin red curves denote the solutions of ${\cal{D}}_{L/R}^{(+)}=0$ and ${\cal{D}}_{L/R}^{(-)}=0$, for positively and negatively charged particles $q=\pm 1$  for $n=1, b=eB/T^{2}=5$ and $\xi=m_{q}/T=0$ in the regime $k_{3}/T\geq 0.2$.
 }\label{fig6}
\end{figure}
\begin{figure}[hbt]
\includegraphics[width=8cm,height=6cm]{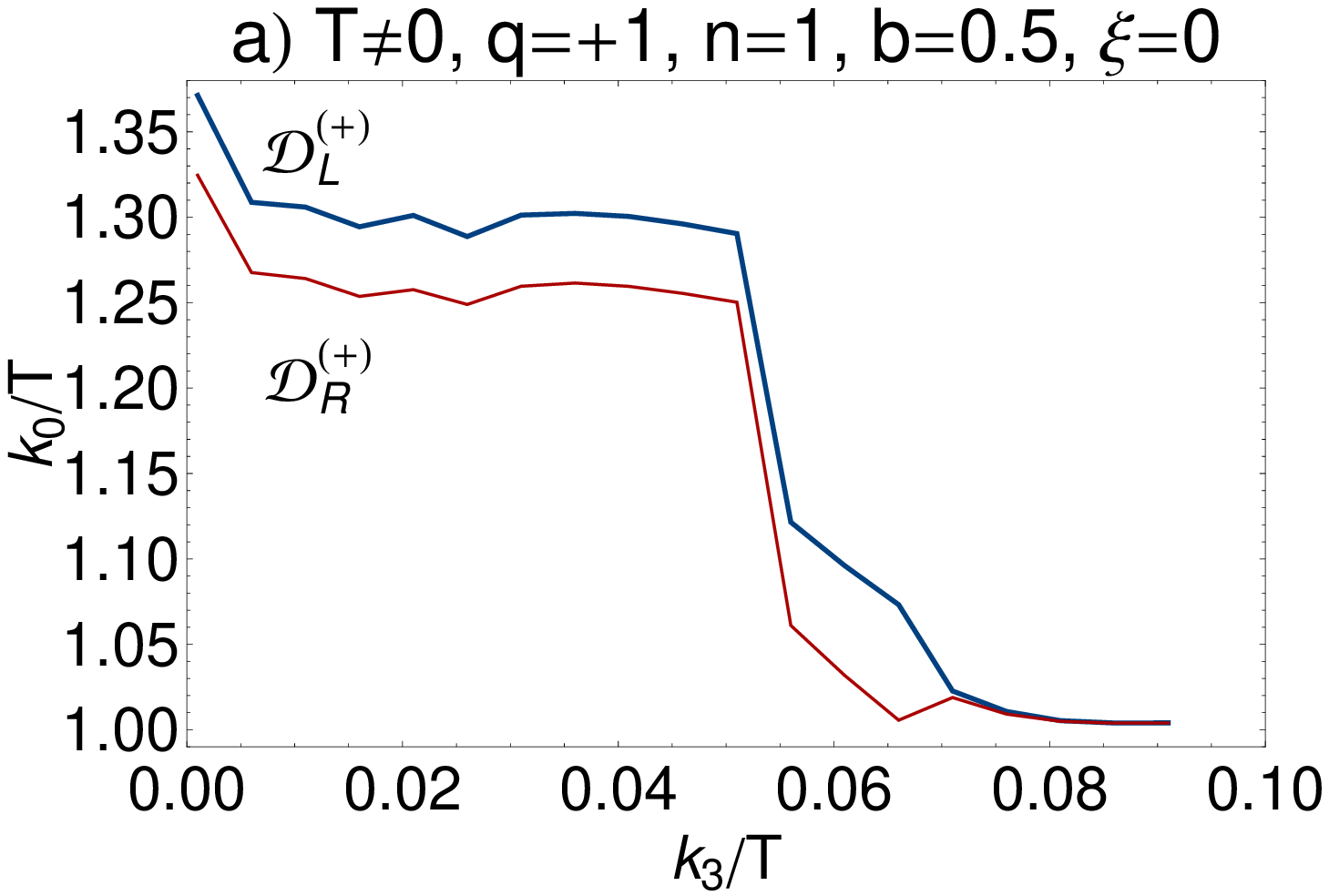}
\includegraphics[width=8cm,height=6cm]{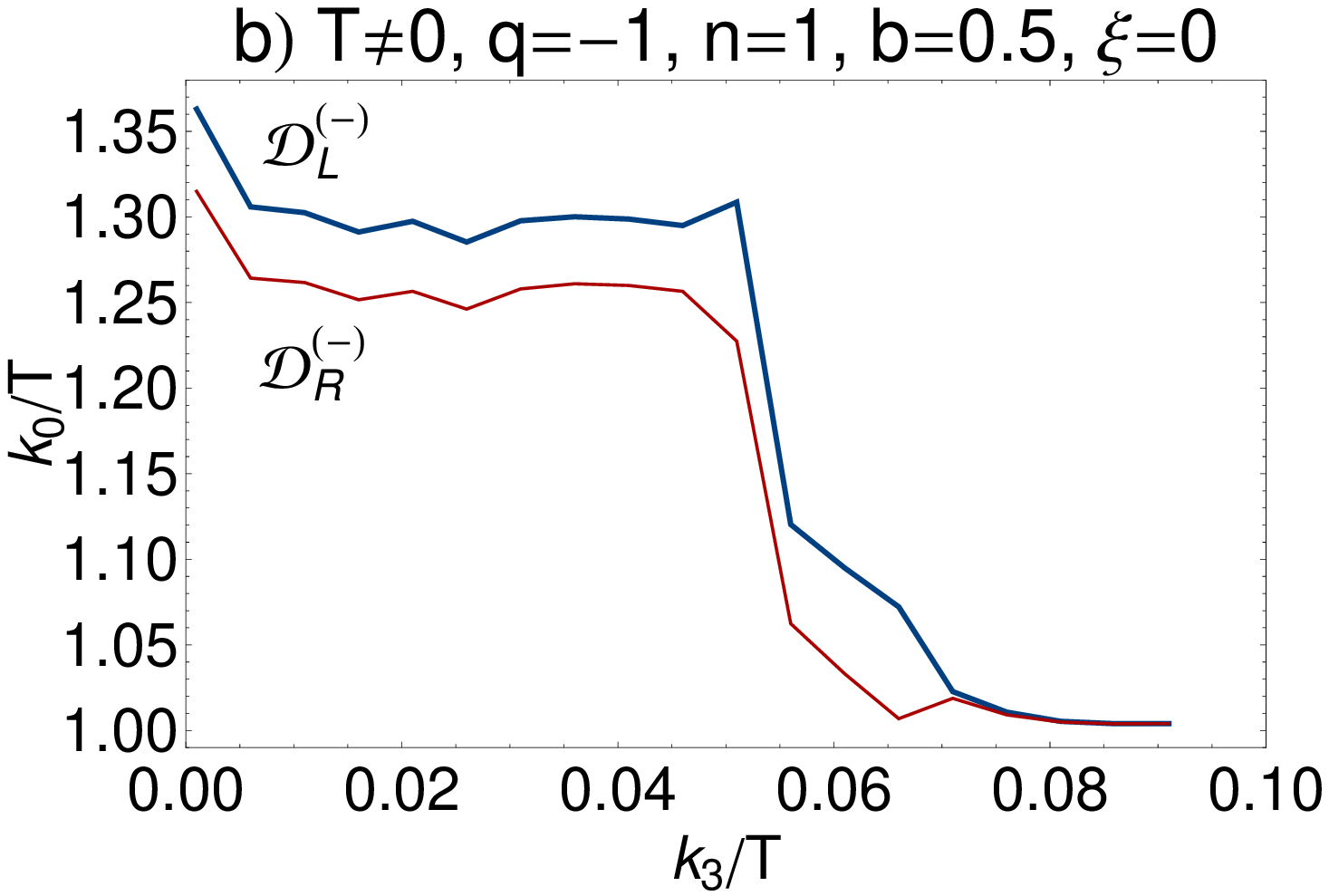}
\caption{(color online). The HTL-approximated $k_{3}/T$ dependence of $k_{0}/T$ for positively (panel a) and negatively (panel b) charged fermions for $n=1, b=eB/T^{2}=0.5$ and $\xi=m_{q}/T=0$ in the regime $k_{3}/T< 0.2$. The results arise from the dispersion relations (\ref{U5}) and (\ref{U6}) for left- and right-handed massless fermions. Thick blue and thin red curves denote the solutions of ${\cal{D}}_{L}^{(\pm)}=0$ and ${\cal{D}}_{R}^{(\pm)}=0$, respectively.
 }\label{fig7}
\end{figure}
\begin{figure}[hbt]
\includegraphics[width=8cm,height=6cm]{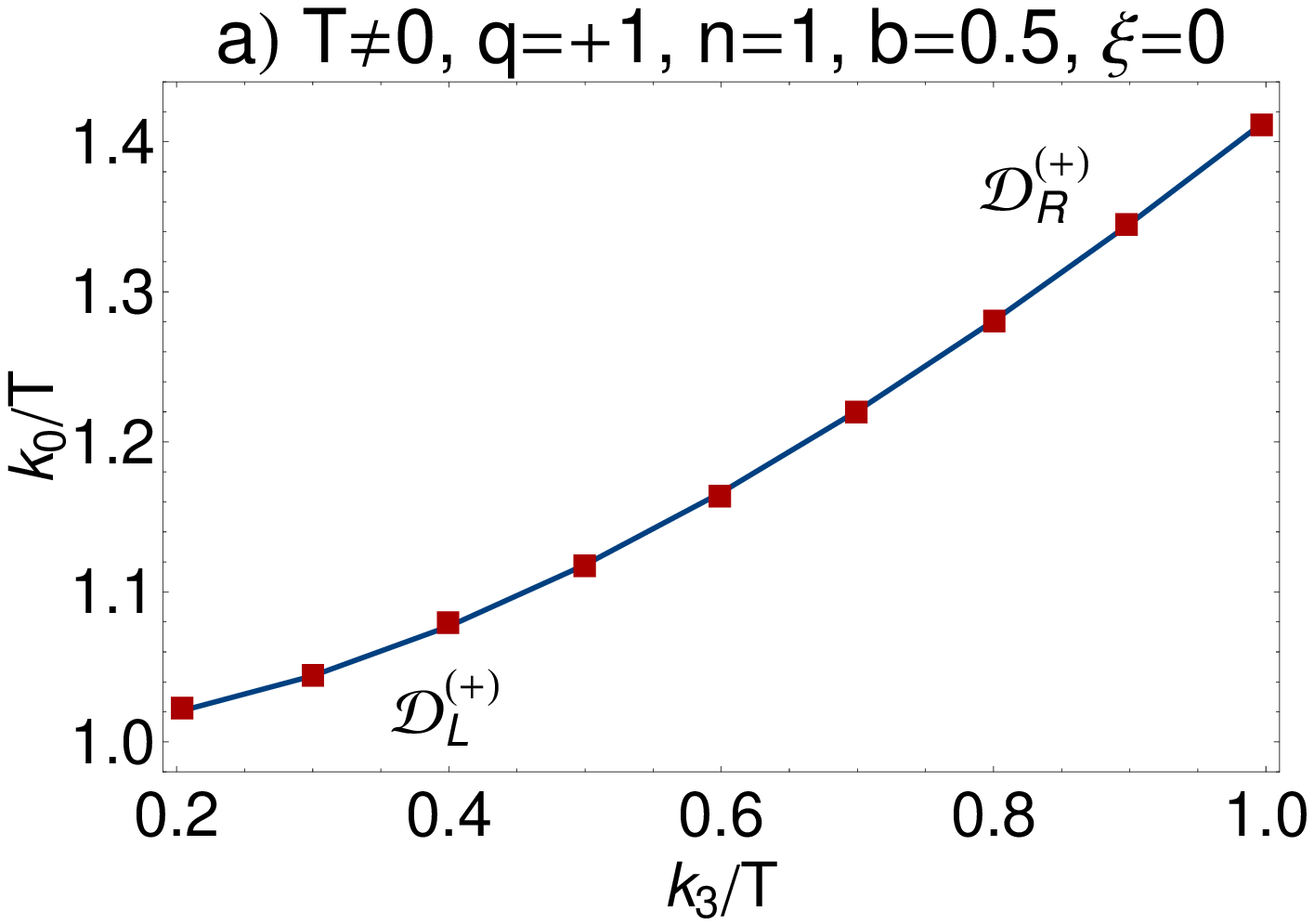}
\includegraphics[width=8cm,height=6cm]{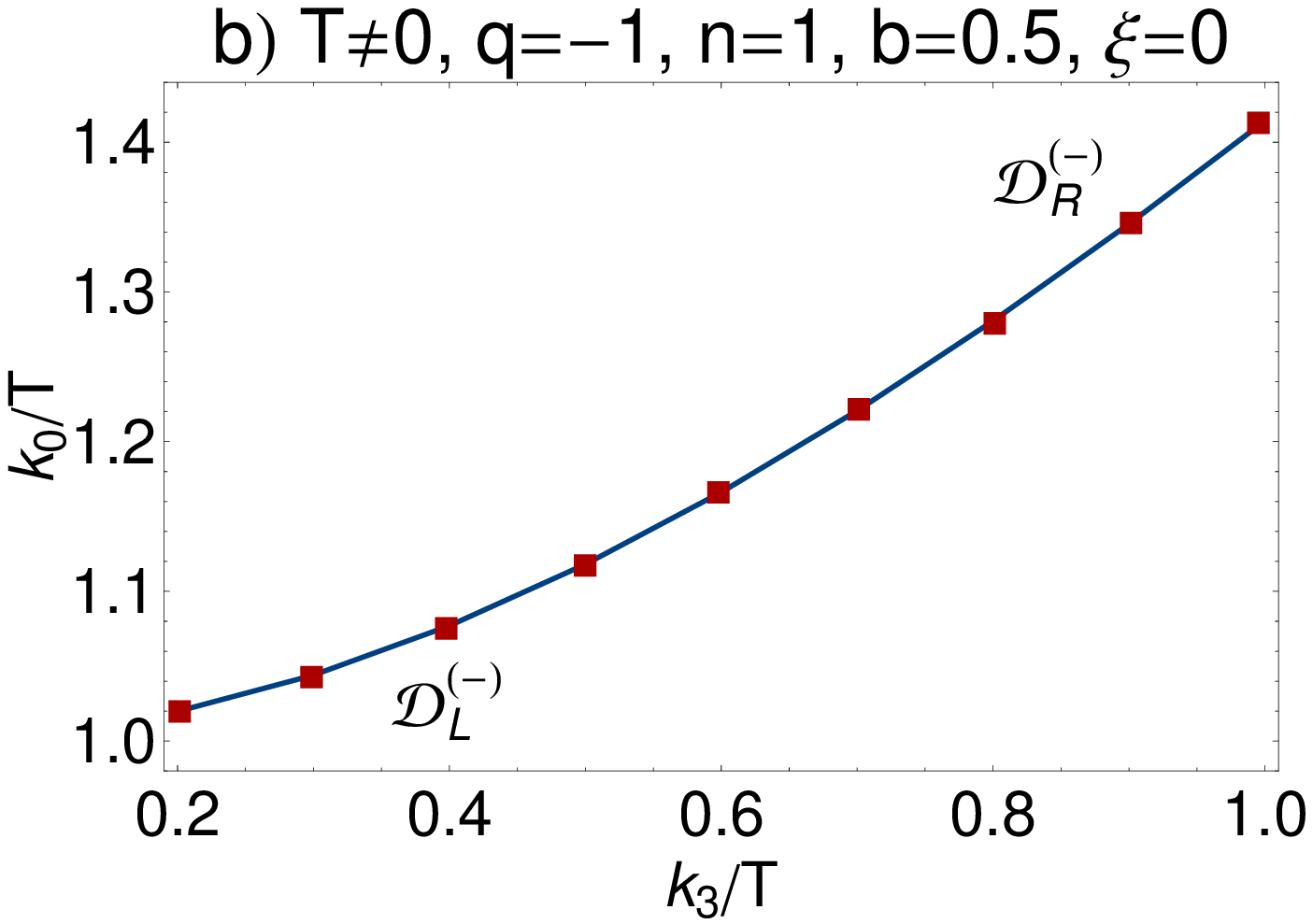}
\caption{(color online). The HTL-approximated  $k_{3}/T$ dependence of $k_{0}/T$ for positively (panel a) and negatively (panel b) charged fermions for $n=1, b=eB/T^{2}=0.5$ and $\xi=m_{q}/T=0$ in the regime $k_{3}/T\geq  0.2$. The results arise from the dispersion relations (\ref{U5}) and (\ref{U6}) for left- and right-handed massless fermions. Thick blue curves and red squares denote the solutions of ${\cal{D}}_{L}^{(\pm)}=0$ and ${\cal{D}}_{R}^{(\pm)}=0$, respectively.
 }\label{fig8}
\end{figure}

Let us start with the dispersion relation of a massive and positively charged fermion in the LLL. It can be obtained by numerically solving $\mbox{det}(\mathbf{\ktildes}_{\|}-m_{q}-\widetilde{\Sigma}_{0}^{(+)})=0$ from (\ref{D4}) or ${\cal{D}}_{0}^{(+)}(\tilde{k})=0$ with ${\cal{D}}_{0}^{(+)}$ from (\ref{D7}). In both cases, it reads
\begin{eqnarray}\label{U1}
k_{0}^{2}(1-A_{+}^{(+)})^{2}-k_{3}^{2}(1-B_{+}^{(+)})^{2}-m_{q}^{2}(1+C_{+}^{(+)})^{2}=0.\nonumber\\
\end{eqnarray}
As concerns the negatively charged massive fermions, the corresponding dispersion relation arises from ${\cal{D}}_{0}^{(-)}=0$,
\begin{eqnarray}\label{U2}
k_{0}^{2}=k_{3}^{2}+m_{q}^{2}.
\end{eqnarray}
Here, the definition of ${\cal{D}}_{0}^{(-)}$ from (\ref{D7}) is used. The numerical results are demonstrated in Fig. \ref{fig3}. In Fig. \ref{fig3}(a), the $k_{3}/T$ dependence of $k_{0}/T$ is plotted for positively (thick blue curve) and negatively (thin red curve) charged massive fermions in the LLL ($n=0$) for $b\equiv eB/T^{2}=5$, $\xi\equiv m_{q}/T=0.25$ and in the regime $k_{3}/T>0.2$.\footnote{At finite temperature, the free parameter $eB$ and $m_{q}$ are scaled with $T$.} To do this, we have used (\ref{A29}) with ${\cal{I}}_{\ell}^{T}$ and ${\cal{J}}_{\ell}^{(i)T}, i=0,3$ from (\ref{A31}), and numerically determined the dependence of $A_{+}^{(+)}, B_{+}^{(+)}$ and $C_{+}^{(+)}$, appearing in (\ref{U1}), on ${k}_{0}$ and $k_{3}$. This results in the thick blue curve in Fig. \ref{fig3}(a). The thin red curve in Fig. \ref{fig3}(a) is the positive energy branch of the dispersion relation of negatively charged massive fermions from (\ref{U2}) for $b=5$ and $\xi=0.25$.\footnote{The negative energy branches of the energy dispersion relations are not plotted in the figures demonstrated in this section.} The regime $k_{3}/T<0.2$ is then explored using  $A_{+}^{(+)}, B_{+}^{(+)}$ and $C_{+}^{(+)}$ from (\ref{A29}) with ${\cal{I}}_{\ell}^{T}$ and ${\cal{J}}_{\ell}^{(i)T}, i=0,3$ from (\ref{A40}) for $b=0.5$ and $\xi=0.25$. In this regime the HTL approximation described in Sec. \ref{sec3b} might be reliable. Similar to Fig. \ref{fig3}(a), the thick blue and thin red curves correspond to the positive energy branches of the dispersion relations (\ref{U1}) and (\ref{U2}), respectively.
Let us notice at this stage, that according to the results from (\ref{D15}) and (\ref{D19}), and the subsequent descriptions, the blue and red curves in Fig. \ref{fig3} are the energy dispersion relations of positively and negatively charged massive fermions with positive and negative spins.
\par
Comparing the energies of fermions with positive and negative spins from Fig. \ref{fig3}(a) [exact results] with the corresponding energies of these fermions from Fig. \ref{fig3}(b) [HTL-approximated results], it turns out that in the regime $k_{3}/T\geq 0.2$, their $T$-scaled energies, $k_{0}/T$, increase with increasing $k_{3}/T$, while in the regime $k_{3}/T<0.2$, the energy of fermions with positive spins (thick blue curve) decreases with increasing $k_{3}/T$, in contrast to the energy of fermions with negative spins (thin red curve).
\par
Performing the above analysis for the case of massless fermions in the LLL, it turns out that the small difference between the energies of fermions with positive and negative spins, appearing in Fig. \ref{fig3}(a) disappears in the limit of vanishing fermionic mass. This can be observed in Fig. \ref{fig4}(a), where the exact $k_{3}/T$ dependence of $k_{0}/T$ is plotted for $n=0$, $b=0.5$ and $\xi=0$ in the regime $k_{3}/T\geq 0.2$. Here, the energy dispersion relations
\begin{eqnarray}\label{U3}
k_{0}^{2}(1-A_{+}^{+})^{2}-k_{3}^{2}(1-B_{+}^{+})^{2}=0,
\end{eqnarray}
for positively charged fermions, and
\begin{eqnarray}\label{U4}
 k_{0}^{2}=k_{3}^{2},
\end{eqnarray}
for negatively charged fermions, are used. In the regime $k_{3}/T<0.2$, however, the qualitative behavior of the $k_{3}/T$ dependence of $k_{0}/T$ for massless fermions is similar to the corresponding results for the massive fermions [see the plots in Fig. \ref{fig4}(b) and compare them with the plots in Fig. \ref{fig3}(b)].
Let us notice, that although the fermionic excitations, whose energy dispersion relations are plotted in Figs. \ref{fig4}(b), are massless, their rest masses (for $k_{3}/T=0$) are nonzero [see, in particular, the rest mass of the positively charged massless fermion from Fig. \ref{fig4}(b)].
This effect is related to the magnetic catalysis \cite{miransky1995}, which is characterized by a dynamical generation of mass in the presence of very strong magnetic field, where the system is dominated by the LLL. As it is well-known, the dynamical mass arises from perturbative corrections to fermion propagator. According to the same argument, the fact that the negatively charged massless fermions have a zero rest mass [see the thin red curve in Fig. \ref{fig4}(b)] is therefore related to the fact that these particles do not receive any contribution from the one-loop self-energy in the LLL.
\subsubsection{Special case: $T\neq 0$, $n=1$ and $m_{q}=0$}
The one-loop corrected fermion self-energy of massless fermions in HLL is analytically computed in the previous section [see (\ref{D20})-(\ref{D22})]. In this case, in analogy to the case of vanishing magnetic fields, described in Sec. \ref{sec2a}, two denominators ${\cal{D}}_{L/R}^{(q)}$ appear for each $q=+1$ and $q=-1$ in (\ref{D20}). We therefore expect two different energy branches for each $q$. The energy branch arising from ${\cal{D}}_{L}^{(q)}=0$ corresponds to a left-handed positively or negatively charged particles, possessing both spin orientations. Similarly, the energy branch arising from ${\cal{D}}_{R}^{(q)}=0$ corresponds to a right-handed positively or negatively charged particles with positive and negative spins. The explicit expressions for the energy dispersion relations of these particles are given by
\begin{eqnarray}\label{U5}
\lefteqn{[k_{0}(1-A_{-}^{(q)})+k_{3}(1-B_{-}^{(q)})]
}\nonumber\\
&&\times [k_{0}(1-A_{+}^{(q)})-k_{3}(1-B_{+}^{(q)})] -2n|qeB|(1-D^{(q)})^{2}\nonumber\\
&&=0,
\end{eqnarray}
and
\begin{eqnarray}\label{U6}
\lefteqn{
[k_{0}(1-A_{-}^{(q)})-k_{3}(1-B_{-}^{(q)})]}\nonumber\\
&&\times [k_{0}(1-A_{+}^{(q)})+k_{3}(1-B_{+}^{(q)})]-2n|qeB|(1-D^{(q)})^{2}\nonumber\\
&&=0.
\end{eqnarray}
They arise from ${\cal{D}}_{L}^{(q)}=0$ and  ${\cal{D}}_{R}^{(q)}=0$, with  ${\cal{D}}_{L/R}^{(q)}$ defined in (\ref{D22}).
To solve these relations, let us first consider the coefficients $A_{\pm}^{(q)}, B_{\pm}^{(q)}$ and $D^{(q)}$ from (\ref{A29}) for $q=+1$ and (\ref{A30}) for $q=-1$ with ${\cal{I}}_{\ell}^{T}$ as well as ${\cal{J}}_{\ell}^{(i)T}, i=0,3$ from (\ref{A31}). This yields the exact $k_{3}/T$ dependence of $k_{0}/T$ for massless left- and right-handed fermions. For $n=1, b=eB/T^{2}=5, \xi=m_{q}/T=0$, the results are demonstrated in Figs. \ref{fig5}(a) and \ref{fig5}(b), in the regime $k_{3}/T\geq 0.2$. They correspond to $q=+1$ and $q=-1$, respectively. As it turns out, there is no difference between the solutions of ${\cal{D}}_{L}^{(+)}=0$ (blue curve) and ${\cal{D}}_{R}^{(+)}=0$ (red square) for positively charged particles. Similarly, the $k_{3}/T$ dependence of $k_{0}/T$ for negatively charged left- (blue curve) and right-handed (red-square) massless fermions are identical.
\par
To compare the energy dispersion relations of massive fermions for $q=+1$ and $q=-1$,  we have plotted in Fig. \ref{fig6} the exact $k_{3}/T$ dependence of $k_{0}/T$ for massless left- (panel a) and right-handed (panel b) fermions. Thick blue and thin red curves denote the solutions of ${\cal{D}}_{L/R}^{(+)}=0$ and ${\cal{D}}_{L/R}^{(-)}=0$, for $n=1, b=5$ and $\xi=0$ in the regime $k_{3}/T\geq 0.2$. As it turns out, the energy increases with increasing $k_{3}/T\geq 0.2$. Moreover, the energy of negatively charged left- and right-handed massless fermions are smaller that their positively charged counterparts.
\par
The regime $k_{3}/T<0.2$ is explored in Fig. \ref{fig7}, where the HTL-approximated solutions of the energy dispersion relations (\ref{U5}) for left-handed and (\ref{U6}) for right-handed massless fermions are demonstrated. To determine these solutions, the coefficients $A_{\pm}^{(q)}, B_{\pm}^{(q)}$ and $D^{(q)}$
from (\ref{A29}) and (\ref{A30}) with the HTL-approximated ${\cal{I}}_{\ell}^{T}$ as well as ${\cal{J}}_{\ell}^{(i)T}, i=0,3$ from (\ref{A40}) are used. They are first numerically computed as functions of $k_{0}$ and $k_{3}$ for $n=1, b=eB/T^{2}=0.5$ and $\xi=0$.
Plugging then the resulting expressions in (\ref{U5}) and (\ref{U6}), these energy dispersion relations are numerically solved. The thick blue and thin red curves in Figs. \ref{fig7}(a) and \ref{fig7}(b) denote the solutions of ${\cal{D}}_{L}^{\pm}=0$ and ${\cal{D}}_{R}^{(\pm)}=0$, respectively. As it turns out, in the regime $k_{3}<0.08 T$, the energy branches for the left- and right-handed positively [Fig. \ref{fig7}(a)] and negatively [Fig. \ref{fig7}(b)] charged massless fermions are split up, in contrast to the results for $k_{3}>0.08 T$ from Figs. \ref{fig7} and \ref{fig8}, where it is shown that left- and right-handed massless fermions have the same energies. In Fig. \ref{fig8}, we have used the above HTL approximation to determine the $k_{3}/T$ dependence of $k_{0}/T$ for left- and right-handed massless fermions in the regime $k_{3}/T>0.2$. Here, we have chosen the same free parameters, $n=1, b=0.5$ and $\xi=0$, as in Fig. \ref{fig7}. The blue curves and red squares in Fig. \ref{fig8} denote the solutions of ${\cal{D}}_{L}^{(\pm)}=0$ and ${\cal{D}}_{R}^{(\pm)}=0$, respectively. The results from Fig. \ref{fig8} coincide qualitatively with the results from Fig. \ref{fig5}, where the \textit{exact} solutions of
(\ref{U5}) and (\ref{U6}) are demonstrated for $n=1, b=5$ and $\xi=0$ in the same $k_{3}/T>0.2$ regime.
\par
Let us notice at this stage, that the appearance of two different energy branches in Fig. \ref{fig7} is in great resemblance with the appearance of two energy branches for fermionic particle and plasmino excitations at finite temperature and zero magnetic fields. The latter case is discussed in Sec. \ref{sec2a}, where it was shown that two energy branches in Fig. \ref{fig2} correspond to fermionic excitations with positive (particle) and negative (plasmino) helicity to chirality ratios.
At finite temperature and for nonzero magnetic fields, discussed in the present section, however, the massless fermionic excitations, whose energy dispersions are demonstrated in Fig. \ref{fig7}, are, in contrast to the ordinary particle and plasmino modes at finite $T$, eigenstates of the chirality and the spin operators, ${\cal{P}}_{L/R}$ and $\Sigma_{3}$ [see Sec. \ref{sec4} for a proof].
Despite this difference, and because of the similarity between the production mechanism of new excitations in the case of nonvanishing $T$ and $B$ with the mechanism leading to particle and plasmino excitations at finite $T$ and zero $B$, the new excitations will be referred to as \textit{hot magnetized plasminos}. The crucial point is that, according to our findings, they seem to appear only in the limit of soft momenta $k_{3}\ll T$ and weak magnetic fields $eB\ll T^{2}$, where a HTL approximation is reliable, and where apart from LLL, higher Landau levels are to be taken into account.
\par
In the rest of this section, we will show that similar excitations appear at zero temperature and for nonvanishing moderate magnetic fields.
\subsection{Plasminos in a cold and magnetized QED plasma}\label{sec5b}
As we have described in the previous section, according to the notations used in the present paper, the zero temperature case can be regarded as a special case of the finite temperature case. This is because the similarity/difference between the general structure of one-loop self-energy of fermions at zero and nonzero temperatures from (\ref{A18}) and (\ref{A32}). Thus, in the zero temperature case, the one-loop corrected propagator of fermions in the presence of a constant magnetic field is, as before, given by (\ref{D6})-(\ref{D7}) for $n=0$ and (\ref{D8})-(\ref{D11}) for $n\geq 1$. The only difference with the finite temperature case is that the coefficients $a_{\pm}^{(q)}, b_{\pm}^{(q)}, c_{\pm}^{(q)}$ and $d^{(q)}$, appearing in these equations, are to be redefined as
\begin{eqnarray}\label{U7}
a_{\pm}^{(q)}&\equiv& k_{0}(1-A_{\pm}^{(q)}),\nonumber\\
b_{\pm}^{(q)}&\equiv& k_{3}(1-A_{\pm}^{(q)}),\nonumber\\
c_{\pm}^{(q)}&\equiv& m_{q}(1+C_{\pm}^{(q)}), \nonumber\\
d^{(q)}&\equiv&s_{q}\sqrt{2n|qeB|}(1-D^{(q)}),
\end{eqnarray}
with $A_{\pm}^{(q)}, C_{\pm}^{(q)}$ and $D^{(q)}$ presented in (\ref{A13}) for $q=+1$ and (\ref{A14}) for $q=-1$. Similar to what is performed in the previous section, we have determined numerically the latter coefficients for a large number of $k_{0}$ and $k_{3}$. In this way, we were able to find the best fits for  $A_{\pm}^{(q)}, C_{\pm}^{(q)}$ and $D^{(q)}$ as functions of $k_{0}$ and $k_{3}$. Plugging then the corresponding expressions in $a_{\pm}^{(q)}, b_{\pm}^{(q)}, c_{\pm}^{(q)}$ and $d^{(q)}$ from (\ref{U7}), we have numerically solved the energy dispersion relations for cold fermions in a magnetized QED plasma. In what follows, we will separately consider two different cases of $n=0$ and $n=1$ for massive fermions, and will present the $k_{3}/m_{q}$ dependence of $k_{0}/m_{q}$ for $b\equiv eB/m_{q}^{2}=5$.\footnote{At zero temperature, the free parameter $eB$ is scaled with the fermionic mass $m_{q}$ instead of with the temperature $T$.}
\subsubsection{Special case: $T=0, n=0, m_{q}\neq 0$}
According to the definitions of ${\cal{D}}_{0}^{(+)}$ and ${\cal{D}}_{0}^{(-)}$ from (\ref{D7}) with $a_{\pm}^{(q)}, b_{\pm}^{(q)}, c_{\pm}^{(q)}$ from (\ref{U7}), the energy dispersion relations of positively and negatively charged massive fermions are given by
\begin{eqnarray}\label{U8}
k_{0}^{2}(1-A_{+}^{+})^{2}-k_{3}^{2}(1-A_{+}^{+})^{2}-m_{q}^{2}(1+C_{+}^{+})^{2}=0,
\end{eqnarray}
and
\begin{eqnarray}\label{U9}
k_{0}^{2}=k_{3}^{2}+m_{q}^{2},
\end{eqnarray}
respectively.
\begin{figure}[hbt]
\includegraphics[width=8cm,height=6cm]{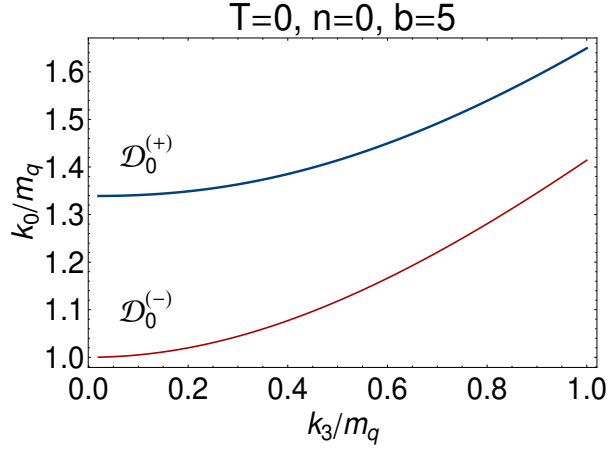}
\caption{(color online). The $k_{3}/m_{q}$ dependence of $k_{0}/m_{q}$ for positively charged (thick blue curve) and negatively charged (thin red curve) massive fermions in the LLL ($n=0$) for $b=eB/m_{q}^{2}=5$. They arise from the dispersion relations (\ref{U8}) for $q=+1$ and (\ref{U9}) for $q=-1$.
 }\label{fig9}
\end{figure}
\begin{figure}[hbt]
\includegraphics[width=8cm,height=6cm]{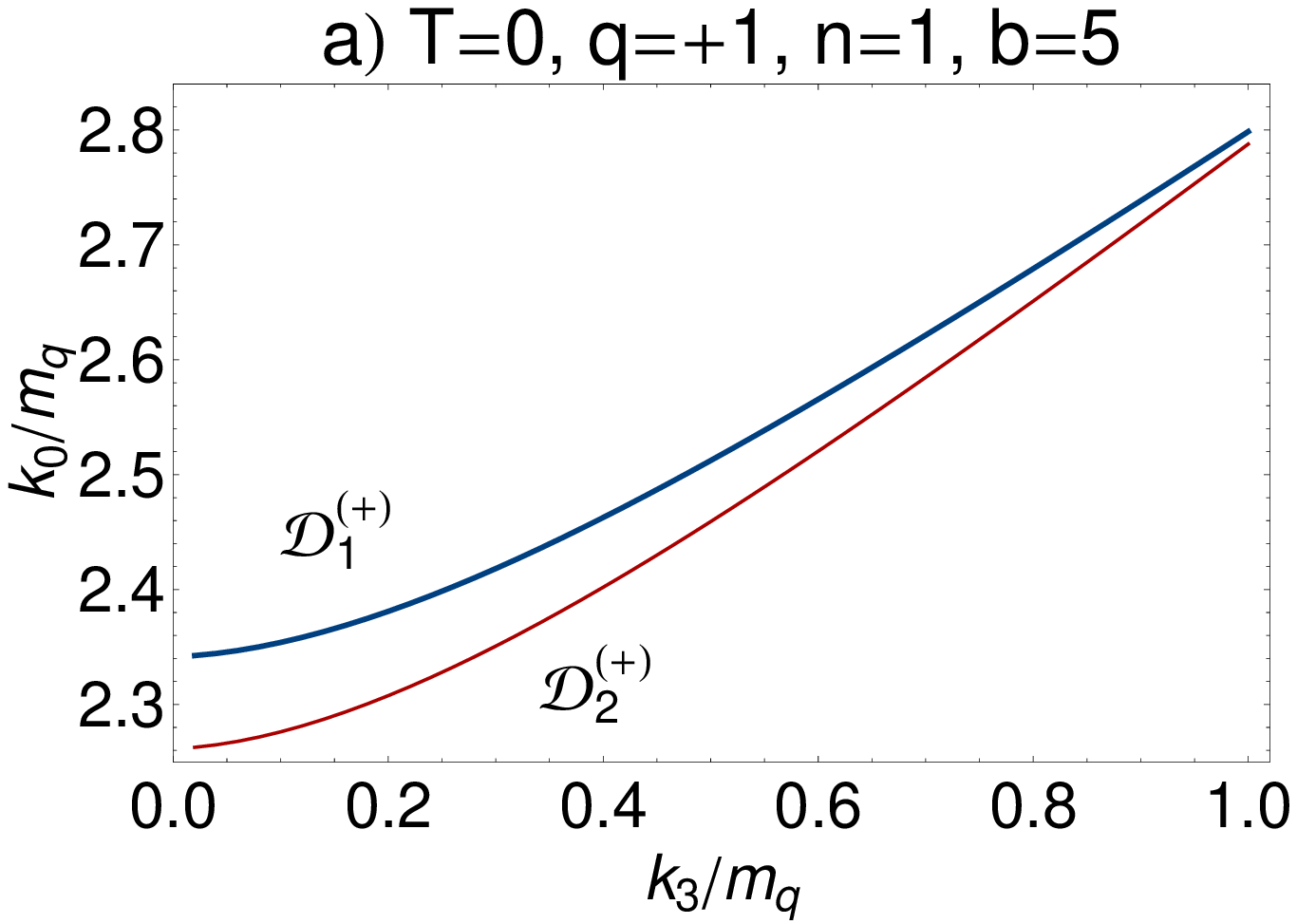}
\includegraphics[width=8cm,height=6cm]{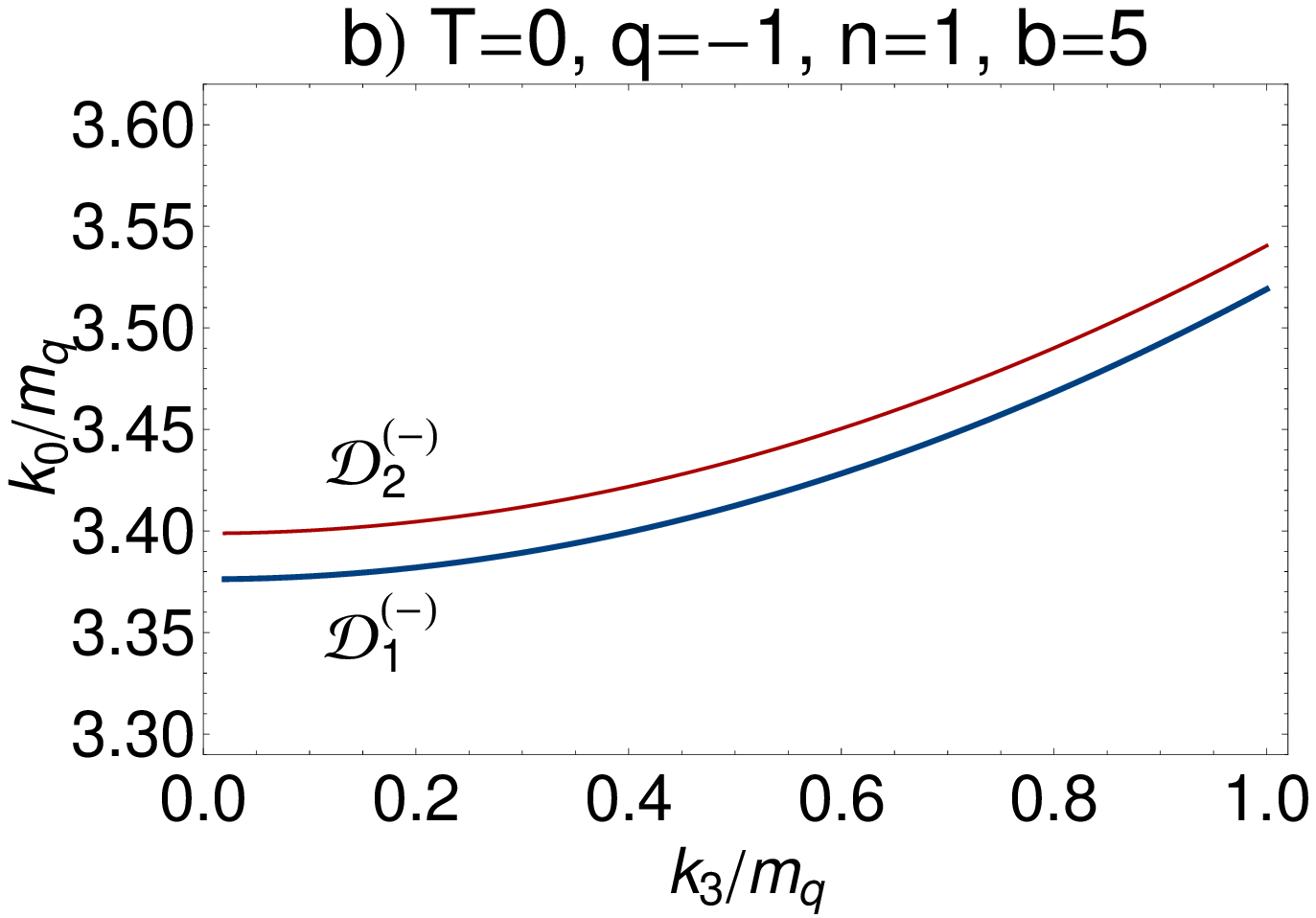}
\caption{(color online). The $k_{3}/m_{q}$ dependence of $k_{0}/m_{q}$ for positively (panel a) and negatively (panel b) charged massive fermions for $n=1$ and $b=eB/m_{q}^{2}=5$. The results arise from the dispersion relations (\ref{U15}) corresponding to ${\cal{D}}_{1}^{(\pm)}=0$ (thick blue curves) and ${\cal{D}}_{2}^{(\pm)}=0$ (thin red curves).}\label{fig10}
\end{figure}
\par
In Fig. \ref{fig9}, we have plotted the $k_{3}/m_{q}$ dependence of $k_{0}/m_{q}$ for these fermions. As we have argued in Sec. \ref{sec4}, two energy branches, appearing in this figure, correspond to positively charged fermions with positive spin (thick blue curve) and negatively charged fermions with negative spin (thin red curve). Since fermions with negative charges receive no one-loop self-energy correction in the LLL, their dispersion relation is identical with the dispersion relation of free fermions with  $k_{0}/m_{q}=1$ for $k_{3}=0$ (see the thin red curve in Fig. \ref{fig9}). On the other hand, as in the finite temperature case, the fact that the rest mass of positively charged fermions $k_{0}/m_{q}>1$, is related with the dynamical creation of mass in the presence of strong magnetic fields, where the system is dominated by LLL. As it turns out from Fig. \ref{fig9}, for each fixed $k_{3}/m_{q}$, the energy of positively charged fermions is larger than the energy of their negatively charged counterparts. This is also related with the same phenomenon of dynamical mass generation through the mechanism of magnetic catalysis \cite{miransky1995}.
\subsubsection{Special case: $T=0, n=1, m_{q}\neq 0$}
As aforementioned the one-loop corrected propagator of massive fermions in HLL is given in (\ref{D8})-(\ref{D11}). To determine the energy dispersion relations of these fermions at zero temperature, we have to solve $\mbox{det}({\ktildes}_{n}-m_{q}-{\Sigma}_{n}^{(q)})=0$ from (\ref{D4}). Alternatively, we can find the roots of the denominator ${\cal{D}}_{n}^{(q)}$ from (\ref{D10}) with  $a_{\pm}^{(q)}, b_{\pm}^{(q)}, c_{\pm}^{(q)}$ given in (\ref{U7}). Here, the denominator ${\cal{D}}_{n}^{(q)}$, being a quartic function in $d^{(q)}$, can be decomposed as
\begin{eqnarray}\label{U10}
{\cal{D}}_{n}^{(q)}={\cal{D}}_{1}^{(q)}{\cal{D}}_{2}^{(q)},
\end{eqnarray}
with
\begin{eqnarray}\label{U11}
{\cal{D}}_{i}^{(q)}&\equiv& (d^{(q)2}-d_{i}^{(q)2}), \qquad i=1,2,
\end{eqnarray}
and
\begin{eqnarray}\label{U12}
d_{1}^{(q)2}&\equiv& {\cal{A}}^{(q)}+ \left({\cal{A}}^{(q)2}+{\cal{B}}^{(q)}\right)^{1/2},\nonumber\\
d_{2}^{(q)2}&\equiv& {\cal{A}}^{(q)}- \left({\cal{A}}^{(q)2}+{\cal{B}}^{(q)}\right)^{1/2}.
\end{eqnarray}
The functions ${\cal{A}}^{(q)}$ and ${\cal{B}}^{(q)}$ in (\ref{U12}) are defined by
\begin{eqnarray}\label{U13}
{\cal{A}}^{(q)}&\equiv&a_{+}^{(q)}a_{-}^{(q)}-b_{+}^{(q)}b_{-}^{(q)}-c_{+}^{(q)}c_{-}^{(q)},\nonumber\\
{\cal{B}}^{(q)}&\equiv&-\left(a_{+}^{(q)2}-b_{+}^{(q)2}-c_{+}^{(q)2}\right)\left(a_{-}^{(q)2}-b_{-}^{(q)2}-c_{-}^{(q)2}\right),\nonumber\\
\end{eqnarray}
with $a_{\pm}^{(q)}, b_{\pm}^{(q)}, c_{\pm}^{(q)}$ from (\ref{U7}). The one-loop corrected fermion propagator in HLL is therefore given by
\begin{eqnarray}\label{U14}
{\cal{S}}_{n}^{(q)}(\tilde{k})=\frac{{\cal{M}}_{n}^{(q)}(\tilde{k})}{{\cal{D}}_{1}^{(q)}(\tilde{k})}+\frac{{\cal{M}}_{n}^{(q)}(\tilde{k})}{{\cal{D}}_{2}^{(q)}(\tilde{k})},
\end{eqnarray}
with ${\cal{M}}_{n}^{(q)}\equiv (d_{1}^{(q)2}-d_{2}^{(q)2})^{-1}{\cal{N}}_{n}^{(q)}$ and ${\cal{N}}_{n}^{(k)}$ defined in (\ref{D9}).
The energy dispersion relations, corresponding to the above two denominators, arise from ${\cal{D}}_{i}^{(q)}=0, i=1,2$, and read
\begin{eqnarray}\label{U15}
d^{(q)2}-d_{1}^{(q)2}&=&0,\nonumber\\
d^{(q)2}-d_{2}^{(q)2}&=&0,
\end{eqnarray}
with  $d^{(q)}$ from (\ref{U7}) and $d_{i}^{(q)}, i=1,2$ from (\ref{D12}).
\par
In Fig. \ref{fig10}, we have numerically determined the $k_{3}/m_{q}$ dependence of $k_{0}/m_{q}$, for $n=1$ and $b=eB/m_{q}^{2}=5$ as well as for $q=+1$ [Fig. \ref{fig10}(a)] and $q=-1$ [Fig. \ref{fig10}(b)]. The thick blue and thin red curves in Fig. \ref{fig10}, denote the solutions for ${\cal{D}}_{1}^{(\pm)}=0$ and ${\cal{D}}_{2}^{(\pm)}=0$, respectively. Here, in analogy to the finite temperature case, described in Sec. \ref{sec2a}, two separate energy branches appear for positively and negatively charged particles.  Their appearance is an indirect consequence of double spin degeneracy in HLL, in contrast to LLL, where the fermions have, depending on their charges, either positive or negative spins (see Fig. \ref{fig9}).
\par
Because of the similarity between the mechanism leading to two energy branches in Fig. \ref{fig2} at finite $T$ and zero $B$, with the case discussed in the present section at zero $T$ and nonzero $B$, the new fermionic excitations whose energy dispersions are demonstrated in Fig. \ref{fig10}, are referred to as \textit{cold magnetized plasminos}. Let us notice at this stage, that, whereas hot magnetized plasminos appear only in the limit of soft momenta $k_{3}\ll T$ and weak magnetic fields $eB\ll T^{2}$ (see Figs. \ref{fig7} and \ref{fig8}), as described in Sec. \ref{sec5}(a), cold magnetized plasminos appear in the presence of moderate magnetic fields for all positive momenta $k_{3}$, and even in the massive case.\footnote{Moderate magnetic fields are necessary, because in a strong magnetic field the fermionic system is solely dominated by the LLL, where, according to our previous arguments, no splitting occurs.}
\section{Summary and Conclusions}\label{sec6}
Plasminos are known to be collective excitations that appear in addition to normal fermionic modes in QED and QCD plasmas at finite temperature. Historically, they are shown to arise as one of the poles of the one-loop corrected fermion propagator at finite temperature $T$. The latter is shown to consist of two poles leading to two different energy dispersion relations. In the limit of vanishing fermionic mass and for small enough momenta of propagating fermions $k$, i.e. for $|\mathbf{k}|\ll T$, these energy dispersion relations belong to two fermionic modes with positive energies but opposite helicity to chirality ratios. The dynamical creation of additional collective modes in hot plasmas can be brought in close relation to the broken Lorentz invariance, induced by the preferred reference frame defined by the heat bath.
\par
A uniform magnetic field aligned in a fixed direction defines also a similar privileged reference frame. One of the consequences of such a frame is the appearance of certain anisotropies in the dynamics of fermions in the longitudinal and transverse directions with respect to the direction of the external magnetic field. Another consequence is the appearance of nontrivial collective modes in the spectrum of Dirac fermions in the presence of such a constant magnetic field. In the present paper, we explored the quasi-particle spectrum of cold and hot QED plasma, by evaluating the general structure of the one-loop corrected propagator of magnetized fermions at zero and nonzero temperatures, and looking for its poles.
\par
To this purpose, we first determined in Sec. \ref{sec2b}, the general structure of the free fermion propagator $S_{n}^{(q)}$ for positively ($q=+1$) and negatively ($q=-1$) charged fermions in the presence of a constant magnetic field $B$. In the momentum space, the free propagator is, in particular, given in terms of the Ritus momentum $\tilde{k}_{n}$, where $n$ labels the Landau levels [see (\ref{S41}) for free fermion propagator in the momentum space]. As it turns out, whereas HLL with $n\neq 0$ are characterized by a double spin degeneracy, in the LLL with $n=0$, the magnetized fermions have \textit{either} positive \textit{or} negative spins. This fact plays an important role in determining the correct spectrum of Dirac fermions in a constant magnetic field.
\par
In Sec. \ref{sec3}, we then determined the general structure of the one-loop self-energy of magnetized fermions, $\Sigma_{n}^{(q)}$, at zero and nonzero temperatures. The main results are presented in (\ref{A18}) for $T=0$ and in (\ref{A32}) for $T\neq 0$. A comparison between these two expressions shows that the isotropy between the components of the longitudinal momentum $\mathbf{k}_{\|}=(k_{0},0,0,k_{3})$ at zero temperature is removed at finite temperature. This is because of the aforementioned breaking of Lorentz invariance at finite $T$. We computed $\Sigma_{n}^{(q)}$ for $q=\pm 1$ in terms of certain coefficients, $A_{\pm}^{(q)}, B_{\pm}^{(q)}, C_{\pm}^{(q)}$ and $D^{(q)}$, and determined these coefficients separately for $T=0$ and $T\neq 0$ up to a number of integrations and a summation over Landau levels. To explore the Dirac spectrum for soft momenta $k_{3}\ll T$ and in the limit of weak magnetic fields $eB\ll T^{2}$, we also evaluated the above coefficients in a HTL expansion in a weak magnetic field.
\par
The one-loop corrected propagator of magnetized fermions ${\cal{S}}_{n}^{(q)}$ is then determined by combining the free fermion propagator $S_{n}^{(q)}$ and the one-loop fermion self-energy $\Sigma_{n}^{(q)}$ in the standard way [see Sec. \ref{sec4}]. Here, we have not distinguished between the zero and the finite temperature case. The main results are presented in (\ref{D6})-(\ref{D7}) for $n=0$ and in (\ref{D8})-(\ref{D11}) for $n\neq 0$.  We considered two special cases $(n=0, m_{q}\neq 0)$ as well as $(n\neq 0, m_{q}=0)$, and determined the spectrum of  fermionic modes by computing the eigenvectors of the numerators of ${\cal{S}}_{n}^{(q)}$. We showed that they are, in particular, eigenvectors of $\Sigma_{3}=\mbox{diag}(\sigma_{3},\sigma_{3})$, defined by the third Pauli matrix $\sigma_{3}$. Hence, in the LLL, the spectrum consists of positively (negatively) charged fermions with positive (negative) spins, while in HLL the expected double spin degeneracy occurs. In addition, massless fermions with $n\neq 0$ turned out to have well-defined left (negative) or right (positive) chiralities, as is described in (\ref{D26}), but they are not eigenstates of the helicity operator, in contrast to the standard $T\neq 0$ and $B=0$ case.
In the massless case, the general structure of ${\cal{S}}_{n}^{(q)}$ in HLL is presented in (\ref{D20})-(\ref{D22}).The appearance of two denominators ${\cal{D}}_{L}$ and ${\cal{D}}_{R}$ in (\ref{D20}) leads to two different energy branches, similar to the case of $T\neq 0$ and $B=0$ [see (\ref{S12})], although the corresponding collective modes have different properties.
\par
In Sec. \ref{sec5}, we numerically determined the energy dispersion relations of these magnetized collective modes at zero and nonzero temperatures. In Sec. \ref{sec5a}, we considered first the case of $T\neq 0$ and $B\neq 0$, and determined the $k_{3}/T$ dependence of $k_{0}/T$ in two different regimes of $k_{3}/T<0.2$ (soft momenta) and $k_{3}/T>0.2$ (hard momenta). To solve the energy dispersion relations arising from the poles of the corresponding propagators, we numerically determined the above mentioned coefficients $A_{\pm}^{(q)}, B_{\pm}^{(q)}, C_{\pm}^{(q)}$ and $D^{(q)}$ in these two regimes, for fixed values of $eB/T^{2}$ and for massive as well as massless fermions. In the LLL, only one energy branch arises in the whole regime of $k_{3}/T$. It belongs to positively or negatively charged fermions with positive or negative spins. In contrast, according to the results from Fig. \ref{fig7}, additional excitations, referred to as hot magnetized plasminos, appear in HLL, where left- and right-handed fermions have both positive and negative spins. In the limit of weak magnetic fields $eB\ll T^{2}$ and soft momenta $k_{3}\ll T$, these collective modes have different energies, while for larger values of $k_{3}/T$ they seem to have the same energy dispersions. This is in contrast to the results arising for nonzero $B$ and zero $T$, discussed in Sec. \ref{sec5b}. In this case, the results for $n=0$ and $m_{q}\neq 0$, are qualitatively the same as in the finite temperature case.  For $n\neq 0$, however, cold magnetized plasminos appear in the presence of moderate magnetic fields and for all positive momenta $k_{3}$, even in the massive case [see Fig. \ref{fig10}].
\par
Let us finally notice, that the group velocities of propagating collective modes can be determined from their energy dispersion relations for $(T\neq 0, B\neq 0)$, and might have applications, e.g., in the physics of heavy-ion collisions. At finite $T$ and zero $B$, the appearance of a minimum at some finite value of the momentum in the energy dispersion of plasminos, leads to a vanishing group velocity for the collective modes. The latter has been interpreted as the appearance of Van Hove singularities \cite{vanhove1953}, e.g. in the low mass dilepton production rate in the QCD plasma \cite{thoma2000}. The sharp structures arising in this quantity are known to provide a unique signature for the presence of deconfined collective quarks in the quark-gluon plasma \cite{pisarski1990}. It would be interesting to determine the dilepton production rate at high temperature and in the presence of moderate background magnetic fields. These are believed to be produced in early stages of heavy-ion collisions \cite{mclerran2007}, and, because of certain medium effects are assumed to be approximately time-independent \cite{rajagopal2014}. We will postpone these kind of phenomenological studies to our future publications.

\section{Acknowledgments}
The authors thank S. A. Jafari for valuable discussions on the application of the results  presented in this paper in the physics of three-dimensional graphene materials.
\begin{appendix}
\section{A useful formula}\label{appA}
\setcounter{equation}{0}
\par\noindent
In this appendix, we will compute the integral
\begin{eqnarray}\label{appA1}
I_{n,\ell}=\int_{-\infty}^{+\infty}dx_{1}e^{-ip'x_{1}}f_{n}^{+}(x_{1},k_{2})f_{\ell}^{+}(x_{1},p_{2}).
\end{eqnarray}
Here, according to (\ref{S29}),
\begin{eqnarray}\label{appA2}
f_{n}^{+}(x_{1},k_{2})=a_{n}e^{-\frac{(x_{1}-\ell_{q}^{2}k_{2})^{2}}{2\ell_{q}^{2}}}H_{n}\left(\frac{x_{1}-\ell_{q}^{2}k_{2}}{\ell_{q}}\right),\nonumber\\
\end{eqnarray}
with $\ell_{q}=|qeB|^{-1/2}$ and $a_{n}=(2^{n}n!\sqrt{\pi}\ell_{q})^{-1/2}$. To this purpose, we use the following representation of the Hermite polynomial $H_{n}(x)$
\begin{eqnarray}\label{appA3}
H_{n}(z)=\frac{n!}{2\pi i}\oint\frac{dt}{t^{n+1}}e^{-t^{2}+2tz},
\end{eqnarray}
arising directly from
\begin{eqnarray}\label{appA4}
e^{-t^{2}+2tz}=\sum\limits_{n}\frac{t^{n}}{n!}H_{n}(z).
\end{eqnarray}
Plugging (\ref{appA3}) in (\ref{appA1}), we arrive first at
\begin{eqnarray}\label{appA5}
\lefteqn{
\hspace{-0.8cm}I_{n,\ell}=\frac{a_{n}a_{\ell}n!\ell!}{(2\pi i)^{2}}\oint\frac{dt}{t^{n+1}}\frac{du}{u^{\ell+1}}
}\nonumber\\
&&\hspace{-0.2cm}\times e^{-(t^{2}+u^{2})}e^{-2\ell_{q}(tk_{2}+up_{2})}e^{-\frac{\ell_{q}^{2}(k_{2}^{2}+p_{2}^{2})}{2}}\nonumber\\
&&\hspace{-0.2cm}\times \int_{-\infty}^{+\infty} dx_{1}e^{-\frac{x_{1}^{2}}{\ell_{q}^{2}}+x_{1}\left(k_{2}+p_{2}+\frac{2(t+u)}{\ell_{q}}-ip'_{1}\right)}.
\end{eqnarray}
The integration over $x_{1}$ can be performed by quadratic completing the square, and performing the resulting Gaussian integration over $x_{1}$. This results in \begin{eqnarray}\label{appA6}
I_{n,\ell}&=&\frac{a_{n}a_{\ell}n!\ell!\sqrt{\pi}\ell_{q}}{(2\pi i)^{2}} e^{-[(k_{2}-p_{2})^{2}+{p'}_{1}^{2}]\frac{\ell_{q}^{2}}{4}}e^{-ip'_{1}(k_{2}+p_{2})\frac{\ell_{q}^{2}}{2}}\nonumber\\
&&\times J_{n\ell}(k_{2},p_{2},p'_{1}),
\end{eqnarray}
with
\begin{eqnarray}\label{appA7}
J_{n\ell}\equiv \oint\frac{dt}{t^{n+1}}\frac{du}{u^{\ell+1}}e^{2tu+t\ell_{q}(p_{2}-k_{2}-ip'_{1})+u\ell_{q}(k_{2}-p_{2}-ip'_{1})}.\hspace{-0.3cm}\nonumber\\
\end{eqnarray}
To determine $J_{n\ell}$, we will first perform the integral first over $t$ and then over $u$. Then, integrating first over $u$ and then over $t$, and comparing the two resulting expressions, we will eventually arrive at the generalized formula for $J_{n\ell}(k_{2},p_{2},p'_{1})$.
\par
The integration over $t$ in (\ref{appA7}) can be performed, using the Cauchy formula
\begin{eqnarray}\label{appA8}
\oint\frac{dt}{t^{n+1}}{\cal{F}}(t)=\frac{2\pi i}{n!}\frac{d^{n}}{dz^{n}}{\cal{F}}(z)\bigg|_{z=0},
\end{eqnarray}
which leads to
\begin{eqnarray}\label{appA9}
\oint \frac{dt}{t^{n+1}}e^{tA(u)}=\frac{2\pi i}{n!}A^{n}(u).
\end{eqnarray}
Setting $A(u)\equiv 2u+\ell_{q}(p_{2}-k_{2}-ip'_{1})$, plugging the resulting expression in (\ref{appA7}), using the binomial series identity
\begin{eqnarray}\label{appA10}
\lefteqn{\hspace{-0.8cm}
(2u+\ell_{q}(p_{2}-k_{2}-ip'_{1}))^{n}}
\nonumber\\
&&\hspace{-0.9cm}=\sum_{r=0}^{n}
\left(
\begin{array}{c}
n\\
r
\end{array}
\right)
(2u)^{r}(\ell_{q}(p_{2}-k_{2}-ip'_{1}))^{n-r},
\end{eqnarray}
and eventually replacing $e^{u\ell_{q}(k_{2}-p_{2}-ip'_{1})}$ with
\begin{eqnarray}\label{appA11}
e^{u(k_{2}-p_{2}-ip'_{1})}=\sum_{j=0}^{\infty}\frac{u^j (k_{2}-p_{2}-ip'_{1})^{j}}{j!},
\end{eqnarray}
we arrive first at
\begin{eqnarray}\label{appA12}
\lefteqn{
J_{n\ell}=\frac{2\pi i}{n!}\sum_{r=0}^{n}\sum_{j=0}^{\infty}
\frac{2^{r}\ell_{q}^{n-r+j}}{j!}
\left(
\begin{array}{c}
n\\
r
\end{array}
\right)
}\nonumber\\
&&\times (p_{2}-k_{2}-ip'_{1})^{n-r}(k_{2}-p_{2}-ip'_{1})^{j}\nonumber\\
&&\times \oint \frac{du}{u^{\ell-r-j+1}}.
\end{eqnarray}
Integrating then over $u$,
\begin{eqnarray}\label{appA13}
\oint \frac{du}{u^{\ell-r-j+1}}=2\pi i\delta_{r+j,\ell},
\end{eqnarray}
we obtain
\begin{eqnarray}\label{appA14}
J_{n\ell}&=&\frac{(2\pi i)^{2}}{n!\ell!}\ell_{q}^{n+\ell}(p_{2}-k_{2}-ip'_{1})^{n}(k_{2}-p_{2}-ip'_{1})^{\ell}\nonumber\\
&&\times K_{n\ell}(k_{2},p_{2},p'_{1}),
\end{eqnarray}
with
\begin{eqnarray}\label{appA15}
K_{n\ell}=z^{-n}{\cal{U}}_{\ell-n+1}^{-n}\left(z\right),
\end{eqnarray}
and $z\equiv\frac{\ell_{q}^{2}}{2}[(k_{2}-p_{2})^{2}+{p'}_{1}^{2}]$. Here,  ${\cal{U}}_{a}^{b}(z)$ is the confluent hypergeometric function of the second kind \cite{mathworld}, defined by
\begin{eqnarray}\label{appA16}
{\cal{U}}_{b-a+1}^{-a}(x)\equiv \sum_{j=0}^{a}(-1)^{j}~j!
\left(
\begin{array}{c}
a\\
j
\end{array}
\right)
\left(
\begin{array}{c}
b\\
j
\end{array}
\right)x^{a-j}.
\end{eqnarray}
Let us now consider (\ref{appA7}), and check what would happen if we computed $J_{n\ell}$ by integrating first over $u$ and then over $t$. In the above case, where  the integration over $u$ is performed after the integration over $t$, we have, according to (\ref{appA13}), $\ell=r+j$. For $j\geq 0$, arising from (\ref{appA12}), we get therefore $\ell\geq r$. On the other hand, it is clear from (\ref{appA12}) that $r\leq n$. Combining these results, we arrive at $n\leq \ell$. This fixes the upper limit in the summation over $r$, which is given by $n$ equal to $\mbox{min}(n,\ell)$. If we computed $J_{n\ell}$ by integrating first over $u$ and then over $t$, we would arrive at a summation over $r$ from $r=0$ to $r=\ell$, and, according to the above argument, $\ell$ would be equal to $\mbox{min}(n,\ell)$. In other words, the most general expression for $J_{n\ell}$ is given by (\ref{appA14}) with $n$ and $\ell$ replaced by
 $m\equiv \mbox{min}(n,\ell)$ and $M\equiv \mbox{max}(n,\ell)$. Plugging now this expression for $J_{n\ell}$ in (\ref{appA6}), the most general expression for $I_{n\ell}$ reads
\begin{eqnarray}\label{appA17}
I_{n,\ell}&=&A_{mM} \ell_{q}^{m+M}e^{-ip'_{1}(k_{2}+p_{2})\frac{\ell_{q}^{2}}{2}}\nonumber\\
&&\times(p_{2}-k_{2}-ip'_{1})^{m}(k_{2}-p_{2}-ip'_{1})^{M}\nonumber\\
&&\times e^{-\frac{z}{2}}z^{-m}{\cal{U}}_{M-m+1}^{-m}(z),
\end{eqnarray}
where $A_{n\ell}=(2^{n+\ell}n!\ell!)^{-1/2}$, $z=\frac{\ell_{q}^{2}}{2}[(k_{2}-p_{2})^{2}+{p'}_{1}^{2}]$,  $m= \mbox{min}(n,\ell)$ and $M=\mbox{max}(n,\ell)$. In Sec. \ref{sec3}, the above relation will be used to evaluate the integration over $x_{1}$ and $y_{1}$ in (\ref{A6}) with ${\cal{N}}_{n\ell}^{(q)}(x_{1},y_{1};k_{2},p_{2})$ from (\ref{A10}). Here, we present only the result for one typical combination,
\begin{eqnarray}\label{appA18}
I_{n,\ell-1}^{ }I_{n,\ell-1}^{\dagger}=\frac{1}{n!(\ell-1)!}e^{-z}z^{M-m}[{\cal{U}}_{M-m+1}^{-m}(z)]^{2},\nonumber\\
\end{eqnarray}
with $z=\frac{\ell_{q}^{2}}{2}[(k_{2}-p_{2})^{2}+{p'}_{1}^{2}]$, $m=\mbox{min}(n,\ell-1)$ and $M=\mbox{max}(n,\ell-1)$. All the other integrals appearing in (\ref{A6}) with ${\cal{N}}_{n\ell}^{(q)}$ from (\ref{A7}) are performed in the same way.
\end{appendix}

\end{document}